\documentclass[12pt]{article}
\usepackage[utf8]{inputenc}
\usepackage[margin=2.5cm]{geometry}

\usepackage{amsmath,amssymb}
\usepackage{natbib}

\usepackage{authblk}

\usepackage{hyperref}

\usepackage{tabularx}
\usepackage{xcolor}

\usepackage{float}
\restylefloat{table}

\usepackage{comment}
\usepackage{mathtools}
\usepackage{bm}
\usepackage{lscape}
\usepackage{graphicx}
\usepackage{algorithm}
\usepackage{algorithmic}
\usepackage{subfig}
\usepackage[normalem]{ulem}

\newcommand{\by}{\bm{y}}

\usepackage{algorithm}
\newcommand{\bX}{\bm{X}}

\newcommand{\bS}{\bm{S}}
\newcommand{\bI}{\bm{I}}

\newcommand{\bC}{\bm{C}}

\newcommand{\bV}{\bm{V}}
\newcommand{\bM}{\bm{M}}

\newcommand{\bu}{\bm{u}}
\newcommand{\bbeta}{\bm{\beta}}

\newcommand*\mystrut[1]{\vrule width0pt height0pt depth#1\relax}
\DeclareMathOperator*{\argmin}{arg\,min}
\graphicspath{ {../} }

\title{Sparse Temporal Disaggregation}
\author[1]{Luke Mosley}
\author[2]{Idris Eckley}
\author[2]{Alex Gibberd}
\affil[1]{STOR-i Centre for Doctoral Training, Department of Mathematics and Statistics, Lancaster University, United Kingdom}
\affil[2]{Department of Mathematics and Statistics, Lancaster University,
United Kingdom}
\date{}

\begin{document}
\maketitle

\begin{abstract}
    Temporal disaggregation is a method commonly used in official statistics to enable high-frequency estimates of key economic indicators, such as GDP. Traditionally, such methods have relied on only a couple of high-frequency indicator series to produce estimates. However, the prevalence of large, and increasing, volumes of administrative and alternative data-sources motivates the need for such methods to be adapted for high-dimensional settings. 
    In this article, we propose a novel sparse temporal-disaggregation procedure and contrast this with the classical Chow-Lin method. We demonstrate the performance of our proposed method through simulation study, highlighting various advantages realised. We also explore its application to disaggregation of UK gross domestic product data, demonstrating the method's ability to operate when the number of potential indicators is greater than the number of low-frequency observations.
\end{abstract}


\section{Introduction}

Understanding short-term dynamics of headline macroeconomic variables has become increasingly important in the diverse and digital modern economy we witness today. National Statistics Institutes (NSI) such as the UK's Office for National Statistics (ONS) are increasingly asked to provide more frequent publication of Gross Domestic Product (GDP) or annual industry and household surveys. The typical motivation for providing these more frequent measures is to paint a more timely picture of the current economic health of a country. 
In recent years, several reports \citep[e.g.][]{bean2016independent,pfeffermann2015methodological,euandun2017} have reviewed how official statistics are gathered, and motivated attempts to move away from the traditional survey-centric approach. In particular, there has been a push towards adopting alternative and administrative data-sources that are readily available and measure numerous processes in a timely and reliable manner. \citet{jarmin2019evolving} emphasises that traditional data-sources will remain \emph{critical}, however these will be designed to \emph{complement and improve the measure capabilities} of new higher frequency sources. While a sense of awareness for these new data-sources is being established, for example using VAT return data as an indicator of GDP, or scanned price data in supermarkets as an indicator of inflation, the statistical methodology to calibrate insights from these sources remains under-developed.

This article is motivated by the challenging task of performing high frequency disaggregation for UK national GDP, moving from a quarterly to a monthly resolution. We wish to make use of a variety of monthly data sources, referred to as indicator series, that are believed to indicate the short-term movements of GDP. This information is used to construct a monthly estimate of GDP that will follow similar short-term dynamics to the indicator series, however remain temporally consistent with quarterly observations.

For this task, there are a considerable number of indicator series that one may wish to use. In our application we consider survey based series, such as the monthly business survey (MBS) in both services and production, alongside VAT data, retail sales indices, and several novel indicators such as traffic flows at ports and on roads. In total, we consider 97 indicators, all of which are collected at a monthly frequency. Given the significant interest in fast measurements of economic activity, the ONS has developed a monthly GDP index, and published this statistic since May 2018. This index is constructed using monthly information from the Index of Services, Index of Production and Index of Construction. For details see \cite{ONS2021}. Even though a monthly index exists in this case, there is still great interest in performing temporal disaggregation, the reasons are threefold. Firstly, the monthly index is an output based measure, however economists may also be interested in both expenditure and income based estimates. Since, temporal disaggregation can be applied to any output stream, either expenditure or income based measures could be used. The resulting high-frequency estimate can thus complement the existing output based index. Secondly, due to the construction of the index, publication lags the period of measurement, an issue common to most economic statistics. However, temporal disaggregation can be used to find indicators that are relevant and updated more frequently, potentially enabling the estimation of the output statistic at a more frequent rate than is traditionally reported. This way, we are able to evaluate higher frequency estimates of the economic activity itself, rather than an index, and obtain detailed short-term information at the end of the sample.
NSIs are actively developing so-called fast-indicators for exactly this purpose and in this article we consider several of these in the form of traffic data. Finally, one of the key issues surrounding the fast release of data is in understanding the associated short-term movements. To this end, temporal disaggregation using interpretable indicator series can provide insight by highlighting which indicators are driving movement. 

Early work on temporal disaggregation by \cite{lisman1964derivation} and \cite{denton1971adjustment} set up a constrained quadratic optimisation problem trying to minimise the difference between revised and original series. Generalisations of these approaches were made by \cite{wei1990disaggregation} and \cite{guerrero1995recursive} who used autoregressive moving average (ARIMA) models to interpolate data in the absence of high frequency indicator information. \cite{chow1971best} were the first to suggest a generalised least squares (GLS) regression utilising a set of explanatory variables recorded at the desired high frequency. Their well-established approach is still used today by several NSIs to perform disaggregation and compile national accounts \citep{ess2018}. Extensions of \cite{chow1971best} include \cite{fernandez1981methodological} and \cite{litterman1983random} who allow for complex, non-stationary error structures, \cite{mitchell2005indicator} who derive an approximation using logarithms, \cite{proietti2006temporal} who study a dynamic state-space generalisation and \cite{di1990estimation} who propose a multivariate extension to perform regional temporal disaggregation. See \cite{pavia2010survey} for an extensive literature review of disaggregation procedures and \cite{chen2007empirical} for an empirical comparison using 60 series of annual data from national accounts. More recent work of temporal disaggregation includes \cite{labonne2020temporal} who derive monthly estimates of business sector output in the UK from rolling quarterly VAT returns by employing an unobserved components model to accommodate measurement noise. 

A related but distinct field of work to disaggregation in the econometrics time series literature is that of nowcasting (or real-time forecasting). Nowcasting is an essential tool to overcome the so-called `ragged-edge' problem, namely publication delays of headline variables, which cause missing values at the end of sample. Popular methods include mixed-data sampling (MIDAS) regression \citep{ghysels2004midas}, dynamic factor models \citep{banbura2011look} and mixed-frequency vector autoregression models \citep{koop2018regional}. These methods can be used, for example, to predict the next quarter of UK GDP using monthly indicators, to prevent long waits for publication lags. As previously mentioned, our focus is on an alternative and quite distinct,  unsupervised learning task, where we attempt to estimate unobserved monthly GDP between each quarterly sample. As such, nowcasting within a temporal disaggregation context such that we can obtain nowcasts at the higher frequency remains an interesting avenue for future research. 

\cite{chow1971best} and its extensions have proven to work well when only a couple of indicator series are used for the disaggregation task. However, with the integration of alternative and administrative data sets, the collection of indicator series that can potentially provide important information on the output is vast and we find ourselves operating in a so-called \emph{high-dimensional} setting. In fact, we naturally find ourselves in this setting when we consider the GDP disaggregation challenge previously outlined (97 series over 50 quarters). In this case, classical methods for temporal disaggregation demonstrably fail. It is precisely this gap in the temporal disaggregation literature that we seek to address: how to build a robust and interpretable methodology that can accurately disaggregate key survey based statistics by utilizing large numbers of alternative and administrative data sources. 

Methodologically, our key contribution is to establish a regularised M-estimation framework that extends the Chow-Lin approach, we refer to this framework as \emph{Sparse Temporal Disaggregation} (spTD). A key aspect of the method is the incorporation of a penalty function on the regression parameters which operates alongside the usual GLS cost function. These regularisation functions are able to take a variety of forms, and help ensure the estimator is stable in high-dimensional settings \citep{buhlmann2011statistics}. To illustrate the effectiveness of our methodology we focus on the popular $\ell_1$ penalty (LASSO) \citep{tibshirani1996regression}. This has the twin benefits of computational robustness due to convexity, and interpretation due to its ability to produce sparse sets of regression coefficients. In addition, $\ell_1$ penalisation has rapid, robust algorithms available to compute parameter estimates over a range of sparsity levels \citep{efron2004least}. 

Alternative approaches to dealing with high-dimensionality include \cite{angelini2006interpolation} and \cite{proietti2020nowcasting}, and are principally based on dynamic factor models. Here, factor estimation methods such as principal components analysis are used  to reduce the dimensionality of the large indicator data set before model fitting by combining indicators into a limited number of common factors. Specifically, the large indicator set can be decomposed into the sum of two mutually orthogonal unobserved components. First, the common component that drive the co-movements of the high dimensional indicators through a few common factors, and second, the idiosyncratic component that arises from features that are specific to an individual indicator variable such as measurement error. Standard temporal disaggregation procedures can then be performed using these estimated common factors as predictors. This approach has the joint benefits of being able to handle ragged edge data via filtering techniques when computing factor estimates, and is able to asymptotically remove the idiosyncratic variation in the indicators when estimating the disaggregate series. 
Although such approaches can be effective, the factor model approach possesses difficulties regarding specification and interpretability. There exists uncertainty in how the appropriate factor estimation method is selected, and in the number of factors used \citep{kuzin2013pooling}. A more considerable issue is the loss of interpretation into the individual effects of indicators, as it is the factors or principal components, that are retained instead of original predictors. We therefore tackle the high-dimensional temporal disaggregation problem from the variable selection viewpoint; as our regularised M-estimation framework is able to simultaneously select the relevant indicators and estimate their impact. To our knowledge, this is the first work to tackle the disaggregation problem in this way.

Through extensive simulation studies we investigate the performance of our approach in estimating high frequency disaggregated series in both standard and high-dimensional scenarios. We also compare against the established Chow-Lin method in standard dimensional settings. Finally, we illustrate how the method can be used to solve the high-dimensional GDP disaggregation problem. As we will demonstrate later, our estimated model not only aligns with economic intuition, but also achieves better tracking of the published monthly GDP index when compared against the Chow-Lin method.
The paper is structured as follows. Section 2 introduces the statistical framework for temporal disaggregation and explores shortcomings of existing methods in high dimensions. In Section 3 we establish a sparse modelling framework for high-dimensional temporal disaggregation, before providing details of an estimation strategy and tuning procedure for $\ell_1$ penalisation. Section 4 highlights the advantages of our approach through a simulation study, while Section 5 implements our method on real GDP data. 

\section{Background} 

\begin{figure}
    \begin{minipage}{.33\linewidth}
        \centering
        \subfloat[]{\makebox{\includegraphics[width=\linewidth]{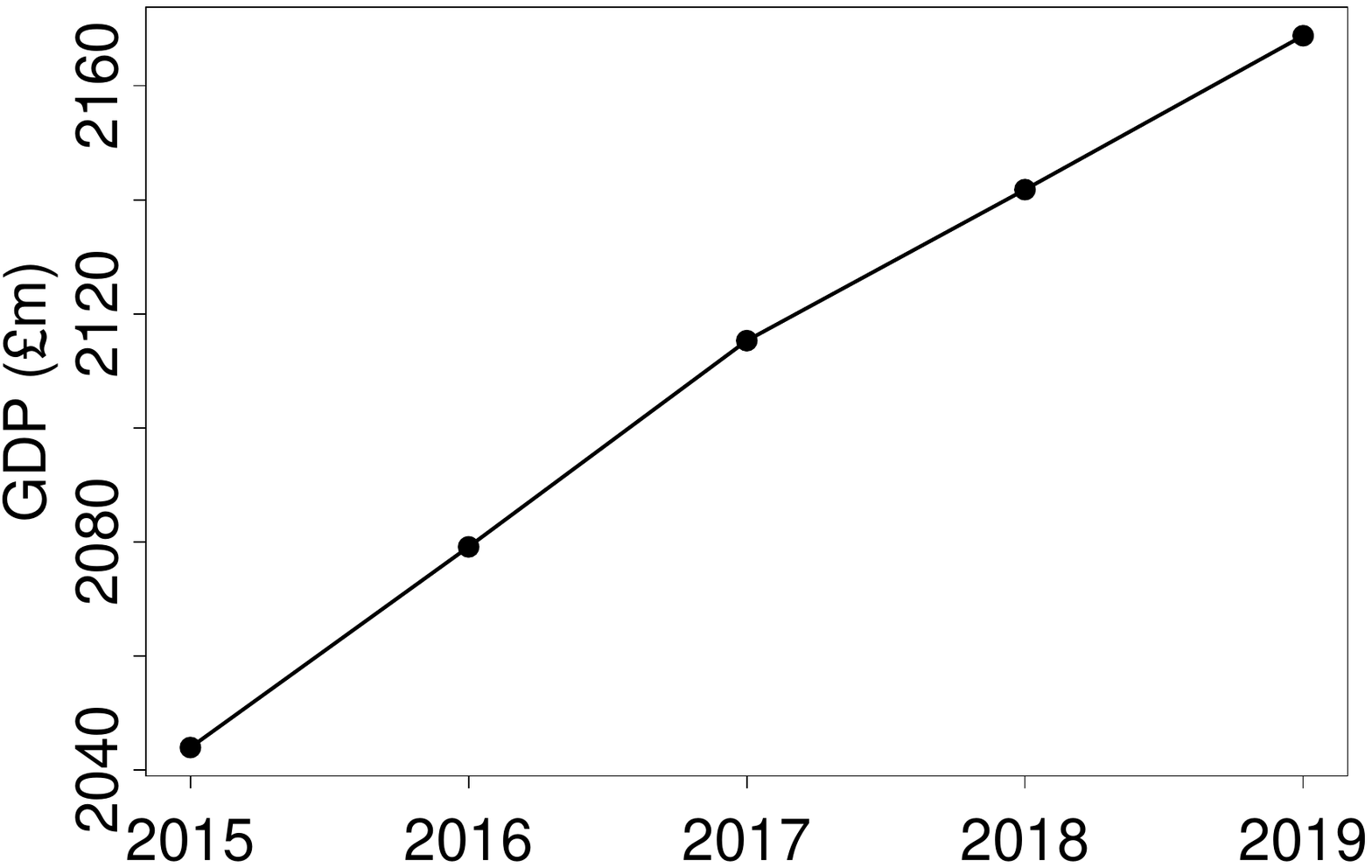}}}
    \end{minipage}
    \begin{minipage}{.32\linewidth}
        \centering
        \subfloat[]{\makebox{\includegraphics[width=\linewidth]{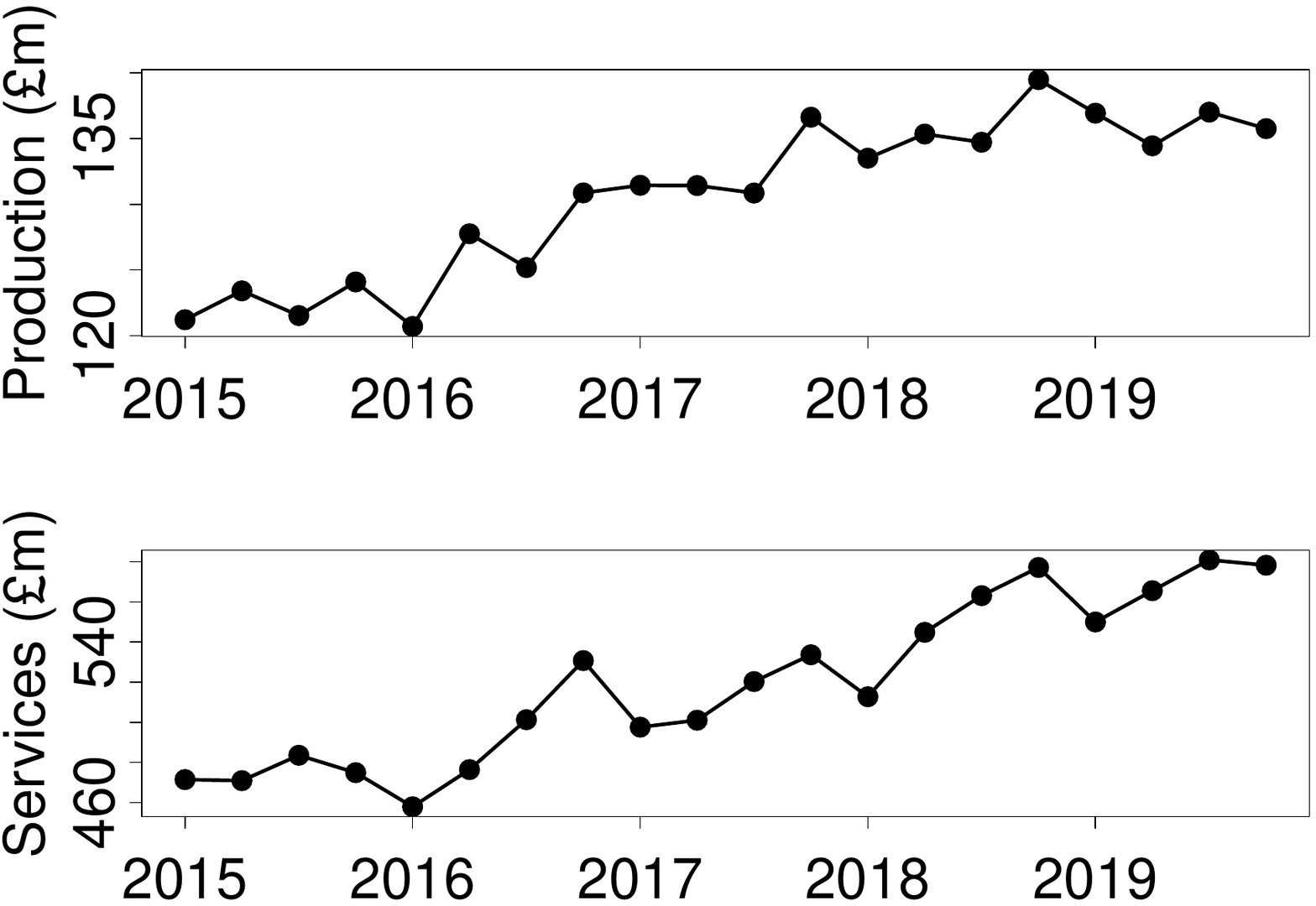}}}
    \end{minipage}
     \begin{minipage}{.32\linewidth}
        \centering
        \subfloat[]{\makebox{\includegraphics[width=\linewidth]{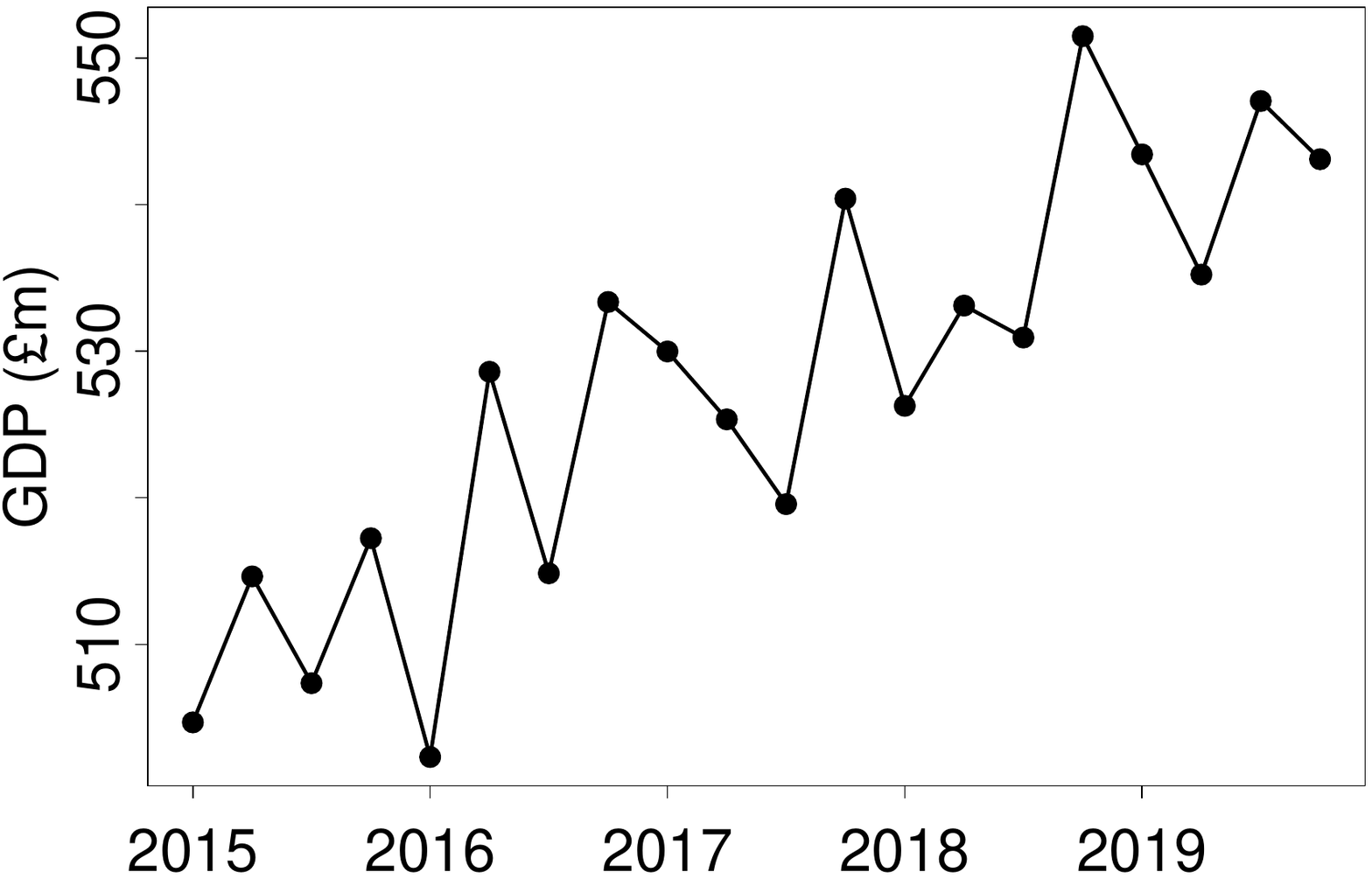}}}
    \end{minipage}
    \caption{\label{fig:TD example} Annual to quarterly temporal disaggregation of GDP from 2015 to 2019 using Business Surveys in Production and Services as quarterly indicator series. (a) Annual GDP; (b) quarterly indicator series; (c) quarterly GDP estimate.  }
\end{figure}

The temporal disaggregation problem is perhaps best viewed from the perspective of a practitioner working in official statistics. The practitioner observes a low frequency data-stream of a key statistic (such as GDP) that they wish to disaggregate to a higher frequency. For the sake of clarity, we can assume that this low frequency data-stream is at an annual scale and denote this as the vector $\by_a \in\mathbb{R}^n$, containing $n$ yearly observations.  The practitioner seeks to construct a disaggregated version of $\by_a$ at a sub-annual time resolution. In addition, it is desirable that the disaggregated series be temporally consistent during each year and not contain spurious jumps between years. Consider, by way of example, a setting where one seeks to produce a quarterly version and denote this by $\by_q\in\mathbb{R}^m$, where $m=4n$ in the case of annual-to-quarterly disaggregation. Note the subscript $q$ to denote a quarterly series and $a$ to represent an annual series. The disaggregation challenge lies in developing a principled approach to interpolate or distribute  between each observed annual data point. A common approach to achieve this goal is to use indicator series recorded at the desired high frequency that are believed to indicate the intra-annual movements. A simple example of temporal disaggregation is shown in Figure \ref{fig:TD example} whereby annual GDP data is disaggregated to a quarterly level by making use of business surveys from production and services industries recorded quarterly. We denote the set of high frequency indicator series as the matrix $\bX_q\in\mathbb{R}^{m\times p}$ where each column is a quarterly time series representing one of $p$ indicators.

We now formulate the canonical temporal disaggregation problem and outline the method proposed by \cite{chow1971best}. To construct the unobserved quarterly time series $\by_q$, the following regression model is assumed at the quarterly frequency
\begin{equation} 
\label{eq:regression}
\by_q = \bX_q \bbeta + \bu_q \, , 
\end{equation}
where $\bbeta\in\mathbb{R}^p$ is the vector of regression coefficients to be estimated, and $\bu_q\in\mathbb{R}^m$ is a residual vector. Note that the matrix $\bX_q$ may contain deterministic terms such as constants and trends. The residual vector $\bu_q$ is mean-zero and has covariance matrix $\bV_q$. \cite{chow1971best} assume the data generating process of $\bu_q$ follows a first-order autoregressive process $u_t = \rho u_{t-1} + \epsilon_t$ with $\epsilon_t \sim N(0,\sigma^2)$ and $|\rho|<1$. This assumption of a stationary residual process allows a co-integrating relationship between $\by_q$ and $\bX_q$ when they are non-stationary, a likely scenario with economic time series. Thus, the regression coefficient $\bbeta$ measures both the long- and short-run effect of $\bX_q$ on $\by_q$. The resulting covariance matrix $\bV_q$ has the well-known Toeplitz form 
\begin{equation*} \bV_q = \frac{\sigma^2}{1-\rho^2}
\begin{pmatrix}
1 & \rho & \dots & \rho^{m-1} \\
\rho & 1 & \dots & \rho^{m-2} \\
\vdots & \vdots & \ddots & \vdots \\
\rho^{m-1} & \rho^{m-2} & \dots & 1
\end{pmatrix} \, ,
\end{equation*}
containing two unknown parameters $\rho$ and $\sigma^2$ that must be estimated.

As the dependent variable in (\ref{eq:regression}) is unobserved, the regression is pre-multiplied by the $n \times m$ aggregation matrix
\begin{equation}
    \bC = \bI_n \otimes (1,1,1,1) \, ,
    \label{eq:aggregation}
\end{equation}
to obtain the observable annual counterpart of (\ref{eq:regression}) given by 
\begin{equation}
    \by_a = \bX_a \bbeta + \bu_a \, ,
    \label{eq:regression2}
\end{equation}
where $\by_a = \bC\by_q$, $\bX_a = \bC \bX_q$ and $\bu_a = \bC\bu_q$ with covariance matrix $\bV_a = \bC\bV_q\bC^\top$. Without loss of generality, we assume we are dealing with flow data where four quarterly figures in a year must sum to their corresponding observed yearly figure, hence the vector of ones $(1,1,1,1)$ appearing in the aggregation matrix. For alternative aggregations see the reviews by \cite{quilis2018temporal} and \cite{sax2013temporal}. 

The regression equation in (\ref{eq:regression2}) is now fully observable and can be solved using standard techniques. As autocorrelation is present in the AR(1) residual process, this is dealt with by using the GLS estimator:
\begin{align}
\label{eq:betaGLS}
    \hat{\bbeta} &= \argmin_{\bbeta \in \mathbb{R}^p} \bigg\{ \left\lVert\bV_a^{-1/2}(\by_a - \bX_a\bbeta)\right\rVert_2^2\bigg\} \\
    &= (\bX_a^\top\bV_a^{-1}\bX_a)^{-1}\bX_a^\top\bV_a^{-1}\by_a \, . \nonumber
\end{align}

The estimator for $\bbeta$ is conditional on the covariance matrix $\bV_a$ which contains the unknown parameters $\rho$ and $\sigma^2$. In fact, the $\sigma^2$ parameter can be factored out of the covariance matrix $\bV_a$ and will cancel itself out in the estimator $\hat{\bbeta}$. Thus, only $\rho$ has to be estimated. It follows that $\hat{\bbeta}$, conditional on $\hat{\rho}$, is a feasible GLS estimator of $\bbeta$. Originally, \citet{chow1971best} proposed an iterative procedure to infer the $\rho$ parameter from the observed autocorrelation of the aggregated residuals, however, it has since been shown this method is not very reliable and suffers when sharp movements are present in the series \citep{chen2007empirical}. \citet{bournay1979reflexions} provide a more intuitive method by first estimating $\hat{\bbeta}$ and $\hat{\bV}_a$ via profile-likelihood maximisation, before searching for the autoregressive parameter over the stationary range of $\rho \in (-1,1)$. 

The feasible GLS estimator $\hat{\bbeta}$ conditioned on the maximum likelihood estimate of $\rho$ is used to construct the target quarterly unobserved series $\by_q$. \cite{chow1971best} show the optimal (best linear unbiased) solution is given by
\begin{equation}
    \hat{\by}_q = \bX_q \hat{\bbeta} + \bV_q\bC\bV_a^{-1}\big(\by_a - \bX_a \hat{\bbeta}\big) \, .
\end{equation}
The first part of this expression $\bX_q\hat{\bbeta}$ is the conditional expectation of $\by_q$ given $\bX_q$. The second is an estimate of the quarterly residual $\bu_q$, obtained by disaggregating the observed annual residuals $\by_a-\bX_a\hat{\bbeta}$ to ensure temporal consistency between the estimates $\hat{\by}_q$ and observations $\by_a$.

\subsection{Shortcomings of the Chow-Lin procedure}

Despite the popularity of \cite{chow1971best} to compile national accounts across Europe \citep{ess2018}, the method runs into several shortcomings when operating in data-rich environments NSIs now find themselves in. Below we outline several of these, prior to introducing our regularised temporal disaggregation approach in Section \ref{sec:sparse_TD}. 

In moderate and high dimensions, the behaviour of the Chow-Lin procedure faces several statistical challenges: a) excessive variance in $\hat{\bbeta}$ impacts interpretation in the weight given to indicator variables; b) unreliable estimation of the AR(1) parameter and variance ($\rho,\sigma$) leads to poor performance in identifying $\bV_q$, and thus the high frequency series; c) interpretation into which indicator series are most relevant is hampered since all indicators will be included in the model by default. To date, there has been limited research to answer this collection of shortcomings in the temporal disaggregation literature. 

Delving into the details of these challenges a little further,  we see the GLS estimator of (\ref{eq:betaGLS}) takes the form:
\begin{equation}
    \hat{\bbeta} = \big[\underbrace{\bX_a^\top\bV_a^{-1}\bX_a}_{\bM}\big]^{-1}\bX_a^\top\bV_a^{-1}\by_a \, .
    \label{eq:GLSestimatorsolution}
\end{equation}
This solution is only uniquely identified if the matrix $\bM$ highlighted in (\ref{eq:GLSestimatorsolution}) is of full rank (invertible). This fails to be the case in high dimensional scenarios as $\text{rank}(\bM)=n$ with $p>n$. In moderate-dimensional scenarios, $\bM$ may have many eigenvalues close to zero, leading to estimates $\hat{\bbeta}$ with high variance and inevitably results in overfitting.
In such settings, \citet{ciammola2005temporal} noted the poor performance of Chow-Lin in estimating the AR(1) parameter $\rho$. Further, a reliable estimate of $\sigma$ is important for quantifying the uncertainty associated with the estimate
$\hat{\bbeta}$. \citet{chow1971best} use the standard estimator: $\hat{\sigma}^2 = || \bV_a^{-1/2}(\by_a-\bX_a\hat{\bbeta})||^2_2/n$, however, it has been shown that even in the classical $n>p$ setting  such an estimator is biased downwards \citep{yu2019estimating}.

Finally, in addition to the challenges of dimensionality, the \citet{chow1971best} approach offers limited model interpretation as no model selection is performed. Since it is sometimes difficult to collect high frequency indicator series, it can be beneficial to identify which are of most importance when monitoring the economic phenomenon of interest. This way, when future estimates are made, irrelevant indicators can be avoided, reducing model complexity and the cost of creating outputs by a great deal. 

\section{Sparse Temporal Disaggregation}\label{sec:sparse_TD}
In this section we introduce a novel sparse temporal disaggregation (spTD) method. Our approach seeks to provide \emph{robust} and \emph{reliable} solutions, with a view to resolving the aforementioned shortcomings of current temporal disaggregation methodologies. In Section 3.1 we provide a general regularised M-estimation framework that allows us to encompass a variety of penalty functions in the Chow-Lin regression framework to accomplish temporal disaggregation in moderate and high dimensions. In Section 3.2 we draw attention to a specific penalty of the general framework, namely the $\ell_1$ penalty and provide a detailed description of the estimation strategy and tuning procedure. Section 3.3 concludes with a discussion on how to extend our model when the indicator series are highly correlated.
We assume throughout that the indicator set $\bX_q$ is fully observed up to the latest low frequency observation and hence does not contain a `ragged edge' structure. It would  be possible to extrapolate indicator variables forward if they contain end-of-sample missingness by assuming the indicators follow a certain time series model but we do not consider that here. 

\subsection{A General Regularised M-estimation Framework for Temporal Disaggregation}

With large scale data sets now becoming popular, developing parsimonious models contains numerous advantages. By imposing  the assumption that only a relatively small subset $K = |\{\beta_r \neq 0\}|$ of the $p$ possible indicators may actually be active in the model, this gives us scope to achieve good performance in increasing the accuracy of estimators by discarding noisy information and helps towards revealing underlying characteristics in the data. The empirical performance of sparse modelling has been examined in various settings by \citet{meinshausen2010stability} and \citet{buhlmann2014high}. 

Mirroring the established \cite{chow1971best} temporal disaggregation approach to construct the high frequency estimator $\hat{\by}_q$, we propose to study estimators of the form:
\begin{equation}
\label{eq:spTDestimator}
    \hat{\bbeta}_\rho = \argmin_{\bbeta \in \mathbb{R}^p} \bigg\{ \underbrace{\mystrut{2ex}\left\lVert \bV_a^{-1/2}(\by_a-\bX_a\bbeta)\right\rVert_2^2}_{\text{Chow-Lin Cost Function}} \; + \underbrace{\mystrut{2ex} P_{\lambda}(\bbeta)}_{\text{Regulariser}}\bigg\} \, ,
\end{equation}
where we explicitly index the solution as a function of $\rho$ to highlight dependence on this parameter. This estimator incorporates a regularising penalty function in conjunction with the classic Chow-Lin cost function to encode the assumption of sparsity. It does this by shrinking coefficients of indicator series, $\beta$, towards zero that cause a large least squares score in the Chow-Lin cost function. By doing so, we simultaneously select important indicator series and estimate their regression coefficients with sparse estimates. This will significantly reduce the variance in moderate dimensions and enable accurate estimators in high dimensions. 
Not only is this novel in the area of temporal disaggregation, working with a GLS cost function in the high dimensional setting has to-date received little attention in the statistics literature. 

The regulariser function, $P_{\lambda}(\bbeta)$ is indexed by the regularisation parameter $\lambda \geq 0$ that controls the degree of shrinkage. This function may take various forms depending on the assumptions required by the user. As indicated by \cite{hastie2015statistical} and \cite{buhlmann2011statistics}, common choices include $\ell_q$ norm penalties $P_{\lambda}(\bbeta) = \lambda \left\lVert \bbeta \right \rVert_q^q = \lambda\sum_{j=1}^p|\bbeta_j|^q $ for $q>0$, and the $\ell_0$ pseudo-norm penalty $\lambda I(\bbeta \neq 0)$ for $q=0$, where $ I(\bbeta \neq 0)$ is the number of non-zero components of $\bbeta$.  Estimates have the parsimonious property with some components being shrunk exactly to zero when $q \leq 1$, and the minimisation problem is convex when $q \geq 1$.

\subsection{The $\ell_1$-spTD method}

Given our interest in performing accurate model selection for temporal disaggregation, we here focus on the LASSO ($\ell_1$) penalty \citep{tibshirani1996regression,hastie2015statistical}. This specification provides an example of our general M-estimation framework whilst enabling us to examine  the benefits afforded by $\ell_1$ shrinkage. Notably, out of the class of $\ell_q$ norms, only the $\ell_1$ penalty has the properties of yielding sparse solutions while maintaining convexity. We refer to this method as $\ell_1$-spTD and utilise the regulariser function
\[
P_{\lambda}(\bbeta) = \lambda||\bbeta||_1 := \lambda_\rho\sum_{j=1}^p |\beta_j|\;.\]

A further useful property of LASSO penalisation is that fast algorithms exist to find estimators for a range of $\lambda$ values. These methods are known as \emph{path algorithms}, with least angle regression (LARS) \citep{efron2004least} perhaps being one of the more widely used approaches. In this work we employ LARS to provide an initial screening of indicator series by constructing unique solution paths $\hat{\bbeta}_\rho(\lambda)$, with varying sparsity levels controlled by $\lambda_\rho$. We suggest an additional re-fitting procedure here to reduce the bias in the LARS solution paths. 

Unlike GLS which generates only one set of coefficient estimates, the loss function in  (\ref{eq:spTDestimator}) will produce a different set of estimates for each value of the regulariser parameter. Correctly selecting the best $\lambda$ is vital. If too small, then not enough shrinkage occurs and we overfit. Conversely, if it is too large then we shrink too much and underfit. Since the parameter $\lambda$, and consequently the coefficient estimate $\hat{\bbeta}_\rho$, depend on the specification of the AR parameter $\rho$, the estimator in (\ref{eq:spTDestimator}) can be seen as maximising a penalised profile likelihood as a function of $\rho$. A final optimisation step searching over the stationary domain of $\rho$ is then used to obtain a final set of parameter estimates.

Our approach proceeds according to the following steps, which we now elaborate on:
\begin{enumerate}
    \item Generating solution paths $\hat{\bbeta}_\rho(\lambda)$ for a range of $\lambda$ using the LARS algorithm;
    \item Reducing the bias of these solution paths using a re-fitting procedure;
    \item Tuning the model for the best $\lambda$ and optimising over $\rho$.
\end{enumerate}
To simplify notation we use $\Tilde{\by}=\bV_a^{-1/2}\by_a$, $\Tilde{\bX}=\bV_a^{-1/2}\bX_a$ and $\Tilde{\bu}=\bV_a^{-1/2}\bu_a$. This represents the GLS rotated aggregate data we find LASSO solutions for.

\subsubsection{Generating Solution Paths}

To compute solution paths $\hat{\bbeta}_\rho(\lambda)$ of the convex problem (\ref{eq:spTDestimator}), we use the LARS algorithm \citep{efron2004least}. This combination of forward stagewise regression \citep{hastie2007forward} and LASSO \citep{tibshirani1996regression} is able to provide full piecewise linear solution paths for many values of $\lambda$. We use LARS as it is a computationally fast algorithm that offers interpretable models, with stable and accurate predictions. Furthermore, the solution path forms a useful graphical output that shows key trade-offs in model complexity \citep{hesterberg2008least}. 

In our approach, we will need to solve Equation (\ref{eq:spTDestimator}) multiple times, over a range of GLS rotations and regulariser parameters. For any fixed $\rho$, the LARS procedure begins with a large value of $\lambda$, and correspondingly sets all regression parameters $\beta=0$. Let the $i$th indicator series be the one that is most correlated with the response $\Tilde{\by}$, this indicator then enters the active set of relevant predictors.  Rather than fit this predictor completely (i.e.\ set to OLS solution), the algorithm moves the coefficient $\beta_{i}$ of this predictor continuously towards its least squares value, causing its correlation with the evolving residual to decrease in absolute value. The value of $\lambda$ is then decreased until another predictor $j$ has as much correlation with the current residual. At this point, the process is then paused and $j$ enters the active set. The two corresponding predictor coefficients $(\beta_{i},\beta_{j})$ now move in the direction of their joint least squares fit of the current residual on the two predictors. The process continues adding predictors in this fashion by gradually decreasing $\lambda$, and is able to remove predictors from the active set (the LASSO step)  if their coefficient hits zero. For this reason, the algorithm usually has more iterations than the number of predictors, $p$.
Each iteration step $l=1,\ldots,k$ of the algorithm has a decreasing $\lambda^{(l)}$ value causing the solution paths to become less sparse as the algorithm moves forward. The final step occurs when the model becomes saturated with $n$ non-zero coefficients. It is interesting to note that the sparsity of this final step contains no more than $\min(n,p)$ non-zero values. This means that the number of different LASSO estimated sub-models is typically $\mathcal{O}(\min(n,p))$, which represents a huge reduction compared to all $2^p$ possible sub-models, especially in the case where $p\gg n$.

After running the LARS algorithm on the regression we get a set of solution paths $\hat{\bbeta}_\rho(\lambda^{(l)})$ for steps $l=1,\dots,k $ that are of the form:
\begin{equation*}
    \begin{cases}
    \text{Step 0: } \qquad \hat{\bbeta}_\rho(\lambda^{(0)}: \lambda^{(0)}>0) = (0,\dots,0) \, , \, \text{with all values zero;} \\
    \text{Step 1: } \qquad \hat{\bbeta}_\rho(\lambda^{(1)}: 0 < \lambda^{(1)} < \lambda^{(0)}) = (--,\beta_{\rho},--) \, , \, \text{one non-zero}; \\
    \text{Step 2: } \qquad \hat{\bbeta}_\rho(\lambda^{(2)}: 0 < \lambda^{(2)} < \lambda^{(1)}) = (--,\beta_{\rho;i},\beta_{\rho;j},--) \, , \, \text{two non-zero}; \\
    \vdots \\
    \text{Step k: } \qquad \hat{\bbeta}_\rho(\lambda^{(k)}) = (--,\beta_{\rho;1},\dots,\beta_{\rho;s},--) \, , \, \text{with } s=\min(n,p) \text{ non-zero}. 
    \end{cases}
\end{equation*}
Figure \ref{fig:larsandbic}(a) shows the LARS algorithm applied to a synthetic data-set with non-stationary indicator series, $n=100$, $p=150$, $\rho=0.5$ and the true $\bbeta$ parameter having the first 10 values equal to 5 and the other 140 equal to 0. This demonstrates the algorithm beginning with the null model, then correctly activating the first 10 predictors towards 5. Then, as the algorithm moves forward, and $\lambda$ decreases, the model becomes less sparse until the model is saturated.

\subsubsection{Reducing the Bias in Solution Paths}

\label{subsec:sptd_rf}

Before tuning the above algorithm to select the optimal $\lambda^{(l)}$ for $l=1,\dots,k$, we propose a preliminary step to reduce the bias in the solution paths. The significance of this preliminary step is assessed in our simulation study in Section 4. It is a recognised drawback of LASSO estimators that they include a small amount of bias, as their absolute value is typically smaller than that of the true parameter; $|\mathbb{E}(\hat{\bbeta}_\rho(\lambda))-\bbeta_\rho|>0$ for $\lambda>0$. This behaviour can be seen in Figure \ref{fig:larsandbic}a) with the parameter estimates for the first 10 predictors being slightly below 5, whereas, the unbiased estimator in this case would have expectation of exactly 5. Many authors \citep{buhlmann2011statistics,zheng2014high} have suggested a simple remedy to this issue is to treat LASSO as a variable screening procedure and to perform a second refitting step on the selected support. \citet{belloni2013least} provide a theoretically justified refitting technique which involves performing a least-squares re-estimation of the non-zero coefficients of the LASSO solution.
We suggest adopting this least-squares re-estimation approach in our algorithm. With each solution path $\hat{\bbeta}_\rho(\lambda^{(l)})$ for $l=1,\dots,k$ obtained from the LARS algorithm, a new design matrix $\Tilde{\bX}'$ is constructed containing only the predictors that have a non-zero coefficient in the corresponding solution path. We then perform usual least squares estimation on $(\Tilde{\bX}',\Tilde{\by})$ and obtain estimates $\hat{\bbeta}_\rho$ with a reduced bias.

\subsubsection{Tuning Model Parameters}
The performance of our estimator relies heavily on the choice of tuning parameter $\lambda$ to select the optimal estimate from $\hat{\bbeta}_\rho(\lambda^{(1)}),\dots,\hat{\bbeta}_\rho(\lambda^{(k)})$. We propose to use the BIC statistic \citep{schwarz1978estimating} to achieve this and choose $\lambda$ such that
\begin{equation}
    \label{BICequation}
    \hat{\lambda}_\rho = \argmin_{\lambda \in \{\lambda^{(1)},\ldots,\lambda^{(k)}\}} \big\{ -2\mathcal{L}(\hat{\bbeta}_\rho(\lambda), \hat{\sigma}^2) + \log(n) K_{\lambda} \big\} \, ,
\end{equation} 
where  $K_{\lambda}=|\{r : (\hat{\beta}_\rho(\lambda))_r \neq 0 \}|$ is the degrees of freedom and $\mathcal{L}(\hat{\bbeta}_\rho(\lambda), \hat{\sigma}^2)$ is the log-likelihood of the GLS regression.

Since we are working with relatively small sample size of low frequency observations, BIC seems more appropriate than resampling procedures such as cross-validation or bootstrap and comes at a much lower computational expense \citep{friel2017investigation}. Furthermore, \citet{efron2004least} show how BIC offers substantially better accuracy than cross-validation and related non-parametric methods in selecting optimal $\lambda$ in LARS. BIC is preferred to other information criteria as parsimony is a primary concern, and BIC generally places a heavier penalty on models with many variables due to the $\log(n)K_{\lambda}$ term.

Assuming Gaussian errors, the log-likelihood of the GLS regression (\ref{eq:betaGLS}) is given by 
$$ \mathcal{L}(\bbeta, \sigma^2) = -\frac{n}{2}\log(2\pi) - \frac{n}{2}\log(\sigma^2) - \frac{1}{2}\log(|\bS|) - \frac{1}{2\sigma^2}(\Tilde{\by}-\Tilde{\bX}\bbeta)^T(\Tilde{\by}-\Tilde{\bX}\bbeta) \, ,$$
where $|\bS|$ is the determinant of the Toeplitz matrix $\bS$ which depends on $\rho$ and is such that $\bV_a = \sigma^2\bS$.
We can then maximise this log-likelihood at the reduced-bias estimator $\hat{\bbeta}_\rho(\lambda)$ from the re-fitted LARS algorithm and using an estimator of the error variance $\hat{\sigma}^2$. Finding a good estimator of $\bbeta$ has received considerably more attention in the literature than finding a good estimator of $\sigma^2$ \citep{reid2016study}. However, constructing a reliable estimator of $\sigma^2$ in finite samples is crucial as it enables one to understand the uncertainty in estimating $\bbeta$ and be able to construct $p$-values and confidence intervals \citep{yu2019estimating}. We utilise the estimator
\[
\hat{\sigma}^2 = \frac{1}{n-K_\lambda}\{\Tilde{\by}-\Tilde{\bX}\hat{\bbeta}_\rho(\lambda)\}^T\{\Tilde{\by}-\Tilde{\bX}\hat{\bbeta}_\rho(\lambda)\} \, , 
\]
which de-biases the residual sum of squares by the degrees of freedom. We use this form of estimator as it has been shown to have promising performance in an extensive simulation study done by \citet{reid2016study}. After substituting the estimators into the log-likelihood to obtain $\mathcal{L}\{\hat{\bbeta}_\rho(\lambda), \hat{\sigma}^2\}$, we simply search along the solution path to find the $\lambda$ that minimises the BIC objective (\ref{BICequation}).

\begin{figure}
    \begin{minipage}{\linewidth}
        \centering
        \subfloat[]{\makebox{\includegraphics[width=0.9\linewidth]{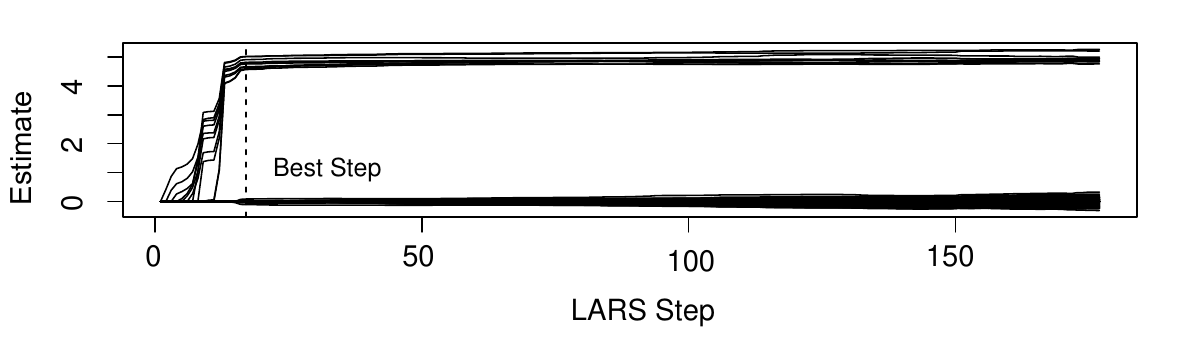}}}
    \end{minipage}
    \begin{minipage}{\linewidth}
        \centering
        \subfloat[]{\makebox{\includegraphics[width=0.9\linewidth]{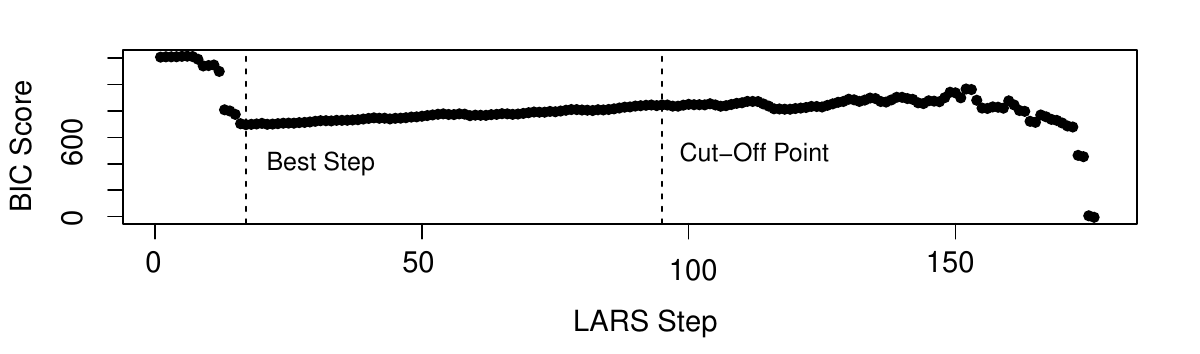}}}
    \end{minipage}
    \caption{Random walk synthetic data example using $n=100$, $p=150$, $\rho=0.5$ and true predictor coefficient $\bbeta$ with the first 10 values equal to 5 and the remaining 140 equal to 0. (a)  Solution paths generated via LARS, plot shows $\hat{\bbeta}(\lambda^{(l)})$ as a function of step $l$; (b) BIC plot for each step of the LARS algorithm indicating the best step and cut-off point.
    }
\label{fig:larsandbic}
\end{figure}

In high dimensional settings this BIC heuristic may not be optimal. Indeed, we find that after a certain number of steps of the LARS algorithm, when $K_{\lambda}$ becomes close to $n$, the BIC starts to behave erratic and drop in value. 
Figure \ref{fig:larsandbic}b) plots the BIC score for each step of the LARS algorithm applied to the high dimensional synthetic non-stationary data-set. After around iteration 100, corresponding to $K_{\lambda}$ being around 70, the BIC starts to exhibit this erratic behaviour. Such behaviour may be due to the expected increase in variance in estimating $\hat{\sigma}^2$ with a small degrees of freedom. To accommodate this shortcoming, we adopt a heuristic approach and only considered solution paths $\hat{\bbeta}(\lambda)$ where $K_{\lambda}$ is less than $n/2$; identified by the \emph{cut-off} line shown. Constructing a well-behaved information criteria in high dimensions is an active research area in the literature. For instance, the extended BIC formulas of \citet{chen2008extended}, or alternative tuning strategies such as \citet{belloni2011square} form directions for future research. 

Once we have found the LARS step minimising BIC, and thus $\hat{\lambda}$, we store the corresponding BIC score and repeat the process with a new initialisation of $\rho \in (-1,1)$. The estimate $\hat{\rho}$ is then given by the $\rho$ that has the lowest overall BIC score in our search. Using $\hat{\rho}$, we then have our final parameter estimate $\hat{\bbeta}_{\hat{\rho}}$. The pseudo code for this procedure can be found in Algorithm \ref{alg:spTD}. Code for implementing the $\ell_1$-spTD algorithm can be found in the R package \texttt{DisaggregateTS}\footnote{Can be downloaded from  \url{https://cran.r-project.org/web/packages/DisaggregateTS/index.html}}.

\subsection{Dealing with Correlated Indicator Series}

It is likely that high-dimensional data sets of economic indicator series will exhibit correlation between series and multicollinearity. It has been shown \citep{fan2001variable,zou2006adaptive}  that in certain correlated design settings the choice of regularisation parameter $\lambda$ is not guaranteed to satisfy the oracle properties and this can lead to inconsistent selection results in high-dimensions. Variable selection consistency is only guaranteed if the design matrix $X$ of indicators satisfies the so-called \emph{irrepresentability condition} \citep[see][for details]{zou2006adaptive}. Intuitively, this condition  says that relevant indicator series in the model are not allowed to be too correlated with irrelevant indicator series. If $X$ violates the irrepresentability condition then an irrelevant variable may enter the active set in the LARS algorithm before all relevant variables have entered, leading to incorrect support recovery. This will cause problems in our re-fitting strategy as an over-fitted support will be re-fit. 

One approach to overcome this difficulty is to apply a different penalty to each predictor variable by re-weighting the usual $\ell_1$ penalty. This is the \emph{adaptive lasso} \citep{zou2006adaptive}. For a linear model, it is defined as a two-stage procedure:
\begin{equation} \label{eq:adaptive}
\hat{\beta}_{\text{adapt}}(\lambda) = \argmin_\beta \big(||y-X\beta||^2_2 + \lambda \sum_{j=1}^p \frac{|\beta_j|}{|\hat{\beta}_{\text{init},j}|} \big) \, ,
\end{equation}
where $\hat{\beta}_{\text{init}}$ is an initial estimator. \cite{zou2006adaptive} proposes to use the OLS estimate $\hat{\beta}_{OLS}$ for $\hat{\beta}_{\text{init}}$, however, in high-dimensional scenarios \cite{buhlmann2011statistics} propose to use the lasso estimate $\hat{\beta}_{lasso}$ for $\hat{\beta}_{\text{init}}$. We propose to use the estimate $\hat{\beta}_{\hat{\rho}}$ from a run of our $\ell_1$-spTD algorithm for $\hat{\beta}_{\text{init}}$. That is, using the aggregated indicator matrix  $\Tilde{\bX}$ that has been GLS rotated by $\hat{\rho}$ obtained from the initial fit, we then re-scale this to get $\bX_{\text{new}} = \Tilde{\bX}|\hat{\beta}_{\hat{\rho}}|$. A second stage regression is then performed using LARS (with BIC tuning) on $(X_{\text{new}},y)$ to get an estimate $\hat{\beta}^*$. Finally, we scale back by $\hat{\beta}_{\text{adapt}}=\hat{\beta}^*|\hat{\beta}_{\hat{\rho}}|$ to obtain the adaptive lasso solution. In non-GLS settings, this has been shown \citep{zou2006adaptive,vdg2011} to have variable selection oracle properties even when the data violates the irrepresentability condition. 

\cite{buhlmann2011statistics} outline scenarios where the the irrepresentable condition holds despite the predictors in $X$ being correlated. These include scenarios when the covariance matrix of $X$ has equal positive correlation, Toeplitz structure or bounded pairwise correlation. In these settings, using LARS without adaptive weighting has been shown to be stable, \citet{bach2011convex} have confirmed this good behaviour in simulations using different levels of correlation. Furthermore, they compare LARS against other lasso-solving algorithms such as coordinate-descent, re-weighted-$\ell_2$ schemes, simple proximal methods, demonstrating that LARS both outperforms and is faster than the other methods for small and medium scale problems. LARS also performed the best when predictors were highly correlated. We confirm this good behaviour in our simulation study in Section 4. 

\begin{algorithm}
\caption{$\ell_1$-spTD\label{alg:spTD}
}
\begin{algorithmic}
\STATE{\textbf{Input:} $\by_a \in \mathbb{R}^n$ and $\bX_q \in \mathbb{R}^{m \times p}$}
\FOR{$\rho \in (-1,1)$} 
    \STATE{\textbf{Initialise:} $\sigma^2=1$, $\bC = \bI_n \otimes \bm{1}$ and $\bV_q := \text{Toeplitz}(\rho,\sigma^2) \in \mathbb{R}^{m \times m}$}
    \STATE{\textbf{Aggregate:} $\bV_a = \sigma^2\bS = \bC\bV_q\bC^\top \in \mathbb{R}^{n \times n}$ and $\bX_a = \bC\bX_q$}
    \STATE{\textbf{GLS Rotate:} $\Tilde{\by} = \bV_a^{-1/2}\by_a$ and $\Tilde{\bX} = \bV_a^{-1/2}\bX_a$}
    
    \STATE{$\mathrm{LARS}(\Tilde{\by} \sim \Tilde{\bX}) \implies \hat{\bbeta}_\rho(\lambda^{(i)})$}
    \FOR{$i=1$ \TO $k$}
        \STATE{Find $K_{\lambda} = \{j: \hat{\bbeta}_\rho(\lambda^{(i)})_j \neq 0 \}$}
        \IF{Performing De-biasing (Sec. \ref{subsec:sptd_rf})}
            \STATE{Define $\Tilde{\bX}' = \Tilde{\bX}[K_{\lambda}]$}
            \STATE{Derive $\hat{\bbeta}_{GLS}(\lambda^{(i)})$ using $\Tilde{\by} \sim \Tilde{\bX}'$}
            \STATE{New non-zero values of $\hat{\bbeta}(\lambda^{(i)})$ are $\hat{\bbeta}_{GLS}(\lambda^{(i)})$}
        \ENDIF
    \ENDFOR
    \FOR{$j$ \textbf{in} $1,\dots,\max\{j : K_{\lambda^{(j)}}< \left \lfloor n/2 \rfloor \right\}$ }
        \STATE{$\hat{\sigma_j}^2 = (\Tilde{\by} - \Tilde{\bX}\hat{\bbeta}_\rho(\lambda^{(j)})^\top(\Tilde{\by} - \Tilde{\bX}\hat{\bbeta}(\lambda^{(j)}))/(n-K_{\lambda})$}
        \STATE{$\mathcal{L}(\hat{\bbeta}_\rho(\lambda^{(j)}), \hat{\sigma}^2) = -n\log(2\pi)/2 - n\log(\hat{\sigma_j}^2)/2 - \log(\det(\bS))/2 - (n-K_{\lambda^{(j)}})/{2}$}
        \STATE{$\mathrm{BIC}(\lambda^{(j)}) = -2\mathcal{L}\{\hat{\bbeta}_\rho(\lambda^{(j)}), \hat{\sigma_j}^2\} + \log(n) K_{\lambda^{(j)}}$}
    \ENDFOR
    \STATE{$\hat{\lambda} = \argmin_{\lambda}(\mathrm{BIC})$}
    \STATE{$\mathrm{BIC}_\rho = \text{BIC}(\hat{\lambda})$}
    
    \STATE{$\hat{\bbeta}_\rho = \hat{\bbeta}_\rho(\hat{\lambda})$}
\ENDFOR
\STATE{$\hat{\rho} = \argmin_\rho(\mathrm{BIC}_\rho)$}
\STATE{$\hat{\bbeta} = \hat{\bbeta}_{\hat{\rho}}$}
\STATE{$\hat{\by}_q = \bX\hat{\bbeta} + \bV_q\bC^\top\bV_a^{-1}(\by_a-\bX_a\hat{\bbeta})$ using $\hat{\rho}$ for $\bV_q$ }
\end{algorithmic}
\end{algorithm}

\section{Simulation Study}

We now provide an in-depth simulation study to asses the performance of our $\ell_1$-spTD algorithm in performing annual-to-quarterly disaggregation in the high dimensional setting, and compare how we do against the well-established Chow-Lin (CL) method \citep{chow1971best} in moderate dimensions. We further examine the benefits of the proposed \emph{re-fit} step (Sec. \ref{subsec:sptd_rf}) to reduce the bias of LASSO estimates. This consists of running Algorithm \ref{alg:spTD} and comparing the results with and without the refitting procedure, these methods are respectively referred to as spTD\_RF and spTD.

To understand the behaviour of the methods under different covariate inputs, we consider two data generating processes for the indicator series in $\bX_q$. Firstly, we consider an example with stationary series by simulating from the standard Normal distribution and secondly, we consider non-stationary series by simulating from a random walk process. That is, we consider (i) $x_{t,j} \sim N(0,1)$ and (ii) $x_{t,j} = x_{t-1,j} + e_{t,j}$, $e_{t,j} \sim N(0,1)$ respectively for $t=1,\dots,m$ and $j=1,\dots,p$, the residual series is given by an AR(1) process $u_j=\rho u_{j-1}+\epsilon_j$ with $|\rho|<1$ and $\epsilon_j \sim \mathcal{N}(0,1)$. We assume a sparse regression coefficient 
\[
\bbeta = (\underset{10}{\underbrace{5,\dots,5}},\underset{p-10}{\underbrace{0,\dots,0}})\;.
\]
This allows us to generate the true `unobservable' quarterly series using $\by_q = \bX_q\bbeta + \bu_q$ and we aggregate this using $\bC=\bI_n \otimes \bm{1}_4$ to get the `observable' annual series $\by_a = \bC\by_q$.

To explore different situations we use the following scenarios of parameters. We fix the low-frequency sample size to be $n=100$ and consider three scenarios for the number of indicator series $p$. These are $p=30, 90$ and $150$ to represent low, moderate and high dimensional settings respectively. We consider three values of the AR(1) parameter $\rho=0.2, 0.5$ and $0.8$ to represent low, medium and high auto-correlation in the residuals respectively. Under each scenario we perform 1000 simulations and assess the performance on the following three outcomes:
\begin{enumerate}
    \item The accuracy of the estimated quarterly disaggregated series $\hat{\by}_q$;
    \item The recovery of the true regression coefficient parameters $\hat{\bbeta}=(\hat{\bbeta_1},\dots,\hat{\bbeta_p})$; 
    \item The distribution of estimates for the true AR(1) parameter $\hat{\rho}$.
\end{enumerate}
The study is similarly extended to include a correlated design setting for the indicators. First, a block structure with equal correlation that satisfies the irrepresentability condition \citep{buhlmann2011statistics}. Second, a random covariance matrix design that does not satisfy this condition. We adopt the adaptive lasso weighting scheme in (\ref{eq:adaptive}) to assess whether this improves spTD\_RF in this second setting.

We implement the simulation in R, making use of the \emph{tempdisagg} package \citep{sax2013temporal} to find CL estimates and the \emph{lars} package \citep{hastie2013lars} to run the LARS algorithm to obtain coefficient paths. 

\subsection{Estimating the True Quarterly Series}

\begin{figure}[htbp]
\centering
    \begin{minipage}{.3\linewidth}
    \centering
    \subfloat[]{\makebox{\includegraphics[width=\linewidth]{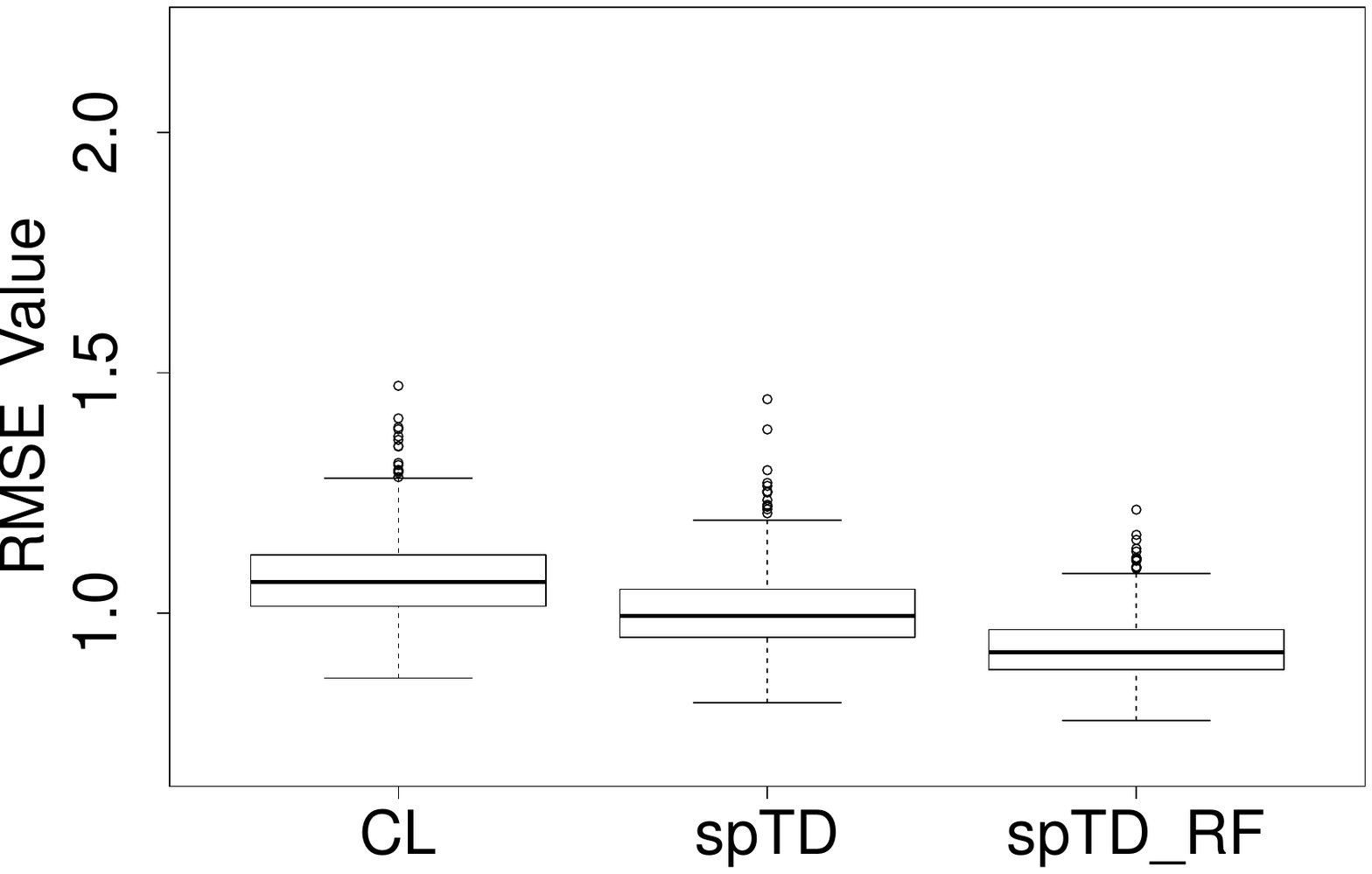}}}
    \end{minipage}
    \begin{minipage}{.3\linewidth}
    \centering
    \subfloat[]{\makebox{\includegraphics[width=\linewidth]{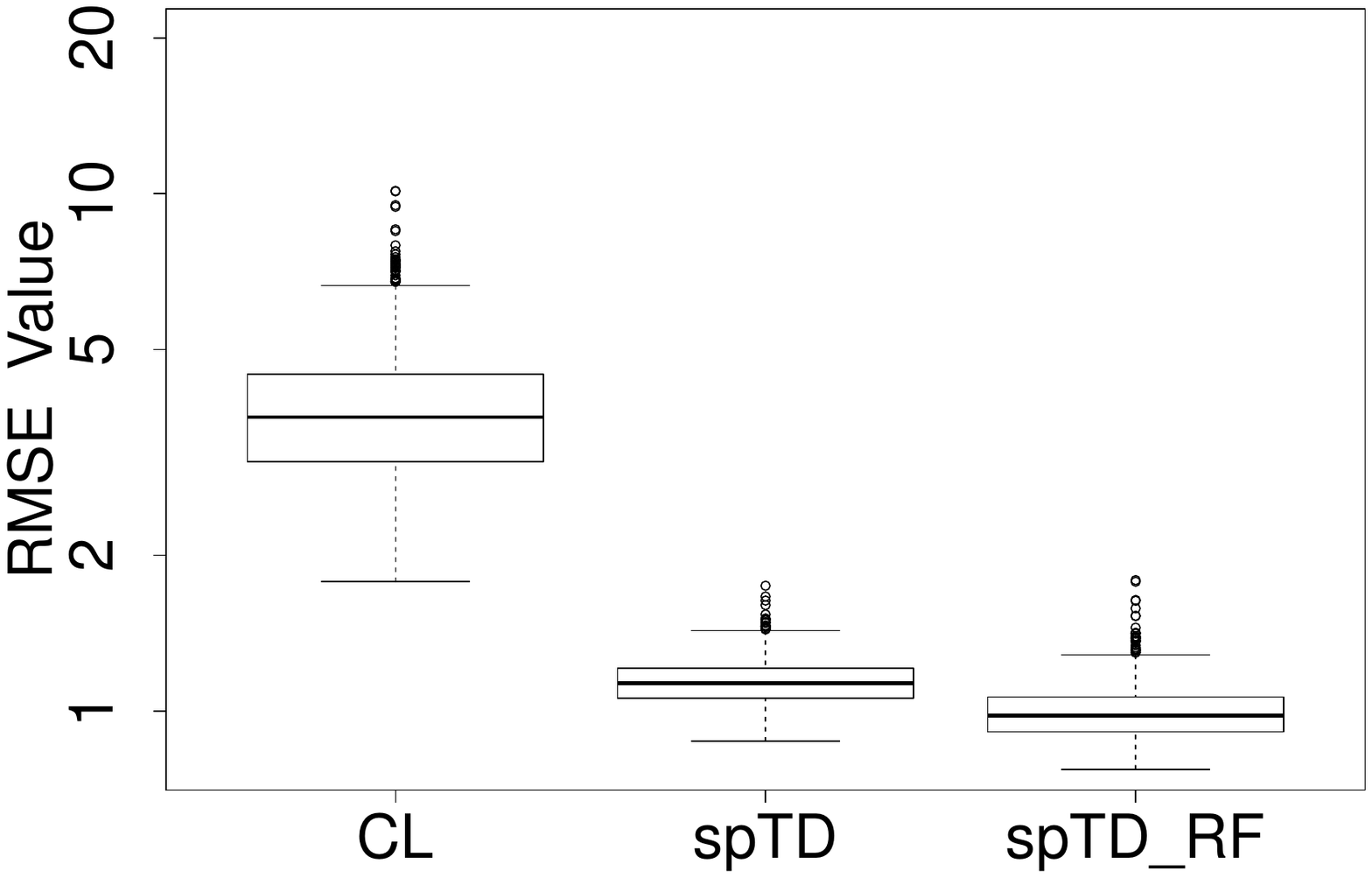}}}
    \end{minipage}
    \begin{minipage}{.22\linewidth}
    \centering
    \subfloat[]{\makebox{\includegraphics[width=\linewidth]{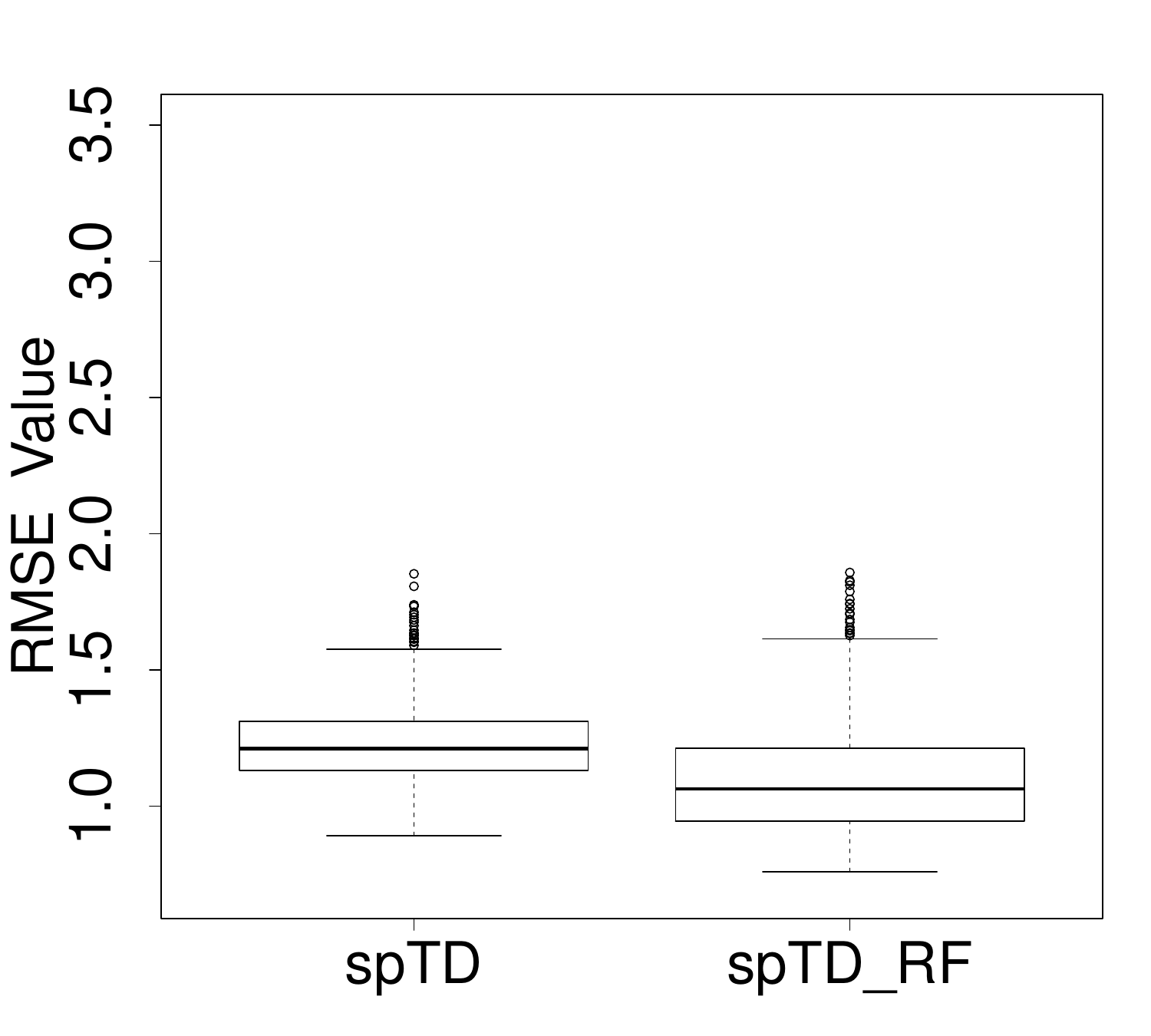}}}
    \end{minipage}\par\medskip
    \begin{minipage}{.3\linewidth}
    \centering
    \subfloat[]{\makebox{\includegraphics[width=\linewidth]{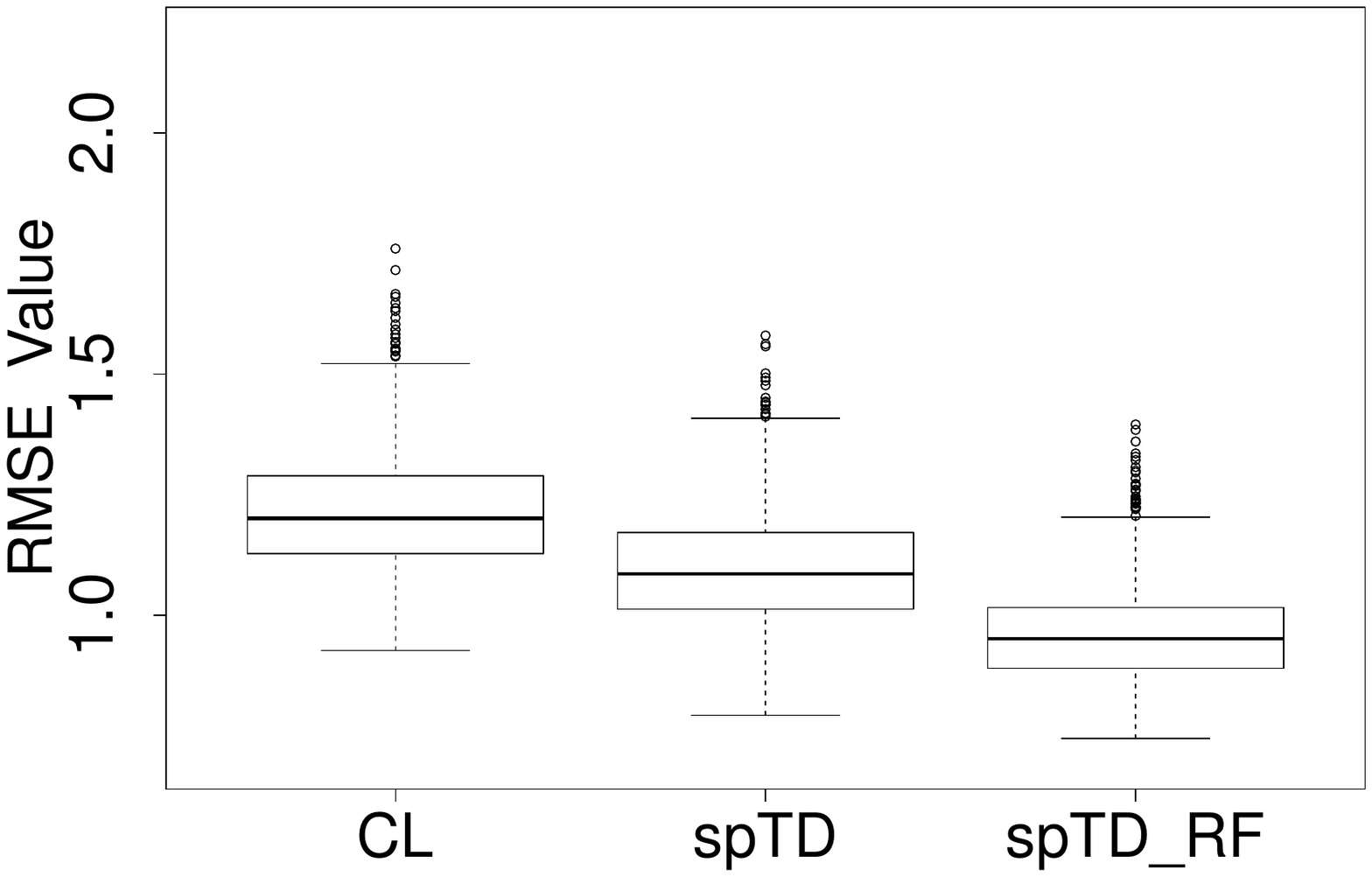}}}
    \end{minipage}
    \begin{minipage}{.3\linewidth}
    \centering
    \subfloat[]{\makebox{\includegraphics[width=\linewidth]{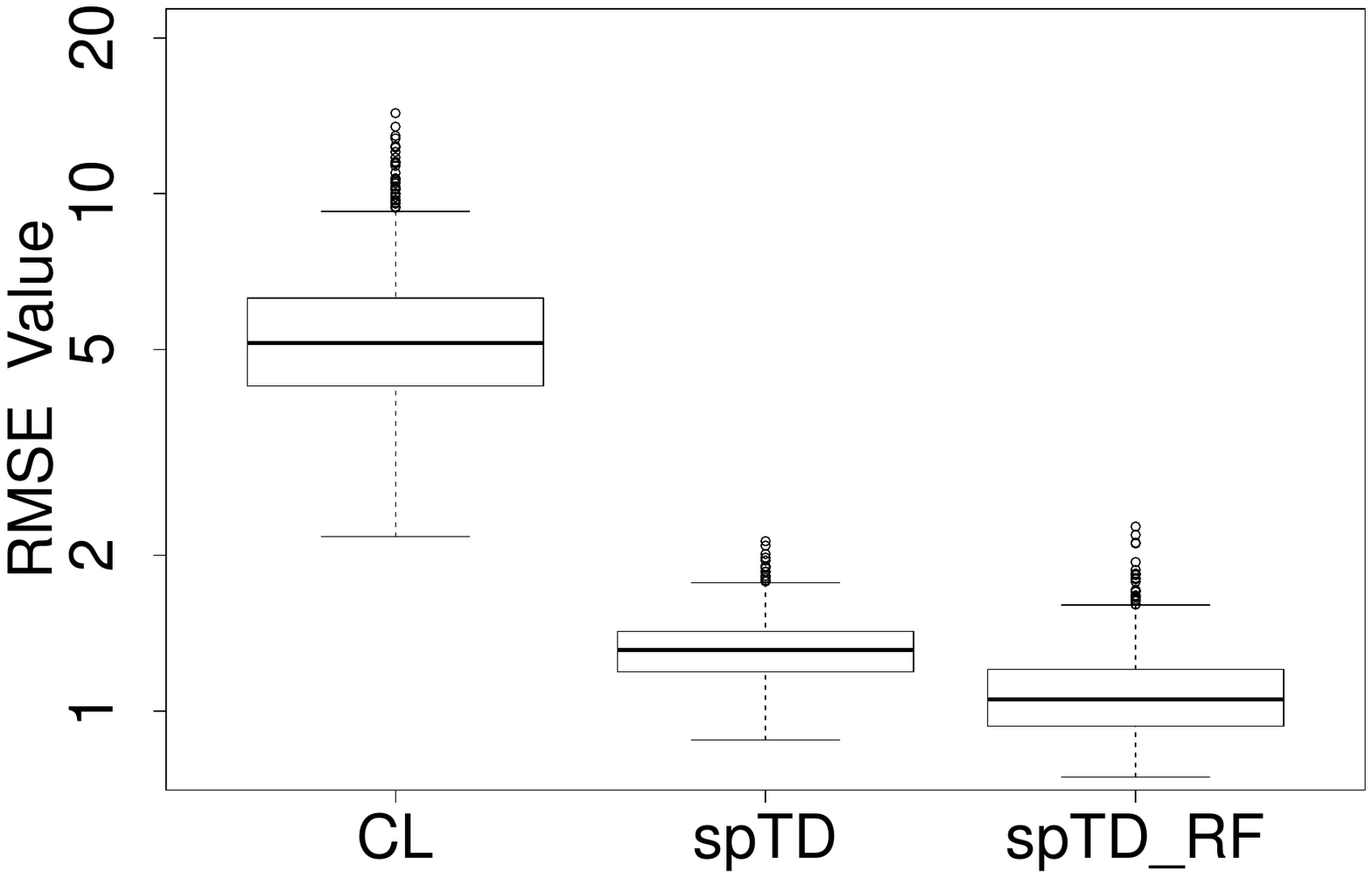}}}
    \end{minipage}
    \begin{minipage}{.22\linewidth}
    \centering
    \subfloat[]{\makebox{\includegraphics[width=\linewidth]{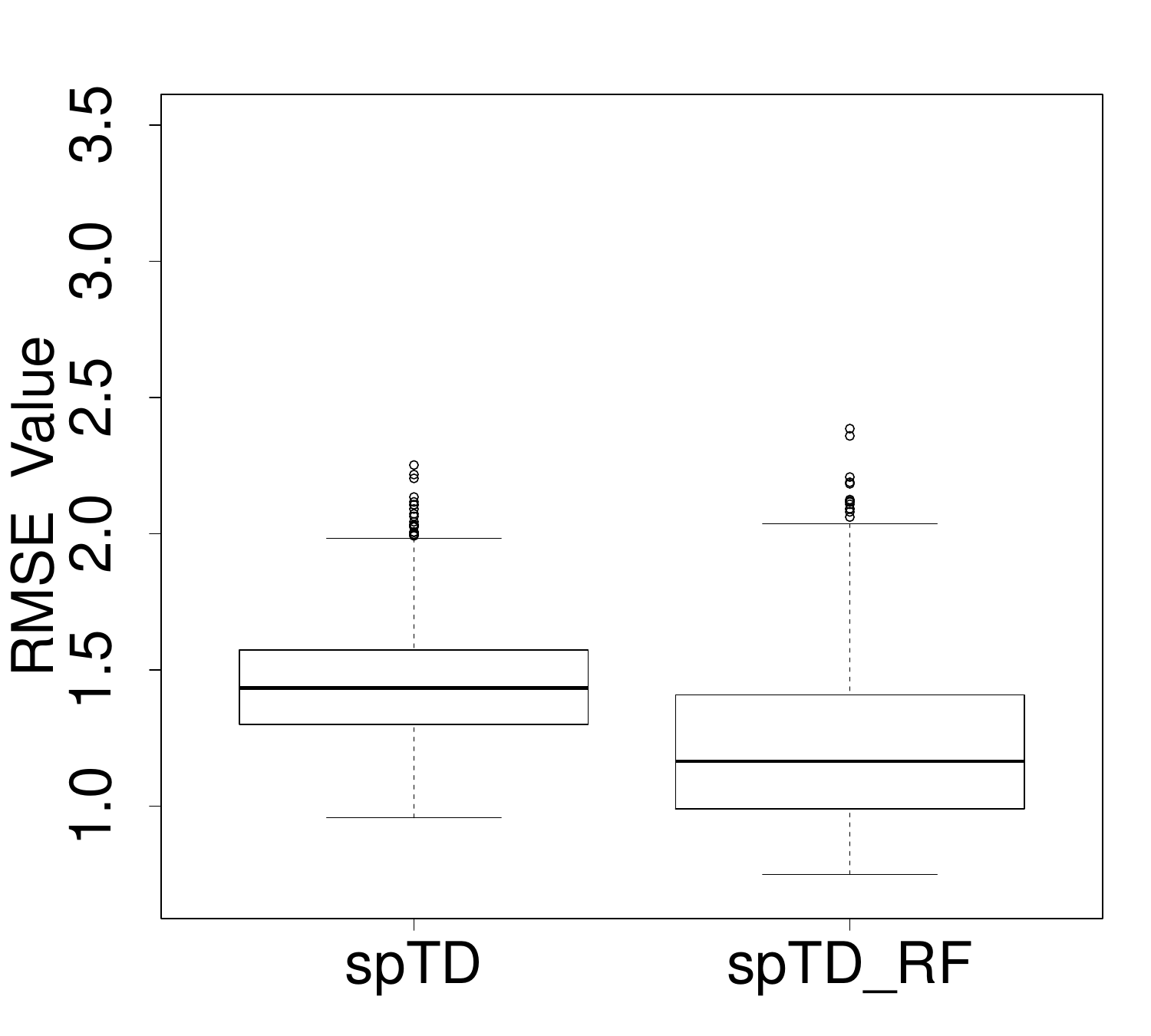}}}
    \end{minipage}\par\medskip
    \begin{minipage}{.3\linewidth}
    \centering
    \subfloat[]{\makebox{\includegraphics[width=\linewidth]{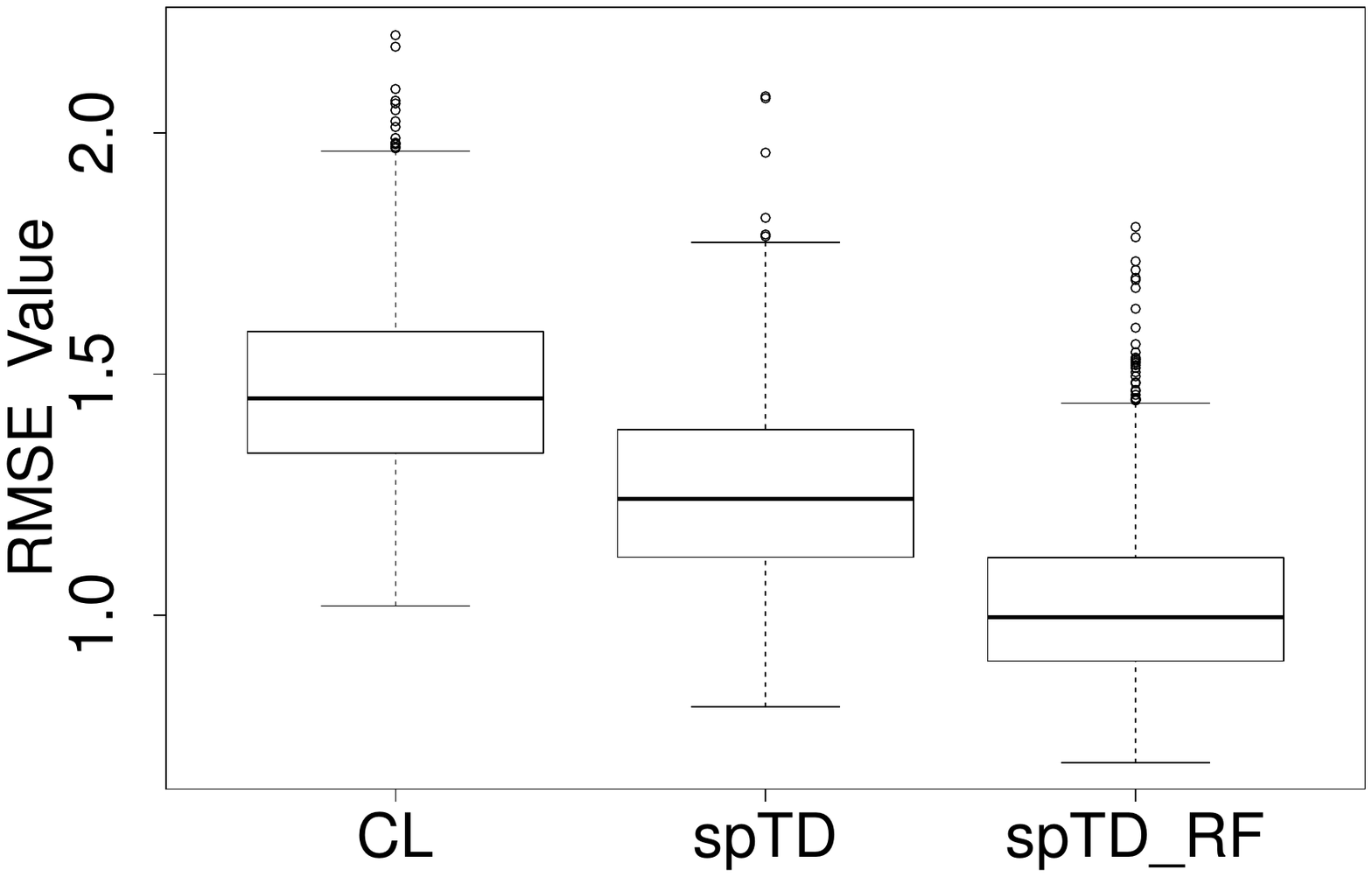}}}
    \end{minipage}
    \begin{minipage}{.3\linewidth}
    \centering
    \subfloat[]{\makebox{\includegraphics[width=\linewidth]{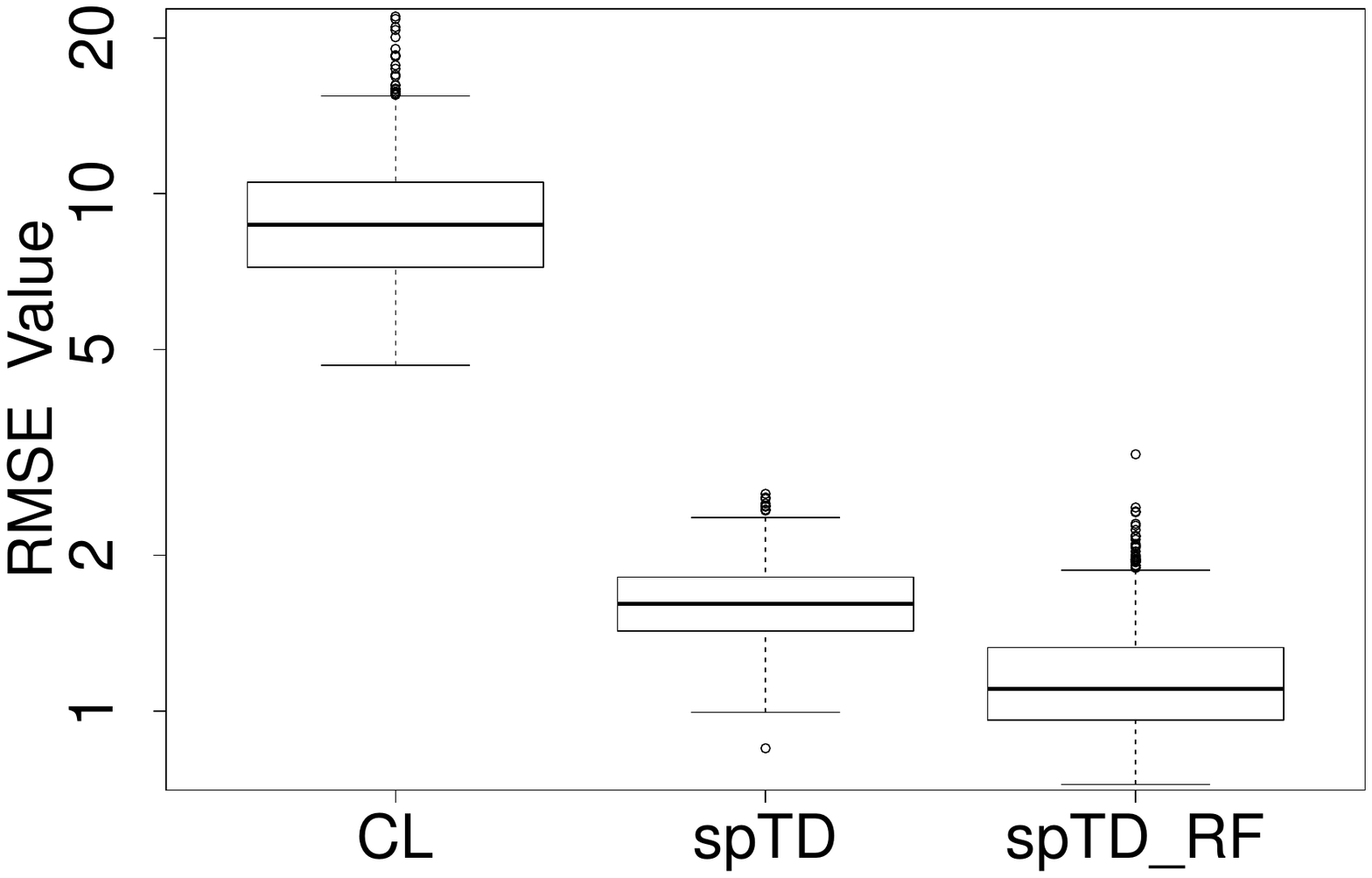}}}
    \end{minipage}
    \begin{minipage}{.22\linewidth}
    \centering
    \subfloat[]{\makebox{\includegraphics[width=\linewidth]{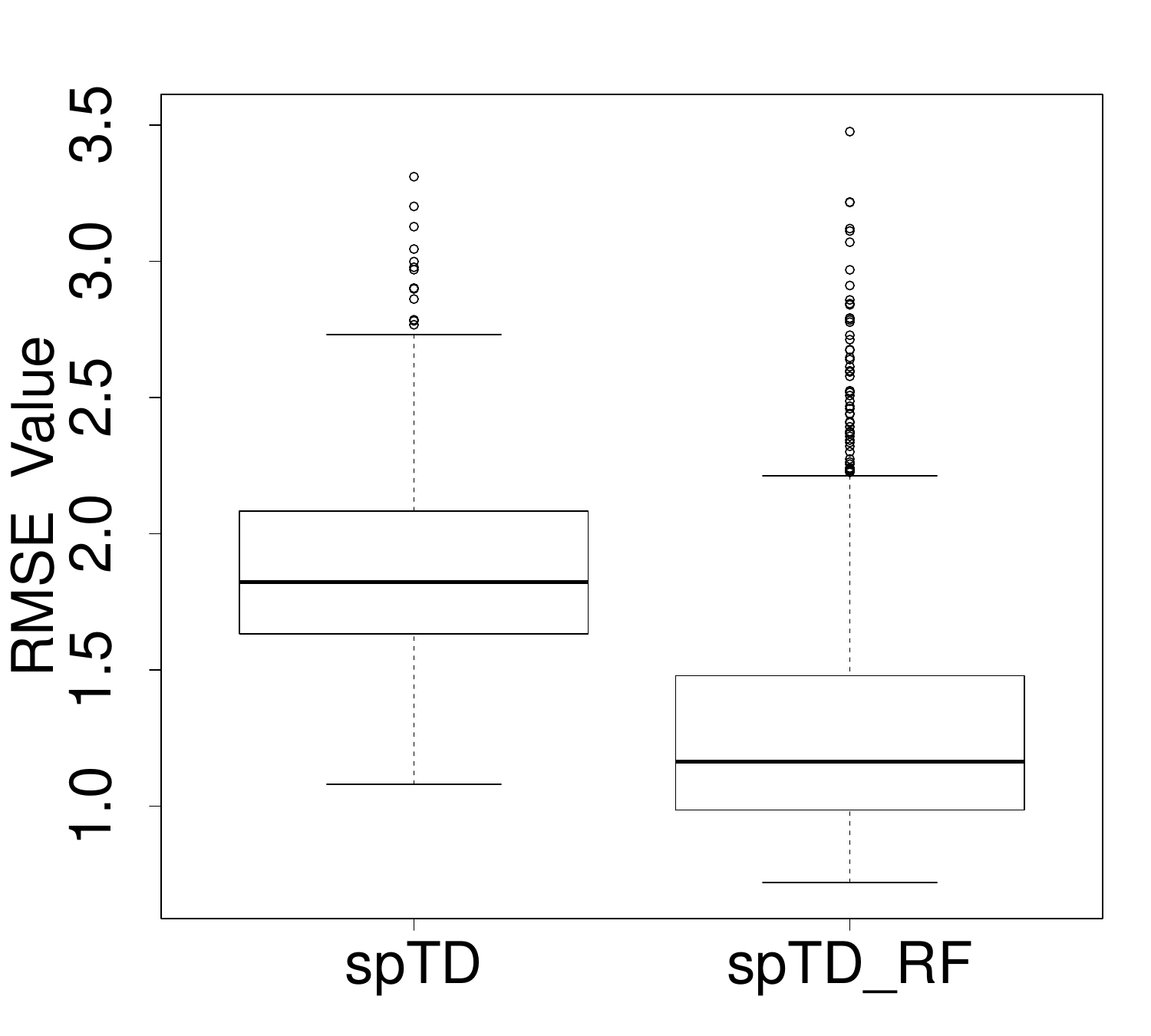}}}
    \end{minipage}\par\medskip
\caption{Boxplots comparing RMSE values of $\hat{\by}_q$ for CL, spTD and spTD\_RF in the stationary i.i.d. standard Normal indicator series experiment. Left, middle and right columns represent $p=30$, $90$ and $150$ respectively. Top, middle and bottom rows represent AR(1) parameter $\rho=0.2$, $0.5$ and $0.8$ respectively. }
\label{fig:IIDRMSE}
\end{figure}

\begin{figure}[htbp]
\centering
    \begin{minipage}{.3\linewidth}
    \centering
    \subfloat[]{\makebox{\includegraphics[width=\linewidth]{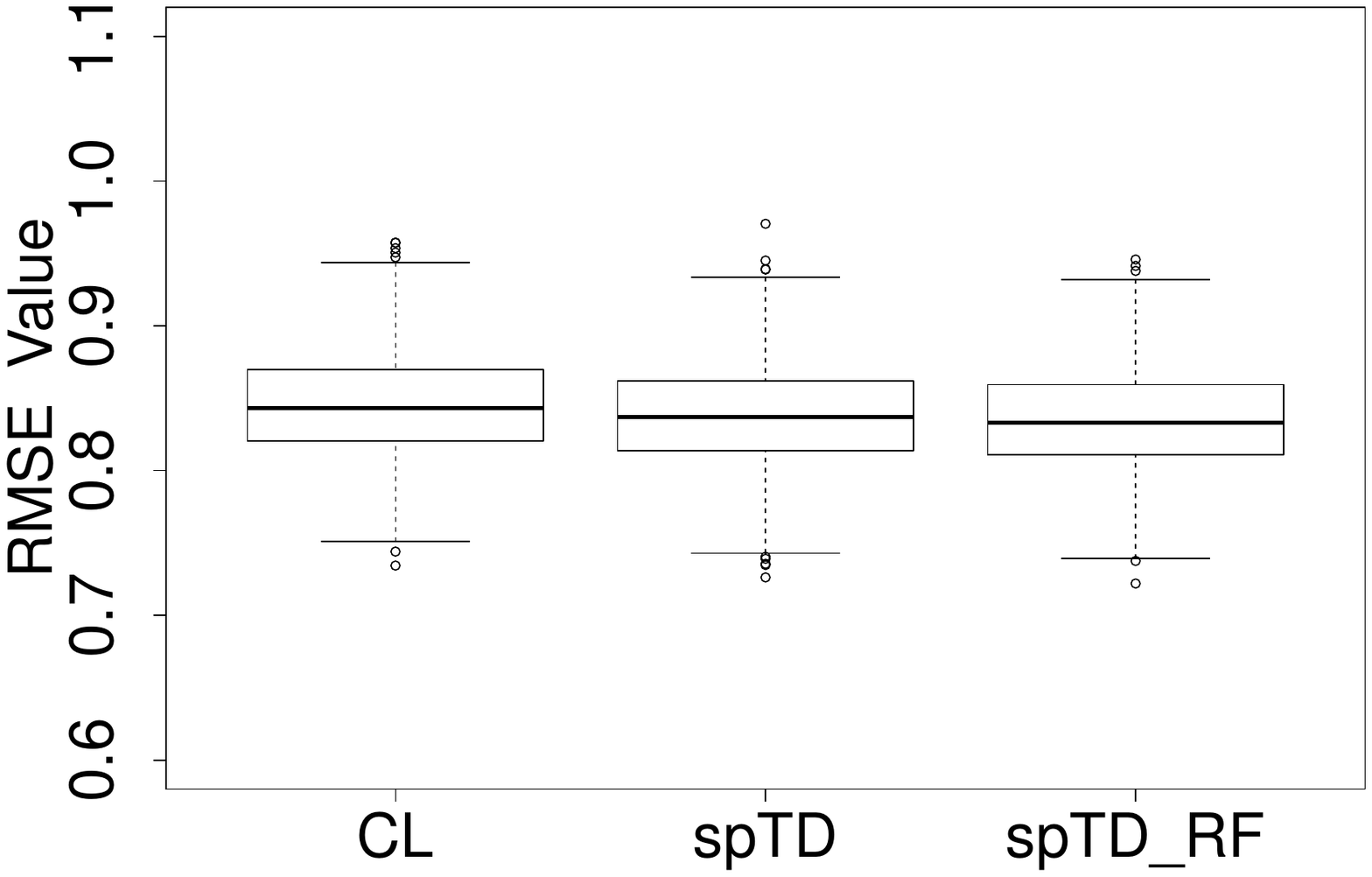}}}
    \end{minipage}
    \begin{minipage}{.3\linewidth}
    \centering
    \subfloat[]{\makebox{\includegraphics[width=\linewidth]{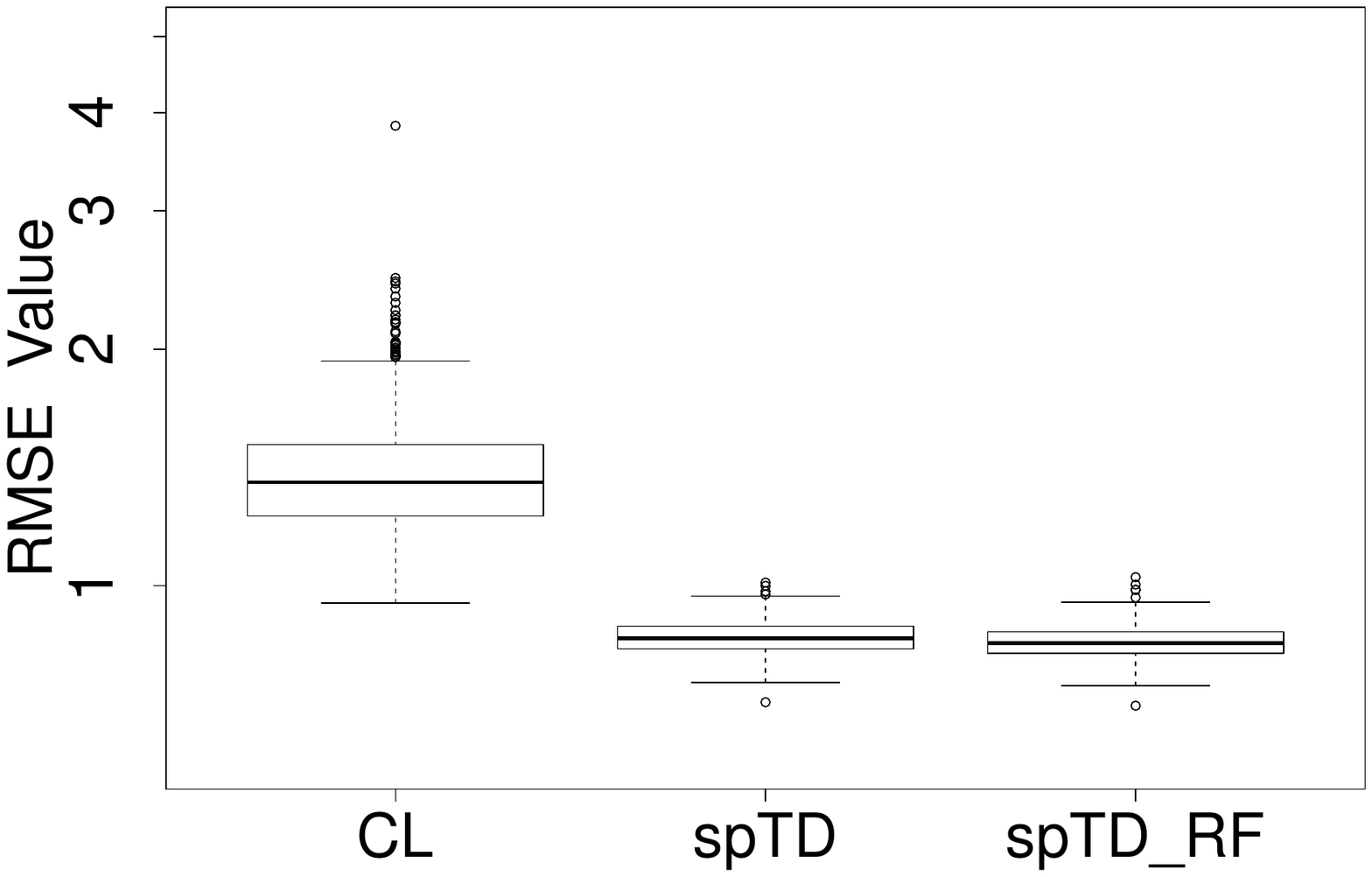}}}
    \end{minipage}
    \begin{minipage}{.22\linewidth}
    \centering
    \subfloat[]{\makebox{\includegraphics[width=\linewidth]{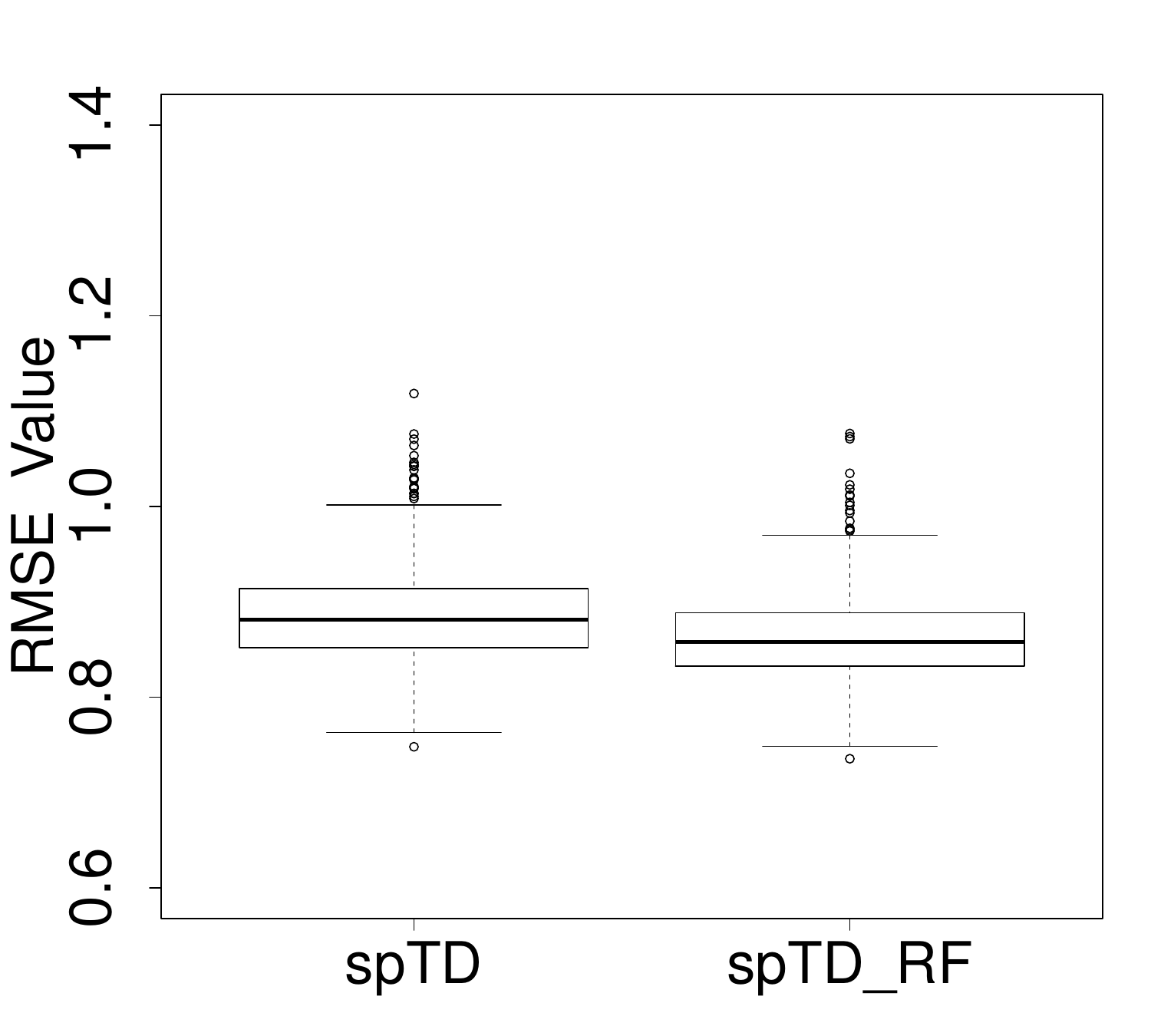}}}
    \end{minipage}\par\medskip
    \begin{minipage}{.3\linewidth}
    \centering
    \subfloat[]{\makebox{\includegraphics[width=\linewidth]{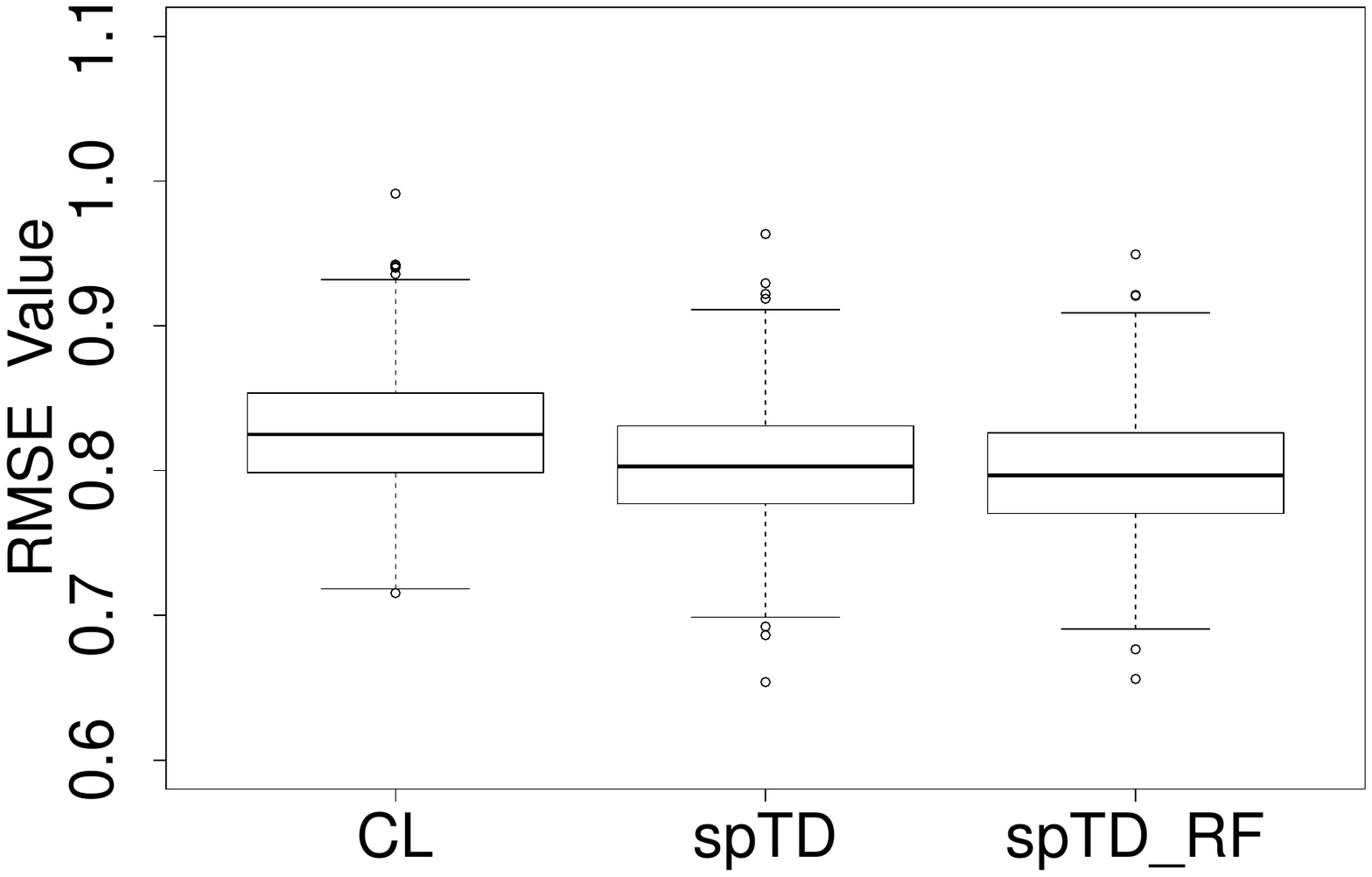}}}
    \end{minipage}
    \begin{minipage}{.3\linewidth}
    \centering
    \subfloat[]{\makebox{\includegraphics[width=\linewidth]{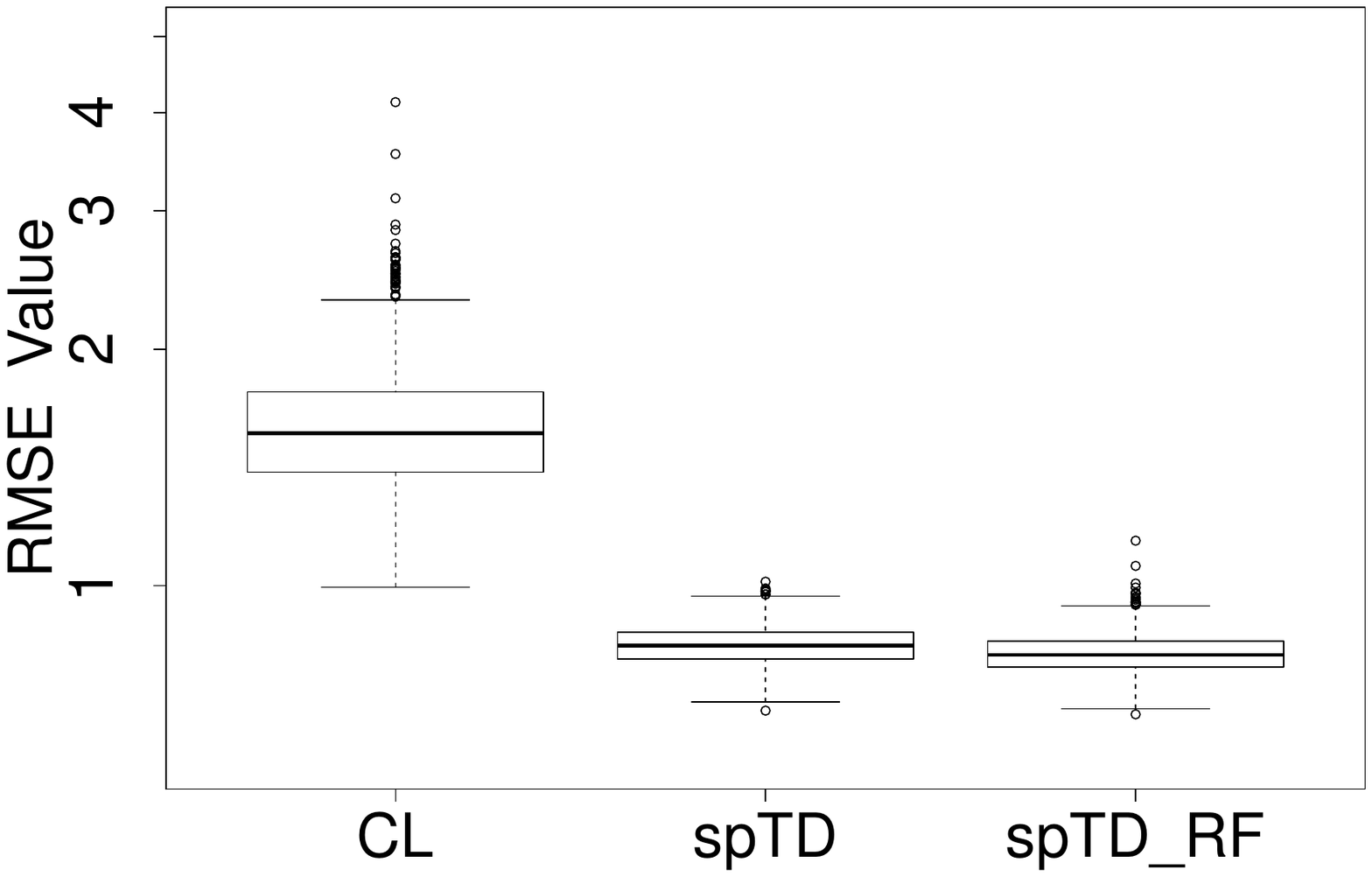}}}
    \end{minipage}
    \begin{minipage}{.22\linewidth}
    \centering
    \subfloat[]{\makebox{\includegraphics[width=\linewidth]{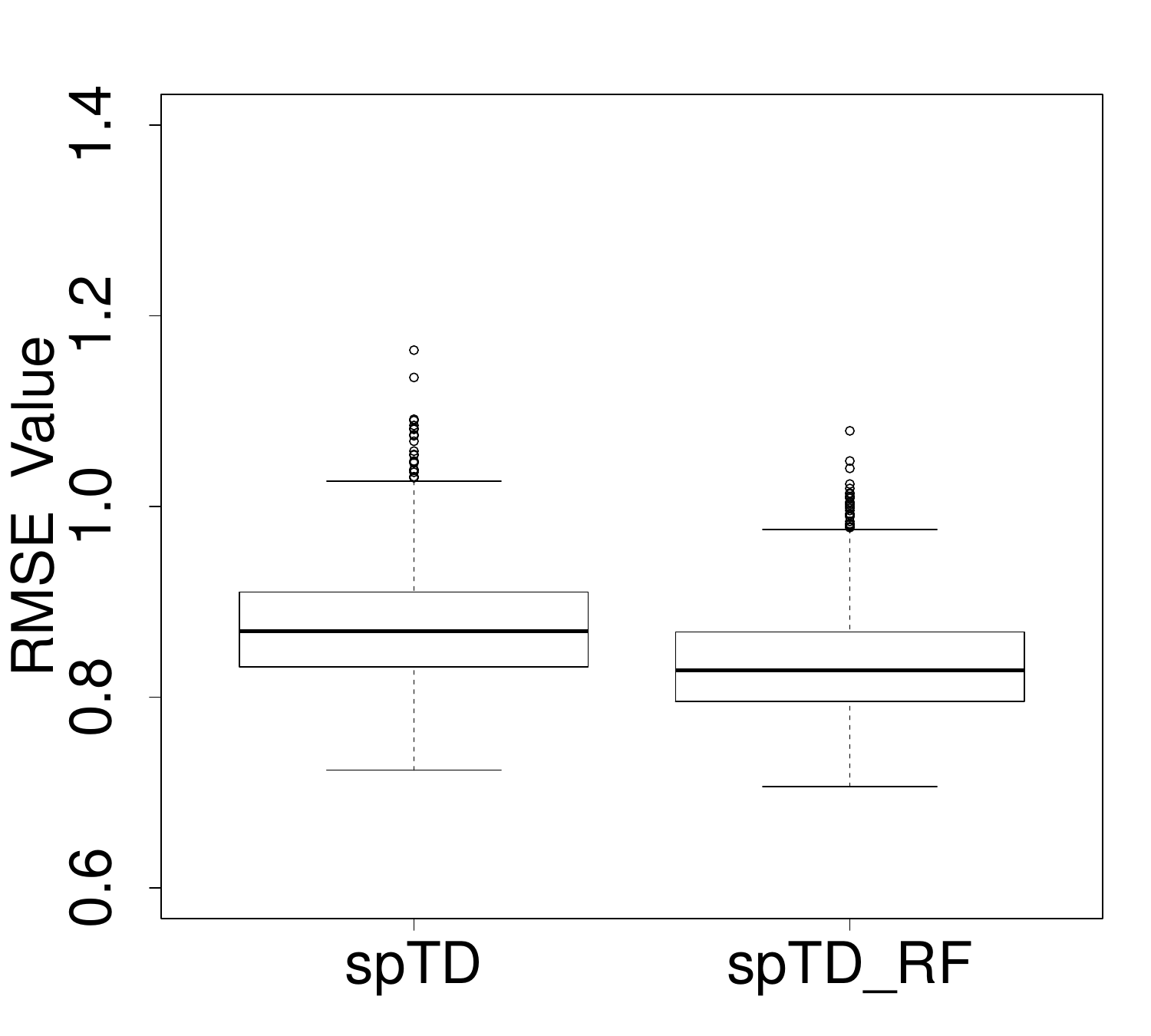}}}
    \end{minipage}\par\medskip
    \begin{minipage}{.3\linewidth}
    \centering
    \subfloat[]{\makebox{\includegraphics[width=\linewidth]{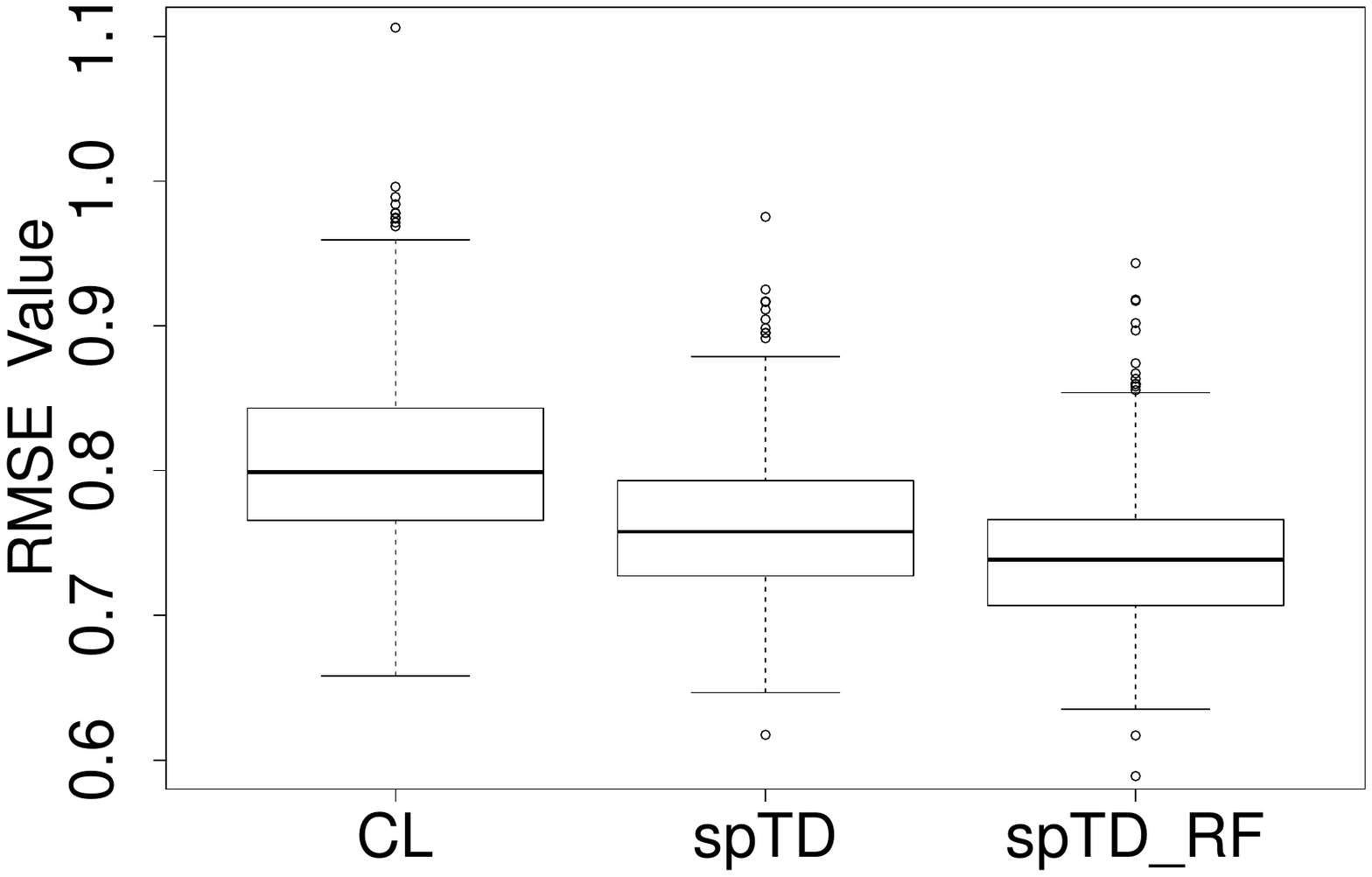}}}
    \end{minipage}
    \begin{minipage}{.3\linewidth}
    \centering
    \subfloat[]{\makebox{\includegraphics[width=\linewidth]{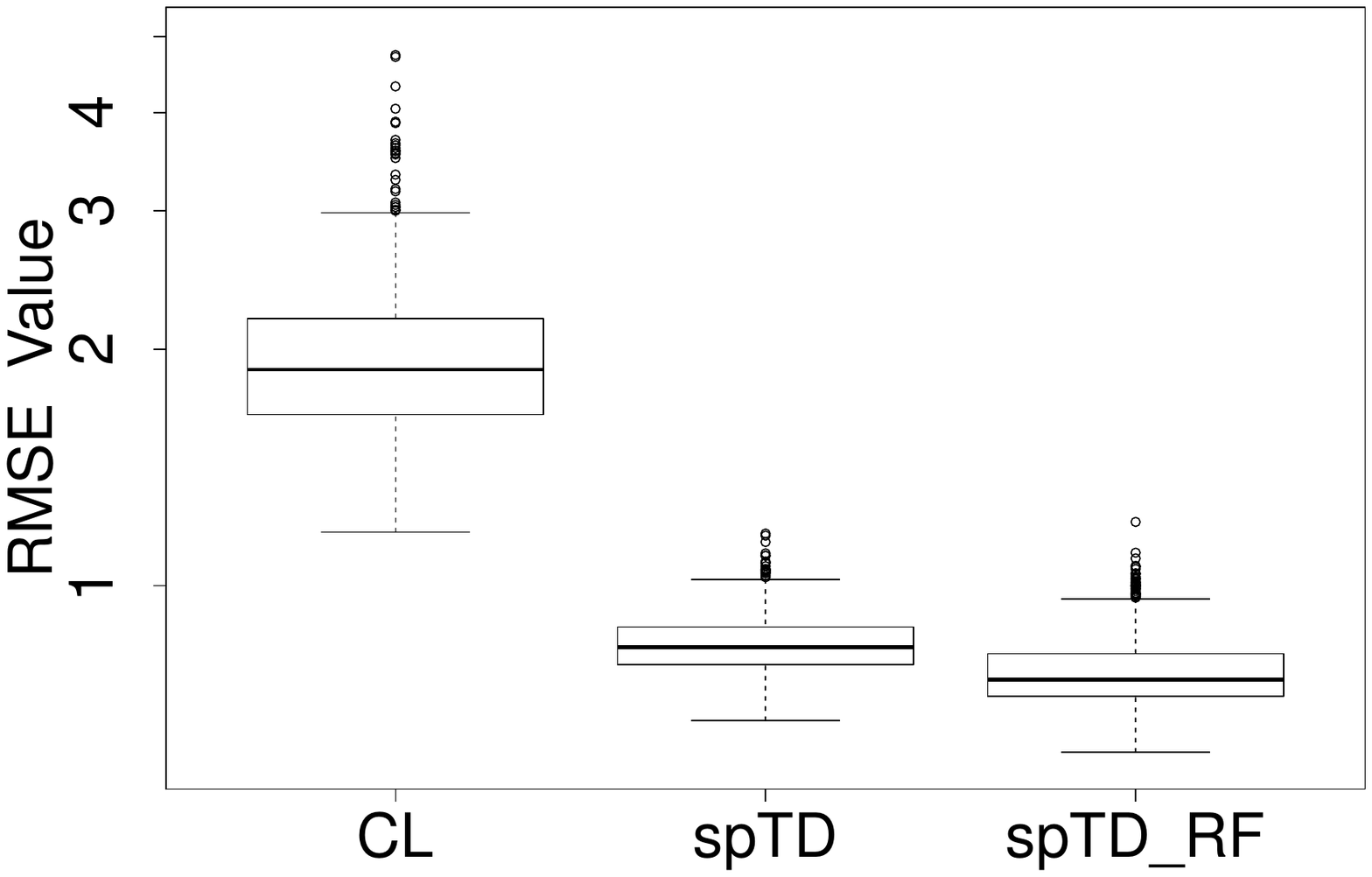}}}
    \end{minipage}
    \begin{minipage}{.22\linewidth}
    \centering
    \subfloat[]{\makebox{\includegraphics[width=\linewidth]{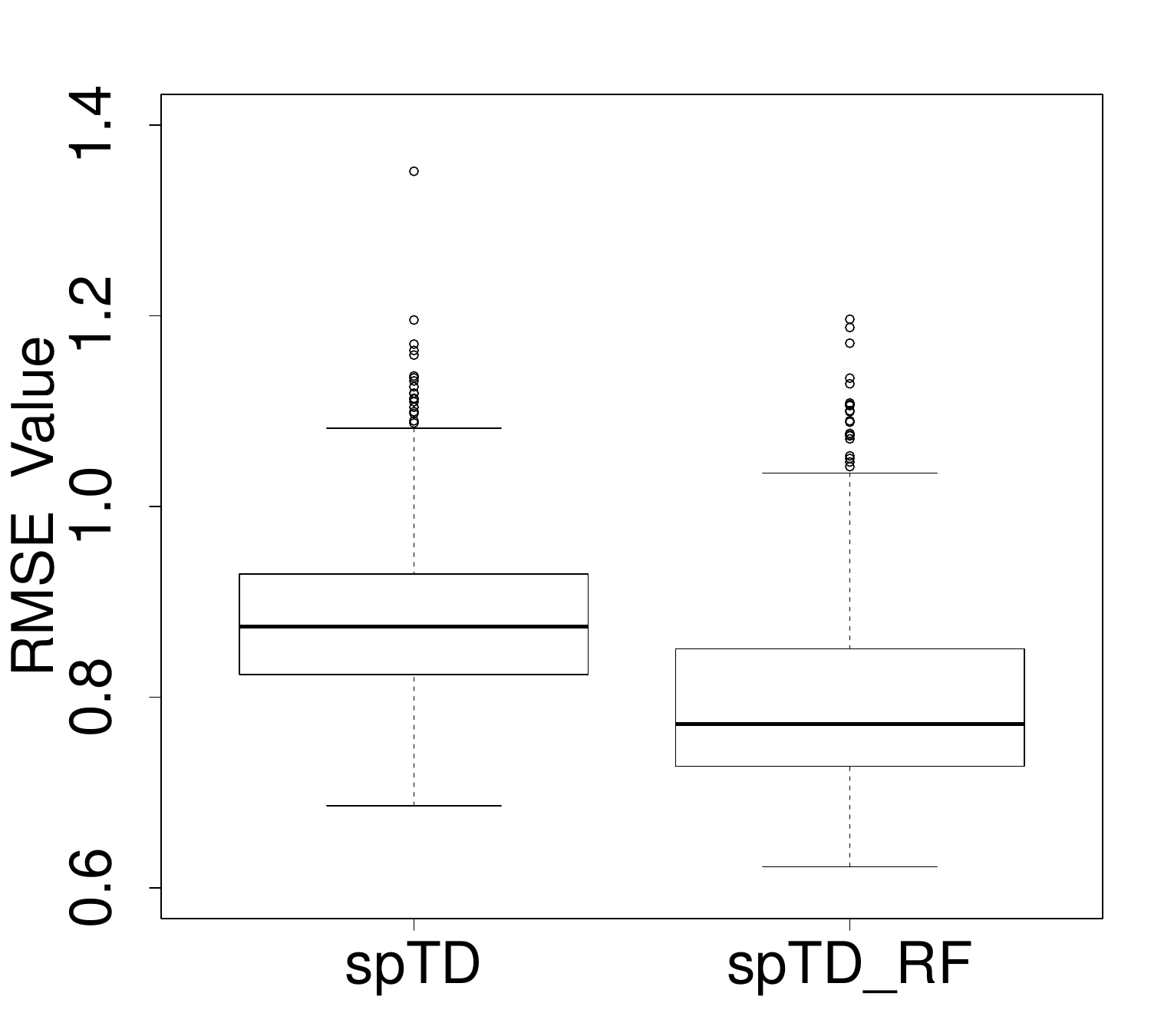}}}
    \end{minipage}\par\medskip
\caption{Boxplots comparing RMSE values of $\hat{\by}_q$ for CL, spTD and spTD\_RF in the non-stationary random walk indicator series experiment. Left, middle and right columns represent $p=30$, $90$ and $150$ respectively. Top, middle and bottom rows represent AR(1) parameter $\rho=0.2$, $0.5$ and $0.8$ respectively. }
\label{fig:RMSE RW}
\end{figure}

To assess how well we estimate $\by_q$, we compute the root mean squared error (RMSE) between the true series and our estimate. Figures \ref{fig:IIDRMSE} and \ref{fig:RMSE RW} display the distribution of RMSE scores over 1000 simulations for $\hat{\by}_q$ in the stationary and non-stationary experiments respectively. The first column in each represents the low dimensional setting with $p=30$. In this setting both CL and the two versions of our method perform similarly well, with spTD\_RF performing slightly better. It is interesting to note that RMSE score is only slightly worse for larger values of the true AR(1) parameter $\rho$, thus, the amount of auto-correlation present in the residuals does have much impact on temporal disaggregation results.

The middle columns represent the moderate dimensional setting, $p=90$, and we start to notice the decrease in performance of CL. To make the comparison visible, we have taken a log-scale of RMSE scores (keeping y-axis values fixed) as the difference in performance is so large. Again with spTD\_RF performing slightly better than spTD, the proposed methods clearly perform a great deal better in moderate dimension scenarios in accurately disaggregating time series.

The columns on the right represent the high dimensional setting of $p=150$. In this setting, CL can no longer be used, whereas, both our methods still perform very well with low RMSE scores across the 1000 simulations. Additionally, experiments with even larger dimension of $p=400$ resulted in similar low RMSE scores for the methods we propose. A further interesting property of these results is that RMSE scores are generally lower in all scenarios for the non-stationary random walk indicator series experiment (Figure \ref{fig:RMSE RW}). This is a consequence of known results in the statistics literature showing least-squares estimators of the parameters for non-stationary but co-integrated series can converge to their true values faster than stationary series \citep{johansen1988statistical}. 

\subsection{Recovery of the True Regression Coefficient}

In addition to the recovery of the true high-frequency series, a further motivation for temporal disaggregation is to understand potential drivers of short term change. To this end, we here assess the recovery of the regression coefficients. Table \ref{tab:beta_metrics} (Appendix) provides the mean and standard deviation (in brackets) across 1000 iterations of our simulation assessing the recovery of the true $\bbeta$ for the three methods CL, spTD and spTD\_RF. It is quite clear from this table that spTD\_RF provides the lowest score on all three metrics used across all scenarios. The improvement of our method on CL in the moderate dimension scenario can again be seen by the RMSE scores. Interestingly, as $p$ increases, the RMSE for our methods decreases, meaning we are closer to recovering the true values of the support of $\bbeta$. The $\ell_\infty := \max_{i=1,\dots,p}|\hat{\beta}_i - \beta_i|$ metric has been included to measure the worst case error, again showing the improvements we make on CL. RMSE and $\ell_\infty$ are on average lower with non-stationary indicator series, however, more false positives (FP) are included as seen in the tables. The refitting step certainly reduces the number of FP and it is interesting to note that as the AR(1) parameter $\rho$ increases, the number of FP on average decreases.  

\begin{figure}[htbp]
    \begin{minipage}{.32\linewidth}
    \centering
    \subfloat[]{\makebox{\includegraphics[width=\linewidth]{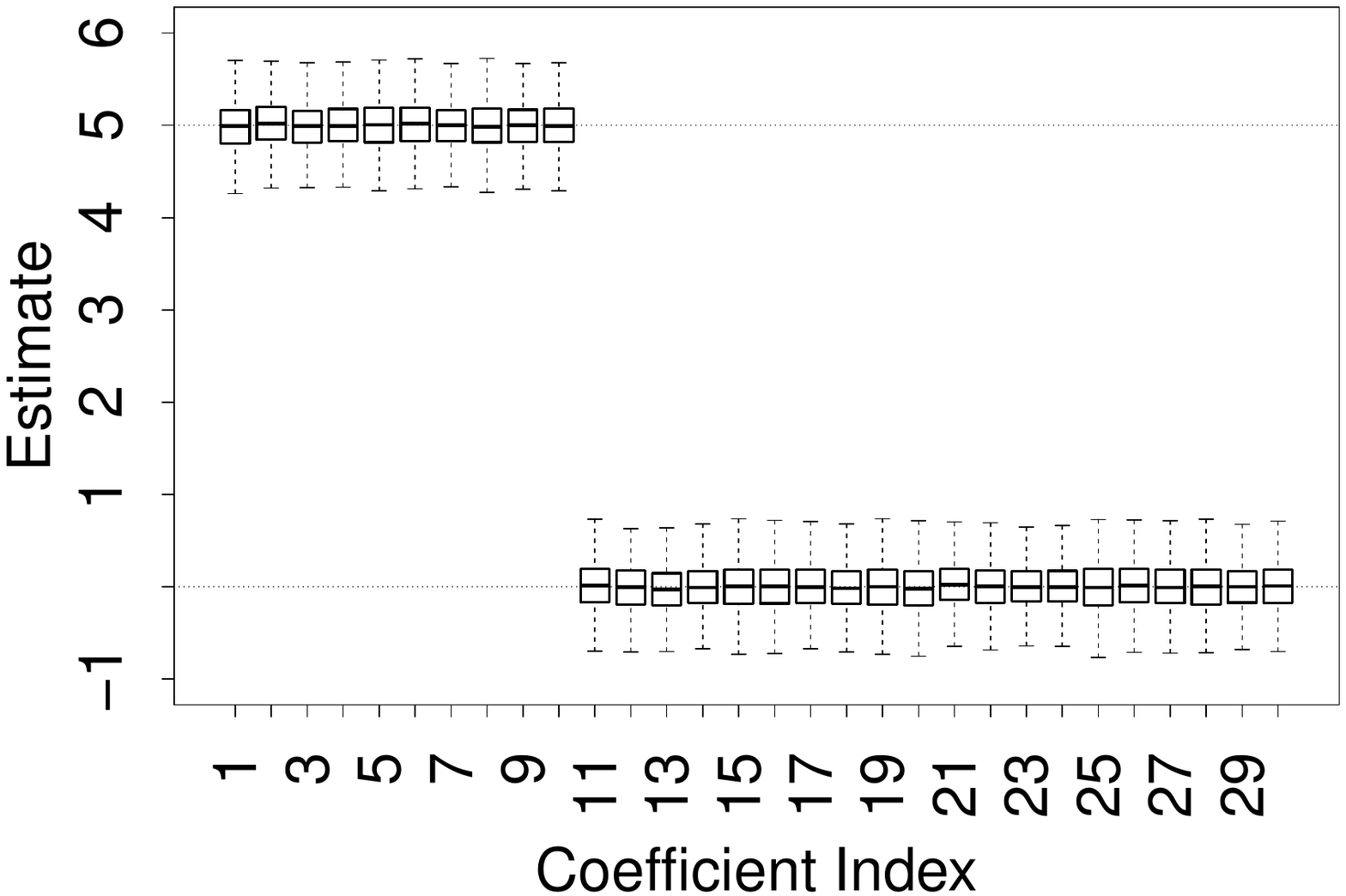}}}
    \end{minipage}
    \begin{minipage}{.32\linewidth}
    \centering
    \subfloat[]{\makebox{\includegraphics[width=\linewidth]{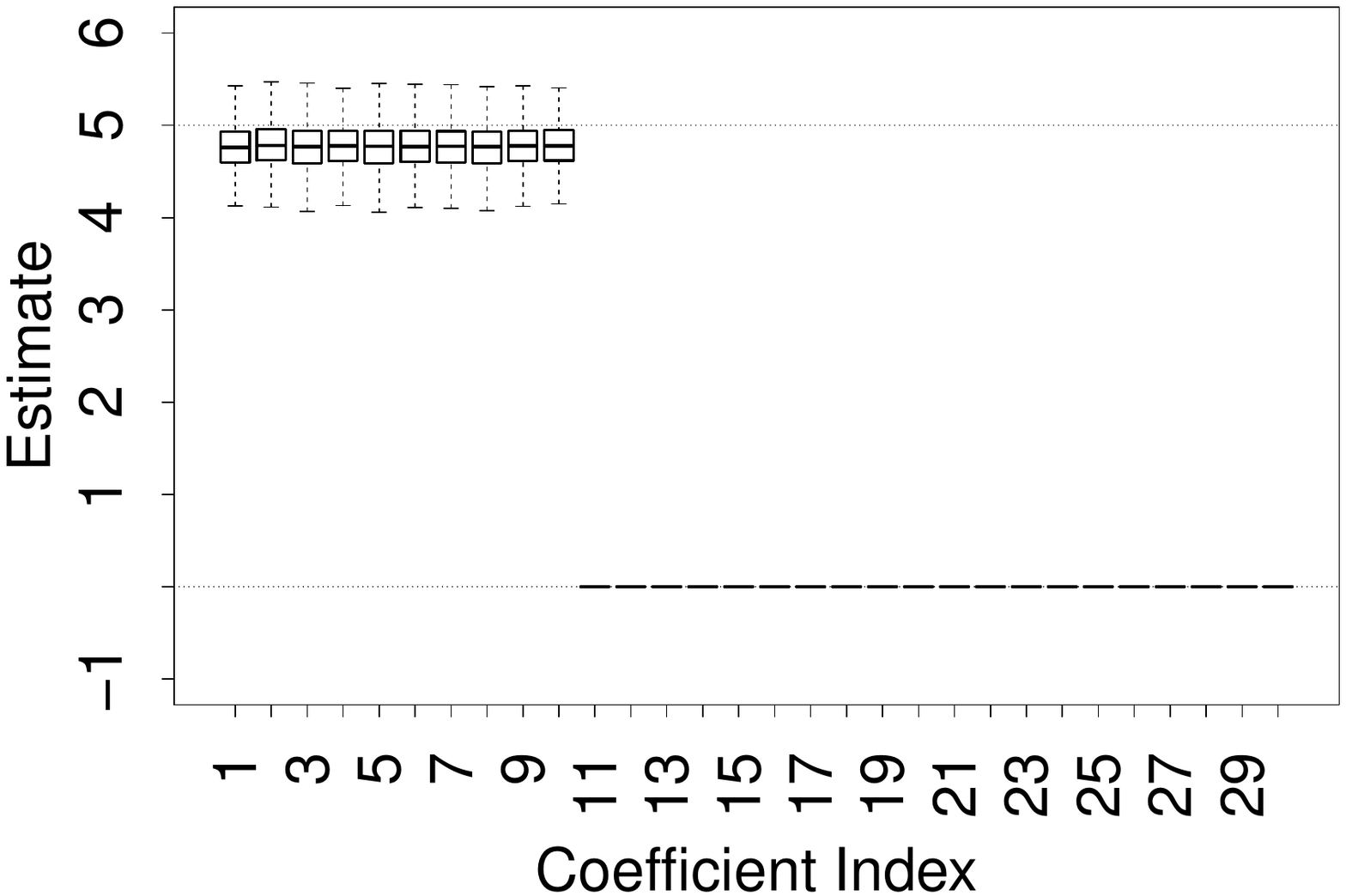}}}
    \end{minipage}
    \begin{minipage}{.32\linewidth}
    \centering
    \subfloat[]{\makebox{\includegraphics[width=\linewidth]{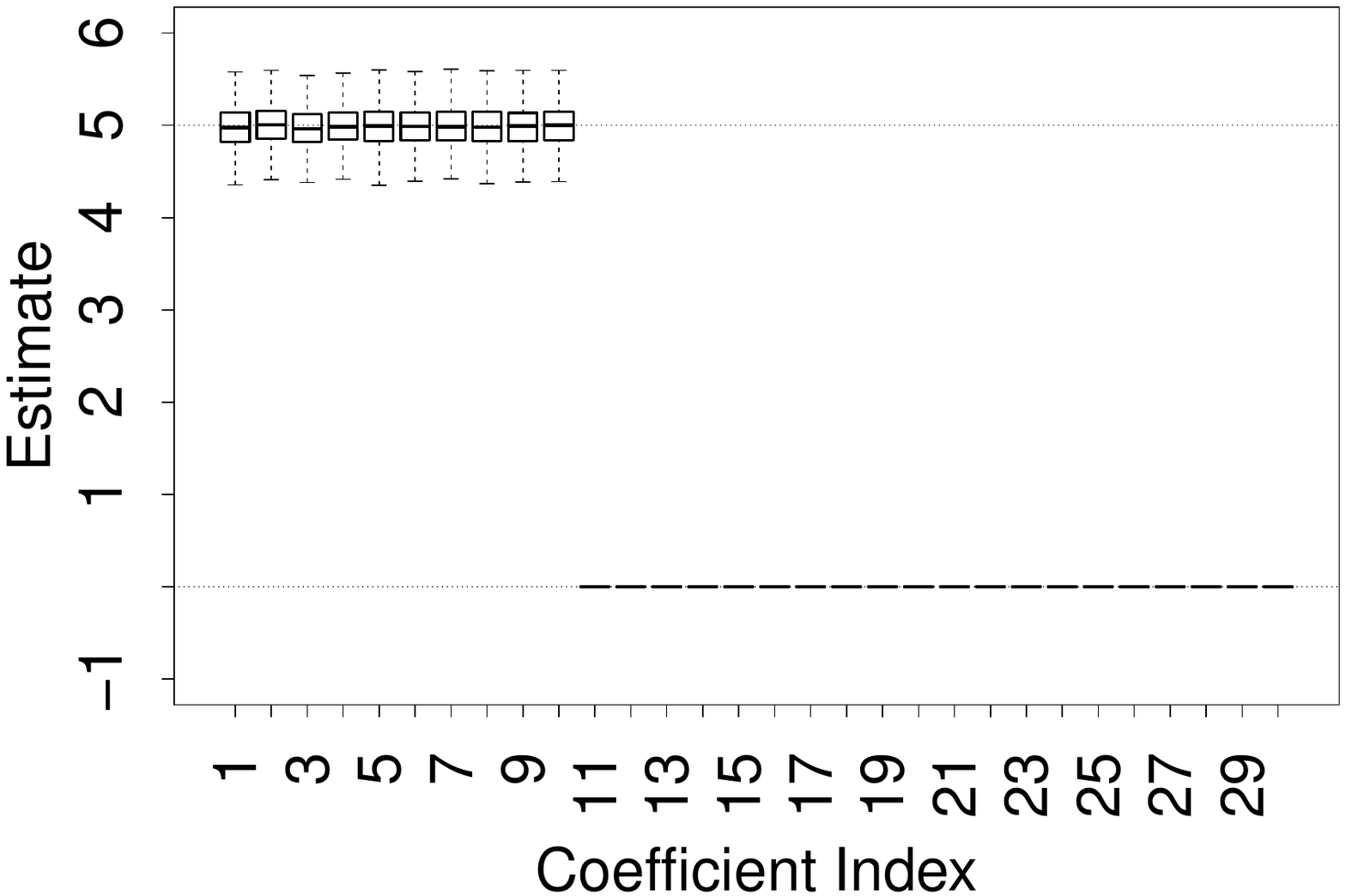}}}
    \end{minipage}\par\medskip
    \begin{minipage}{.32\linewidth}
    \centering
    \subfloat[]{\makebox{\includegraphics[width=\linewidth]{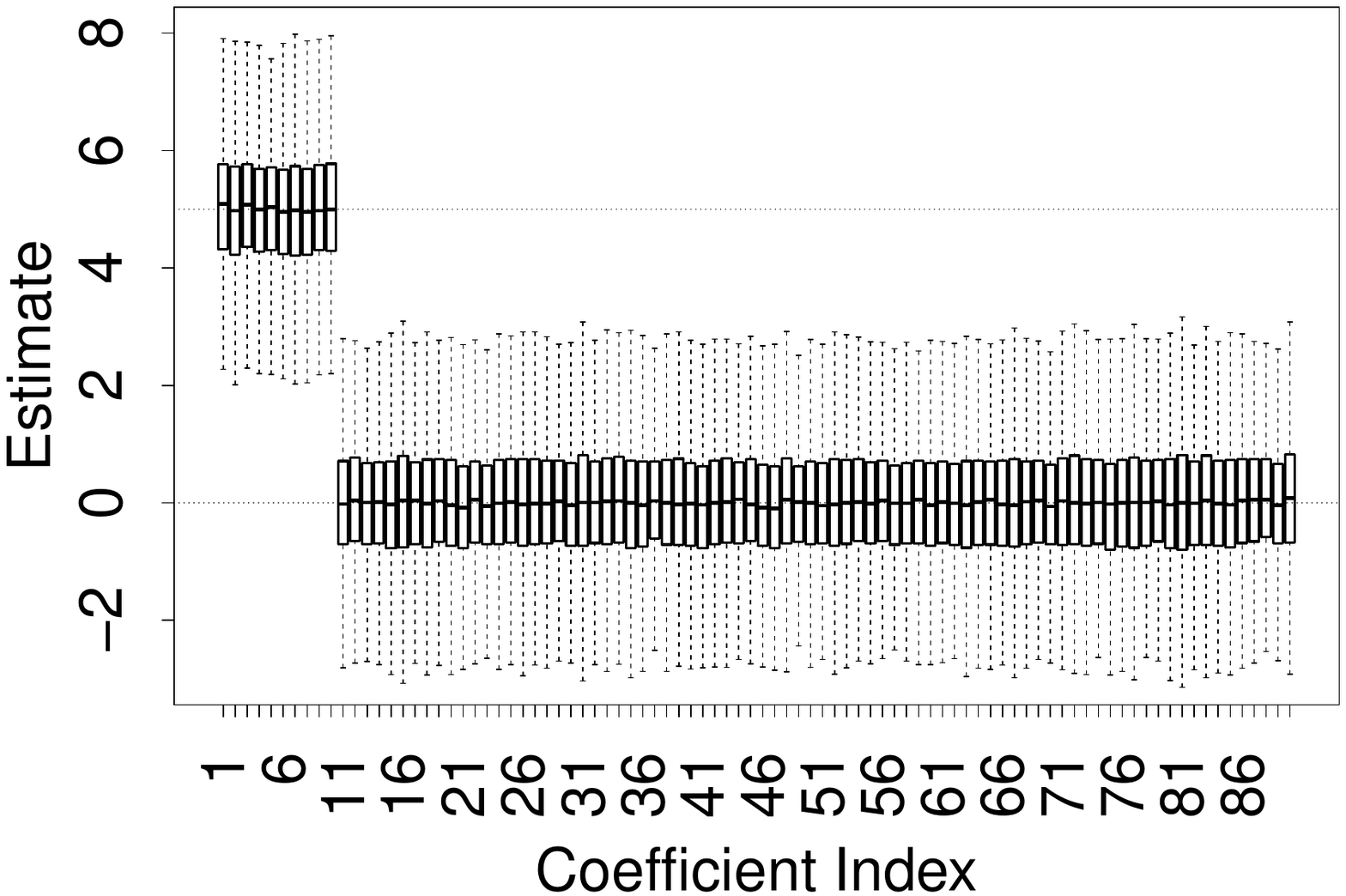}}}
    \end{minipage}
    \begin{minipage}{.32\linewidth}
    \centering
    \subfloat[]{\makebox{\includegraphics[width=\linewidth]{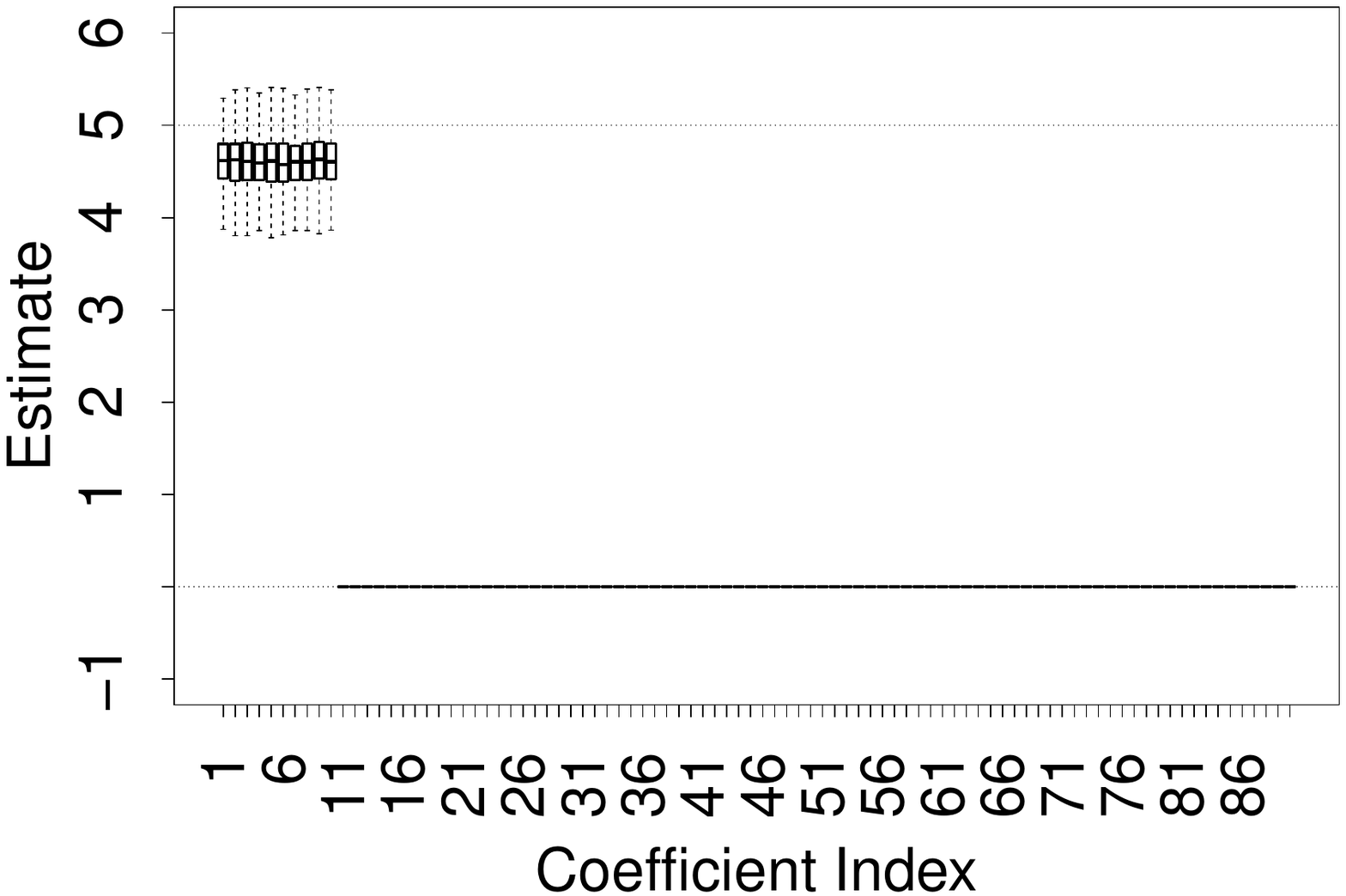}}}
    \end{minipage}
    \begin{minipage}{.32\linewidth}
    \centering
    \subfloat[]{\makebox{\includegraphics[width=\linewidth]{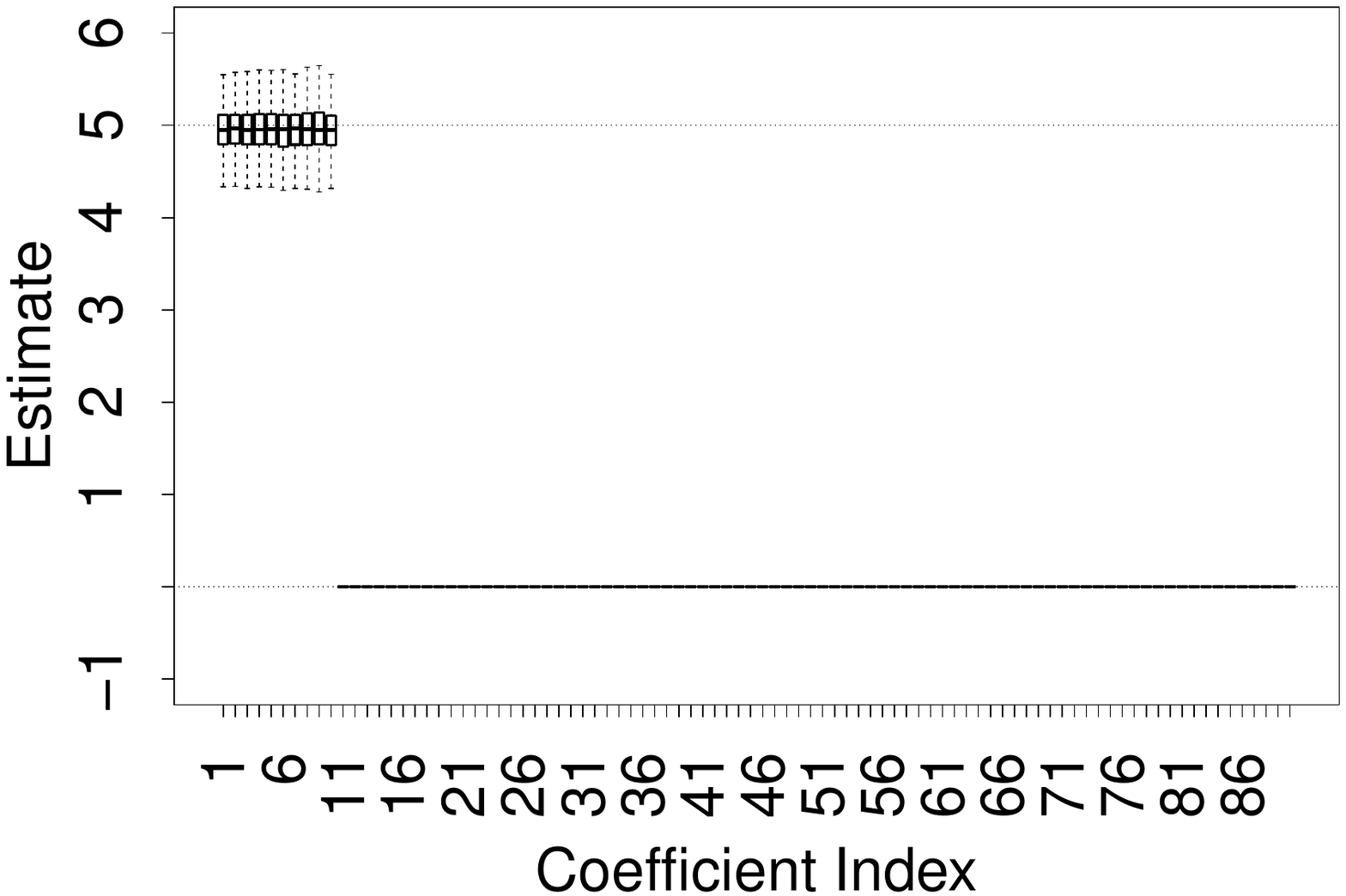}}}
    \end{minipage}\par\medskip
\caption{Boxplots for the estimates $\hat{\beta}_1,\dots,\hat{\beta}_p$ using CL, spTD and spTD\_RF (left to right respectively) with $\rho=0.5$ and $p=30$ (top row) and $p=90$ (bottom row) with stationary indicator series. }
\label{fig:IIDcoefboxplots}
\end{figure}

Figure \ref{fig:IIDcoefboxplots} shows the distribution for each $\bbeta$ parameter over 1000 iterations in the stationary indicator series experiment with CL, spTD and spTD\_RF from left to right and $p=30$ on top and $p=90$ on the bottom. There is very large variance in the CL boxplots, with tails overlapping in the $p=90$ setting (plot d). In comparison to our methods, the variance is very narrow around the true $\bbeta$ values at 5 and 0. It is evident here to see the refitting strategy in action in plots (c) and (f), as the estimates in (b) and (e) have reduced in bias and shifted upwards towards the true support value of 5. Furthermore, the shrinkage effect for our methods are evident by there not being boxes for estimates $\hat{\beta}_6$ onwards. Except for outliers, which have not been shown on this plot, our method is able to correctly set coefficients to zero, which from a practical point of view means analysts/practitioners are able to identify irrelevant indicator series. \vspace{2ex} 

\subsection{Distribution of AR(1) Parameter Estimate}

We restrict our grid search for $\hat{\rho}$ to the positive stationary domain $[0.01,0.99]$. 
Performing the search on the positive domain has been motivated previously in Chow-Lin related studies \citep{sax2013temporal,ciammola2005temporal,miralles2003performance}.

\begin{figure}[htbp]
\centering
    \begin{minipage}{.3\linewidth}
    \centering
    \subfloat[]{\makebox{\includegraphics[width=\linewidth]{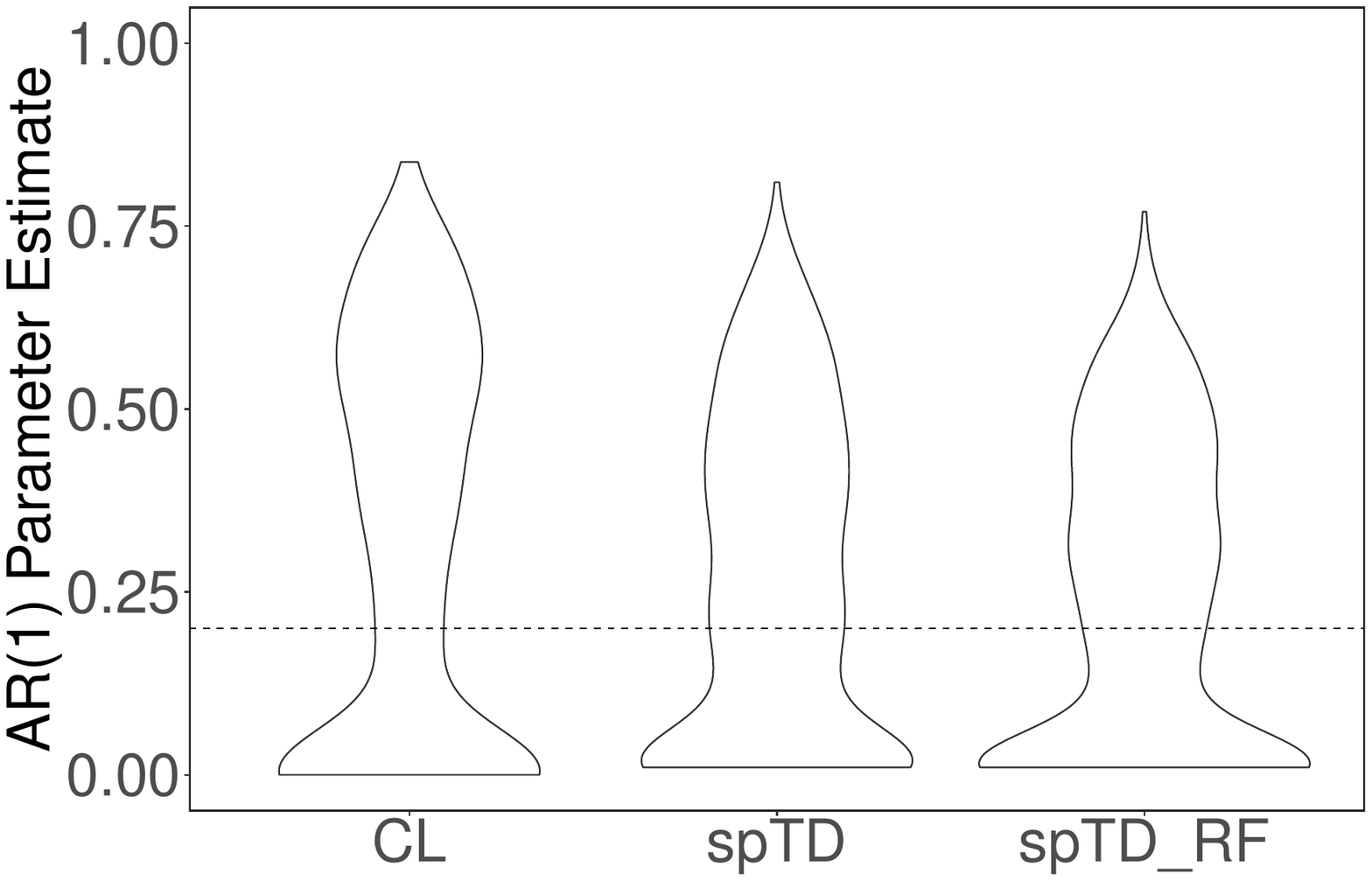}}}
    \end{minipage}
    \begin{minipage}{.3\linewidth}
    \centering
    \subfloat[]{\makebox{\includegraphics[width=\linewidth]{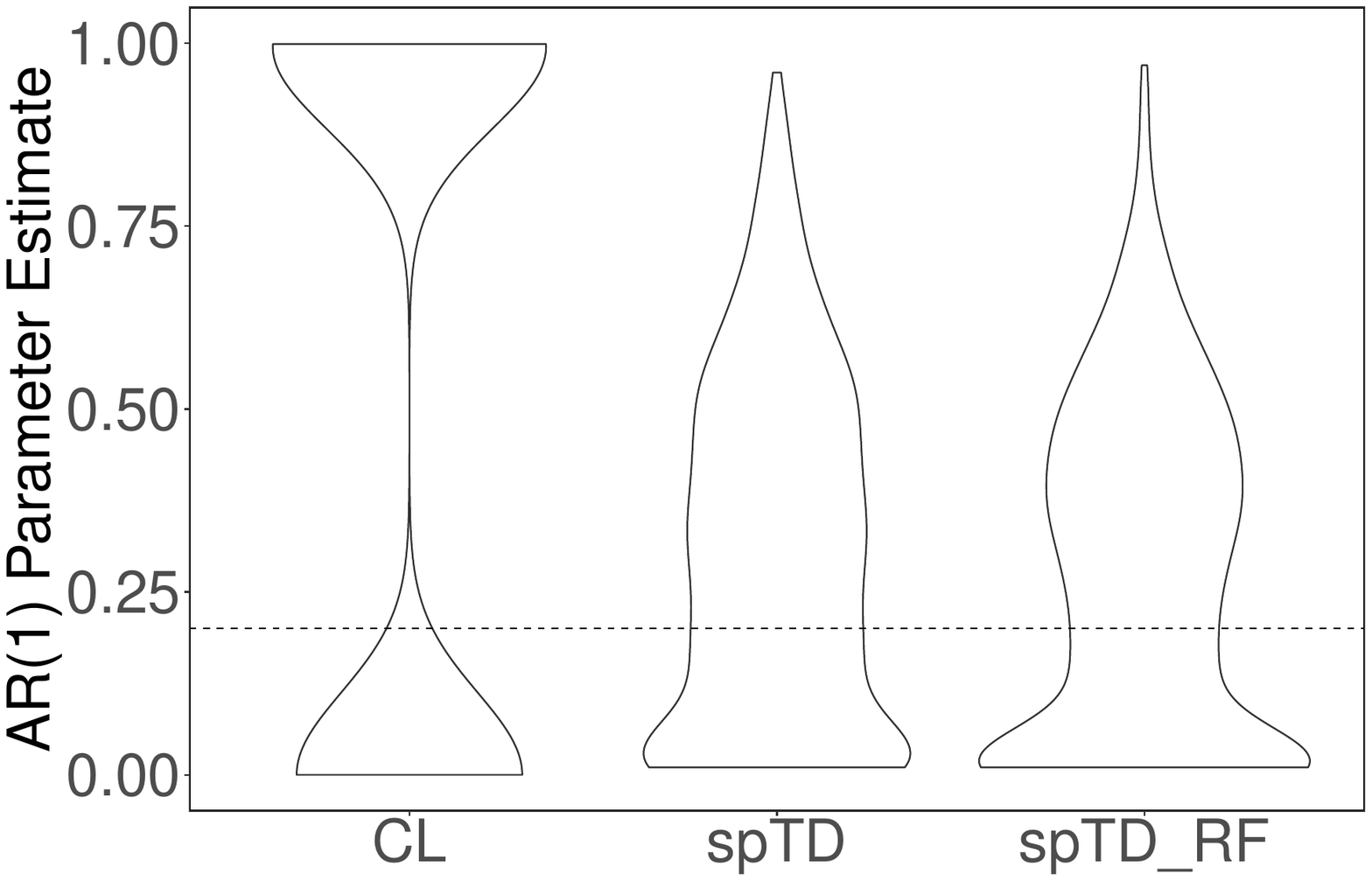}}}
    \end{minipage}
    \begin{minipage}{.22\linewidth}
    \centering
    \subfloat[]{\makebox{\includegraphics[width=\linewidth]{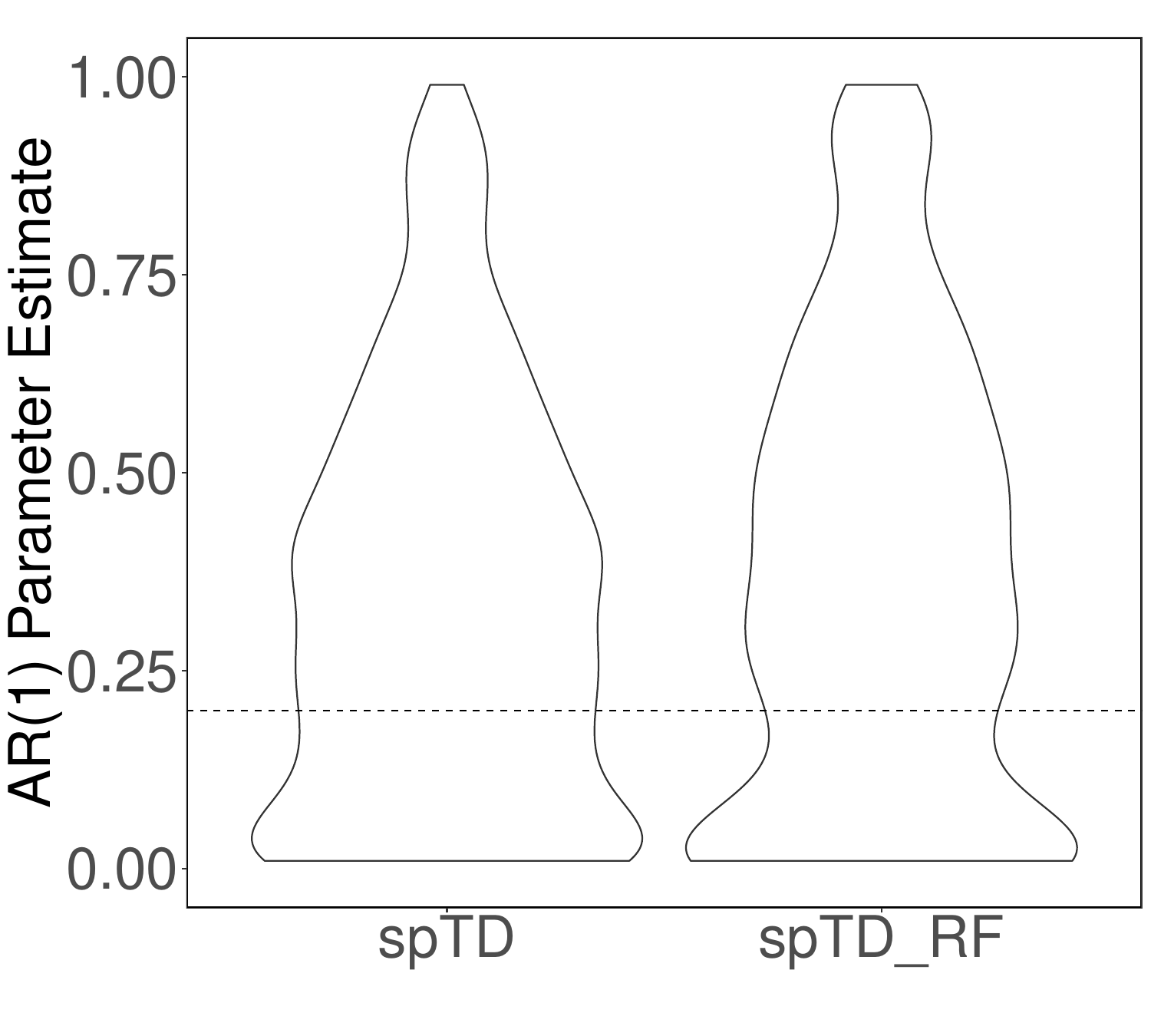}}}
    \end{minipage}\par\medskip
    \begin{minipage}{.3\linewidth}
    \centering
    \subfloat[]{\makebox{\includegraphics[width=\linewidth]{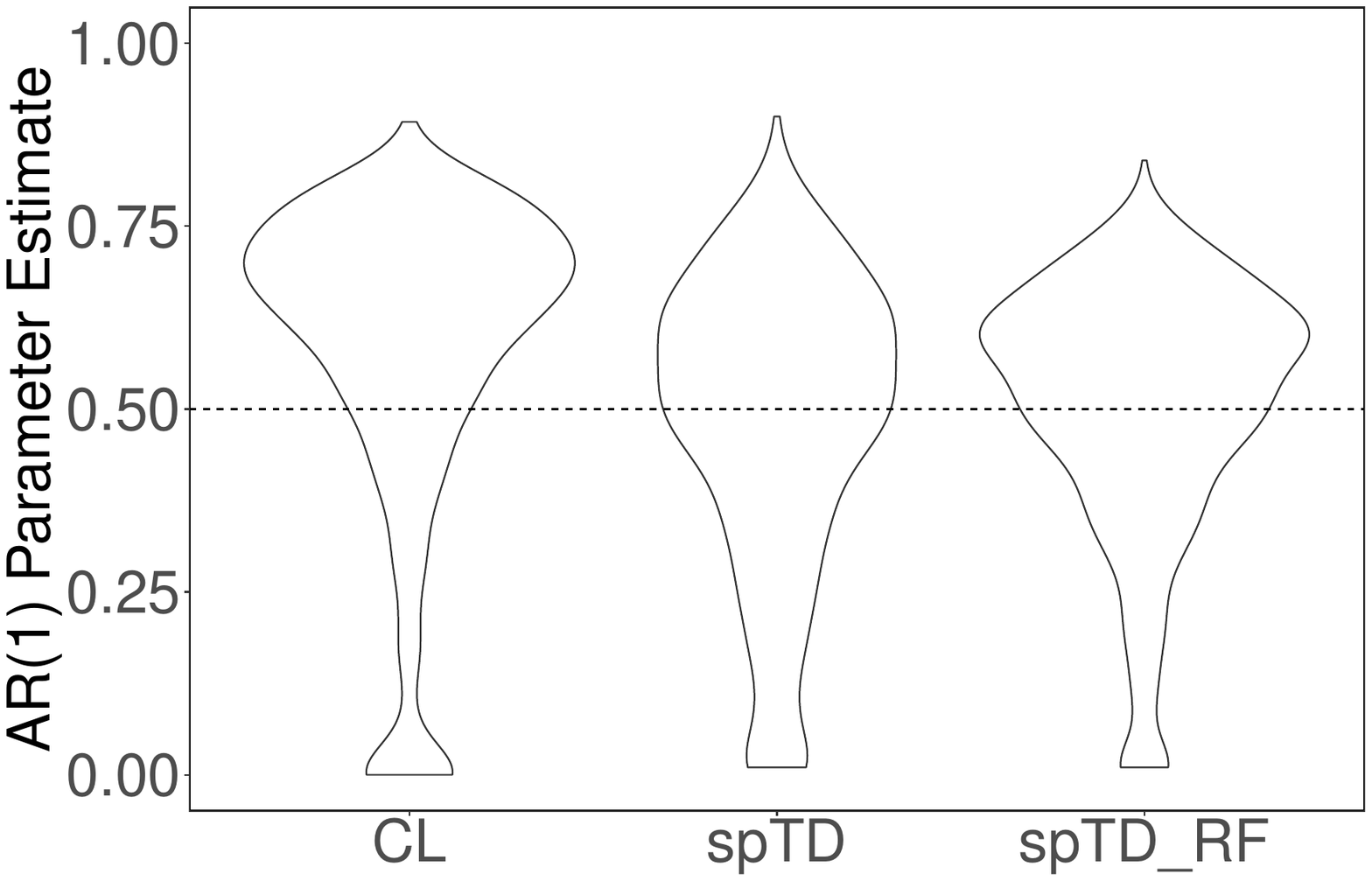}}}
    \end{minipage}
    \begin{minipage}{.3\linewidth}
    \centering
    \subfloat[]{\makebox{\includegraphics[width=\linewidth]{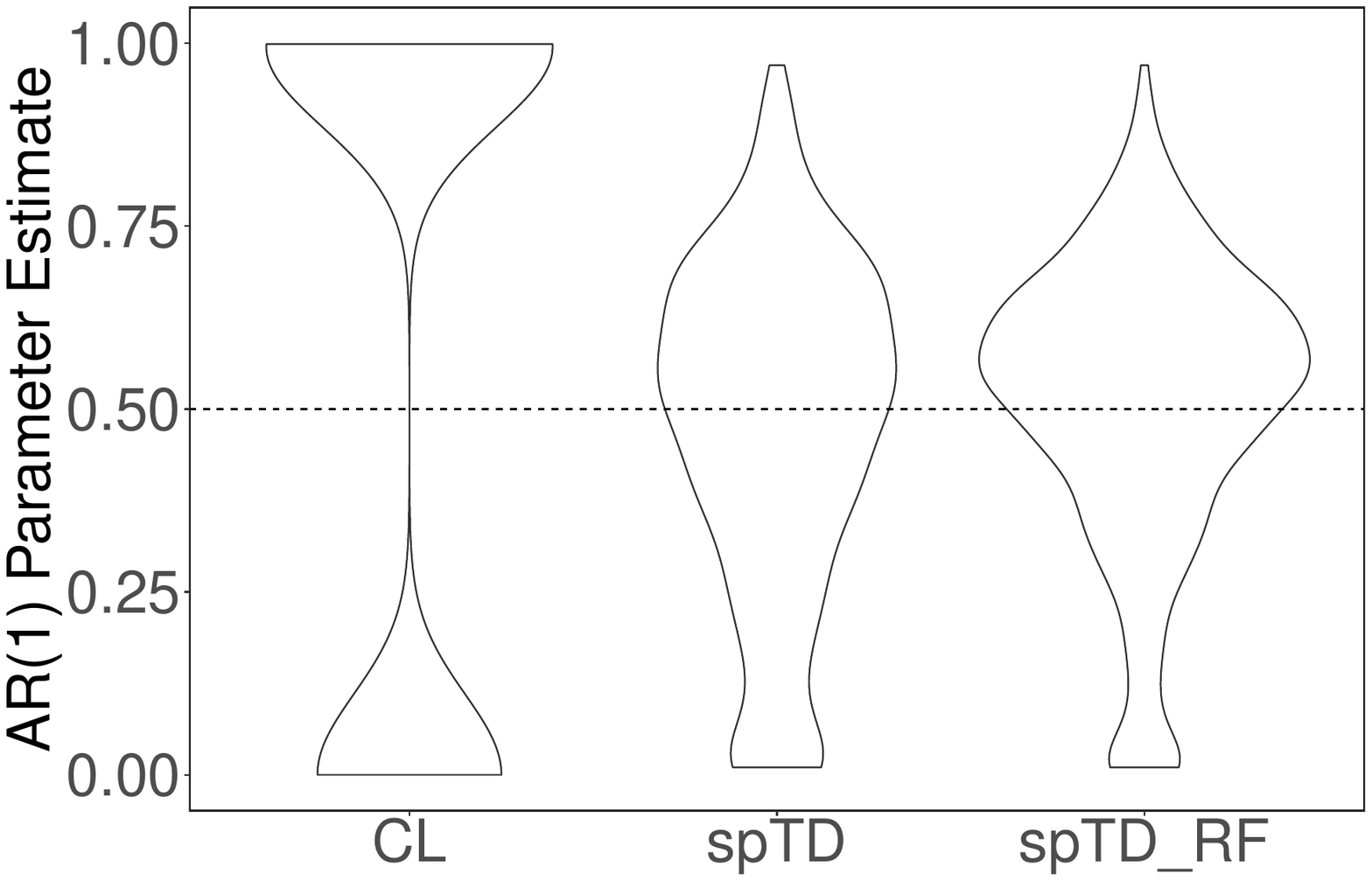}}}
    \end{minipage}
    \begin{minipage}{.22\linewidth}
    \centering
    \subfloat[]{\makebox{\includegraphics[width=\linewidth]{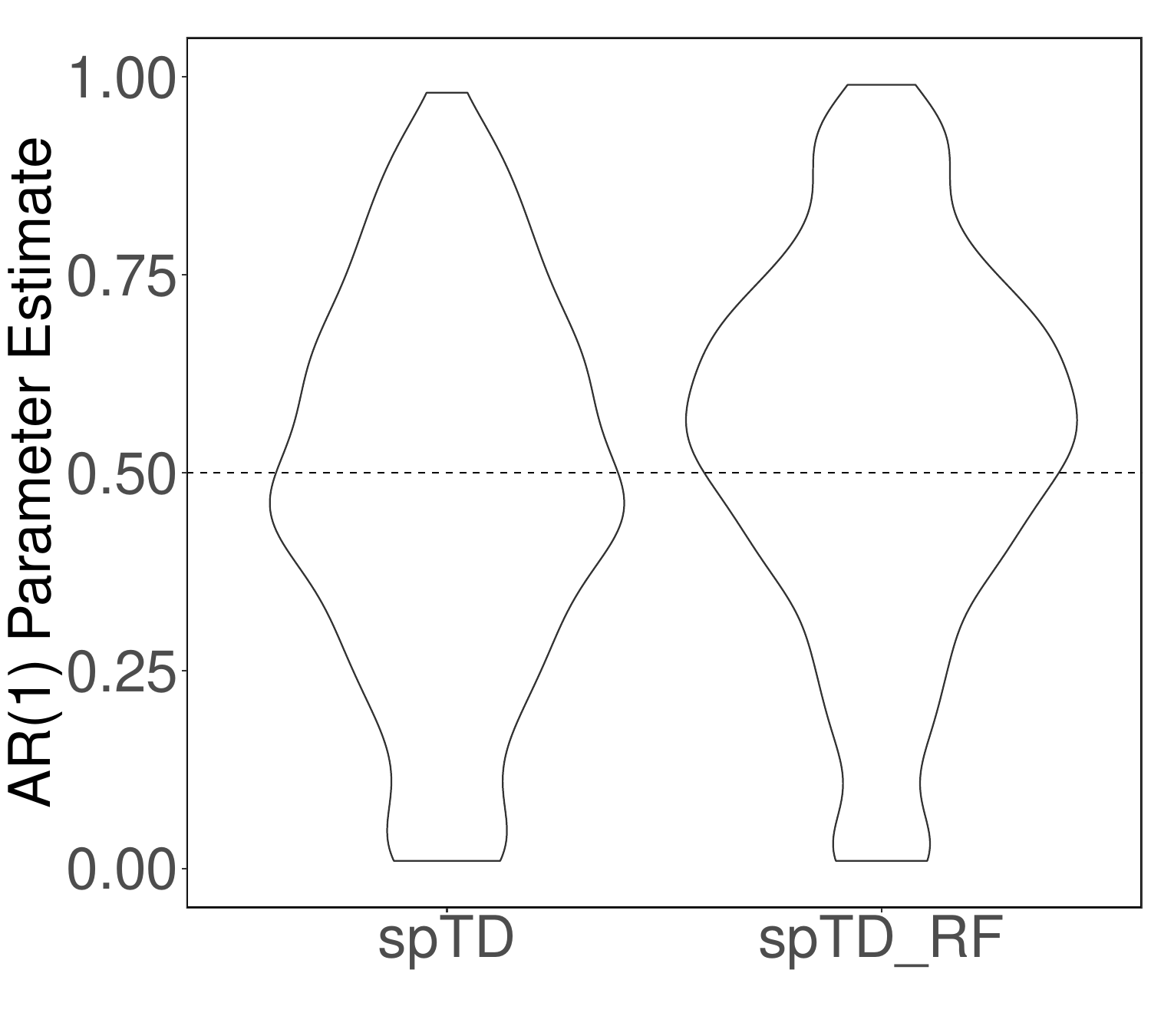}}}
    \end{minipage}\par\medskip
    \begin{minipage}{.3\linewidth}
    \centering
    \subfloat[]{\makebox{\includegraphics[width=\linewidth]{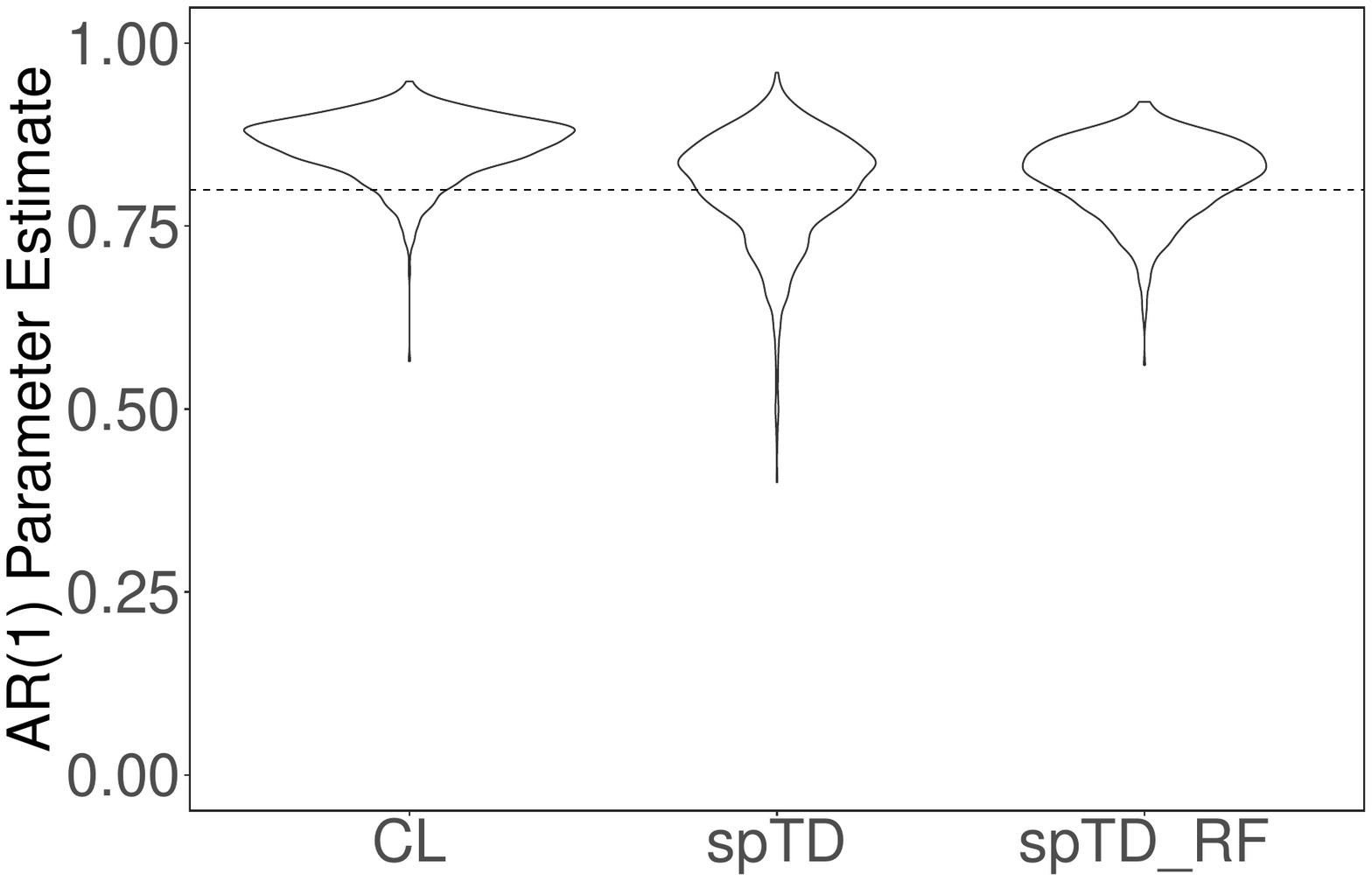}}}
    \end{minipage}
    \begin{minipage}{.3\linewidth}
    \centering
    \subfloat[]{\makebox{\includegraphics[width=\linewidth]{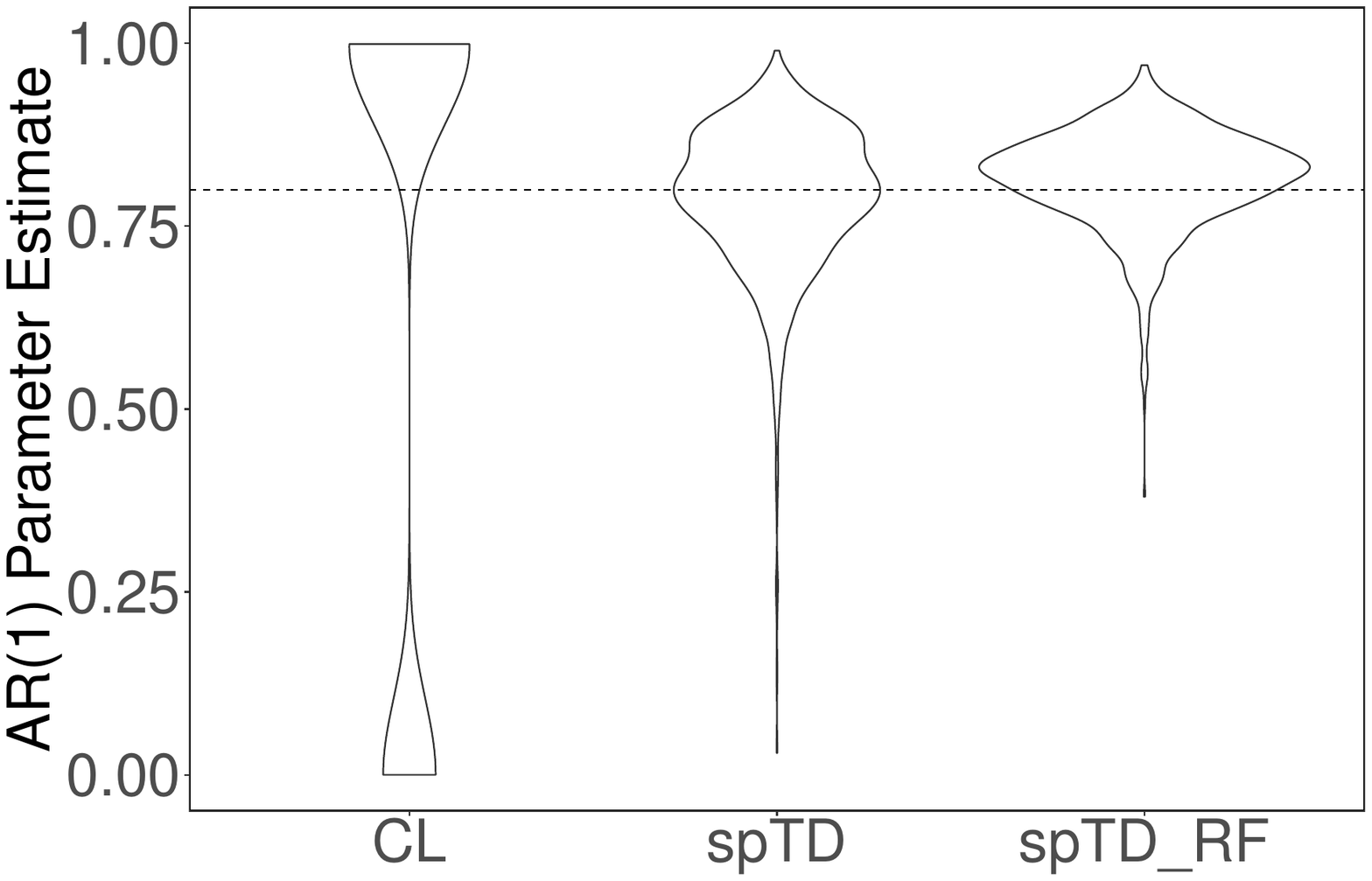}}}
    \end{minipage}
    \begin{minipage}{.22\linewidth}
    \centering
    \subfloat[]{\makebox{\includegraphics[width=\linewidth]{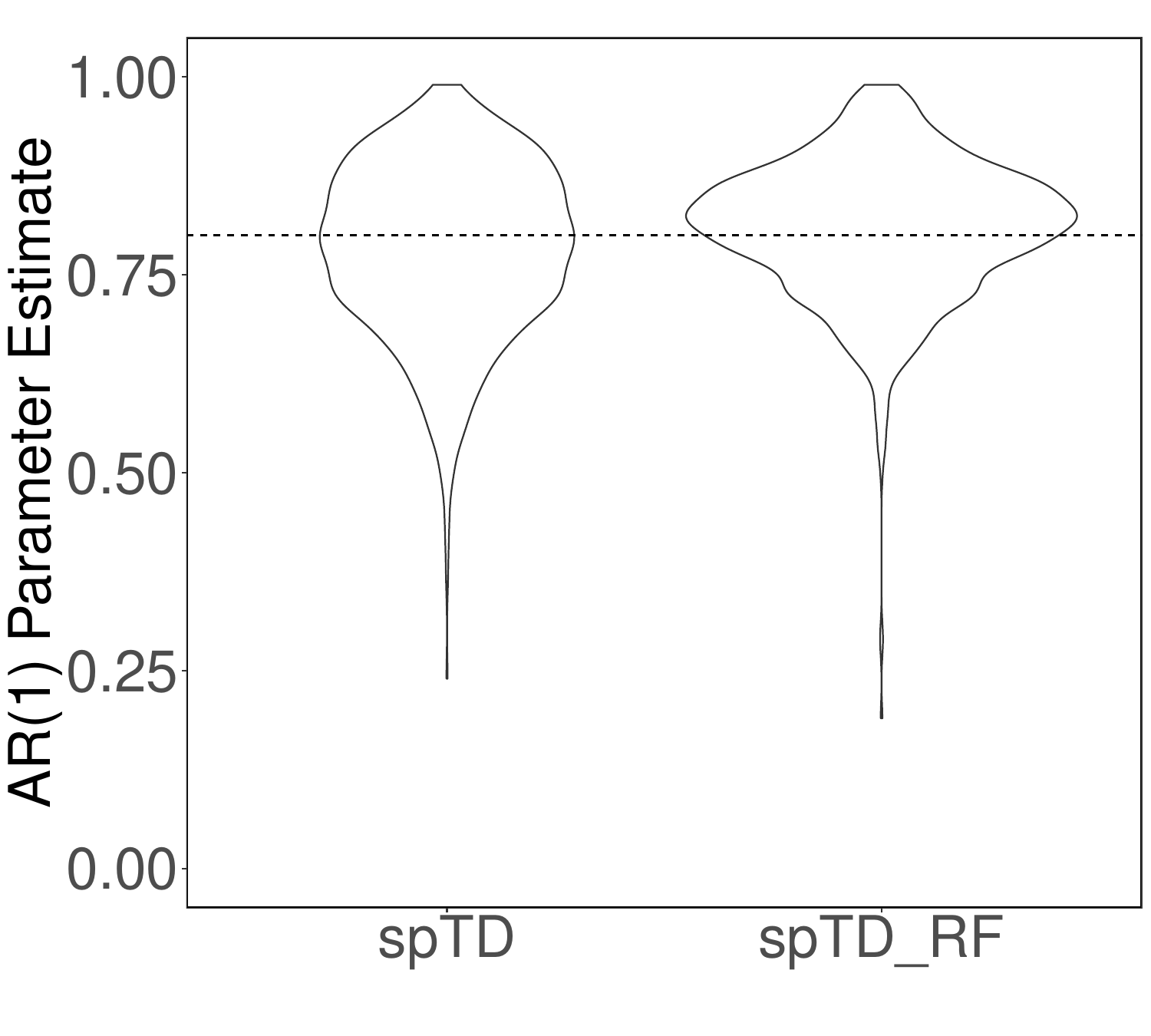}}}
    \end{minipage}\par\medskip
\caption{Violin plots showing the distribution of $\hat{\rho}$ for CL, spTD and spTD\_RF using stationary indicator series. Dimension $p=30$, $90$ and $150$ from left to right and true AR parameter $\rho$ indicated by horizontal dashed line. }
\label{fig:rhoplot}
\end{figure}

Figure \ref{fig:rhoplot} (Appendix) displays the distribution of $\hat{\rho}$ over 1000 iterations for the stationary indicator series experiment. A horizontal line is drawn at the true value of $\rho$ to compare performance. In the first column representing $p=30$ we observe how all three methods tend to over-estimate the true value of $\rho$, with CL over-estimating the most. There is also a large variance associated with the estimator when $\rho=0.2$ and this is again the case when $p=90$ (middle column). In this moderate dimensional setting CL fails to identify the true $\rho$ completely with it seeming to only select $\rho$ at the boundaries of the search space $(0,1)$. The proposed methods seem to perform similarly well with the centre of mass surrounding the correct values. In the high dimensional setting we are again centred around the true values with a little more variance. We believe these estimates will be a lot more accurate with a better way of defining BIC in high dimensions. 

\begin{figure}[htbp]
\centering
    \begin{minipage}{.3\linewidth}
    \centering
    \subfloat[]{\makebox{\includegraphics[width=\linewidth]{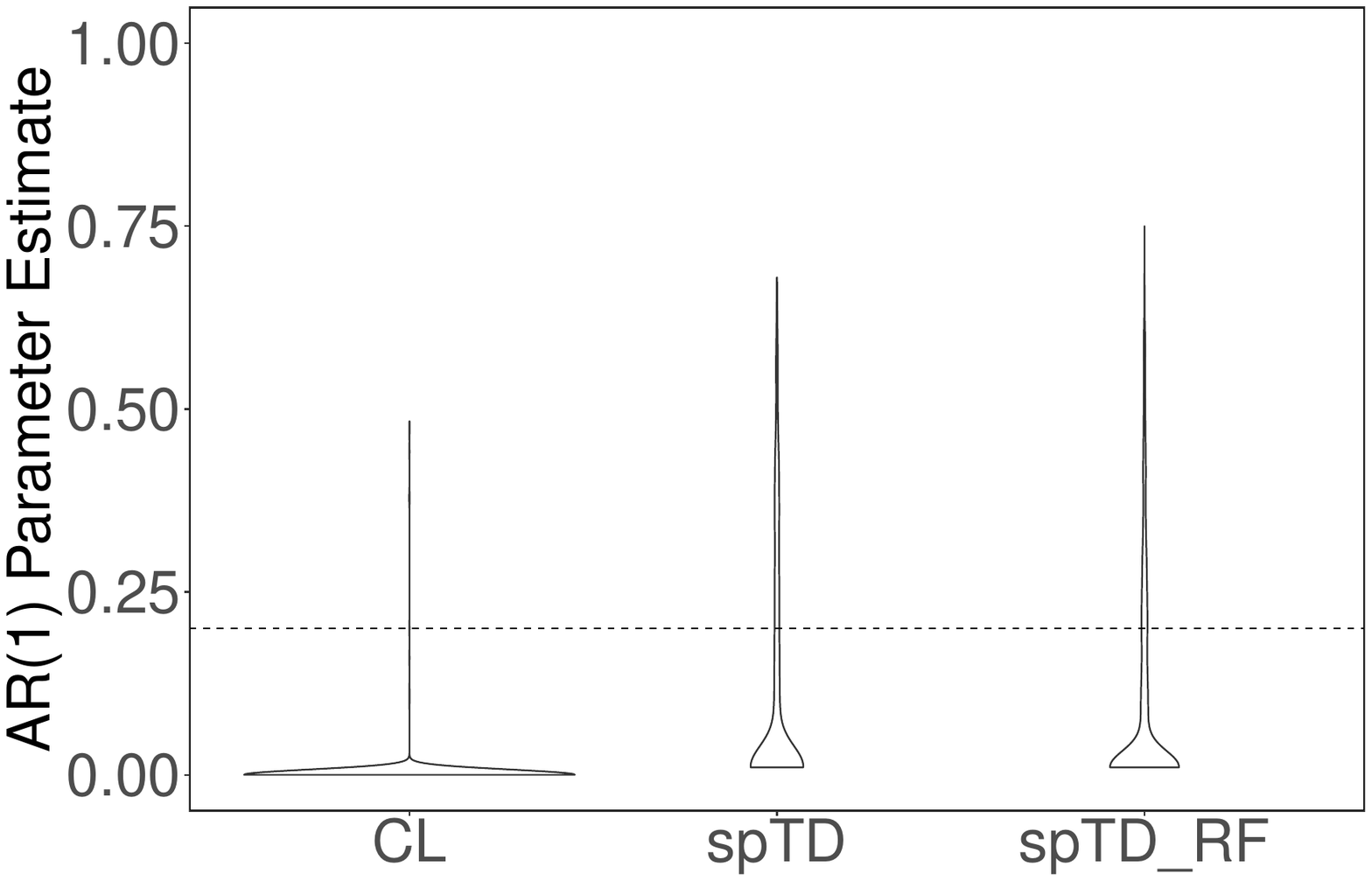}}}
    \end{minipage}
    \begin{minipage}{.3\linewidth}
    \centering
    \subfloat[]{\makebox{\includegraphics[width=\linewidth]{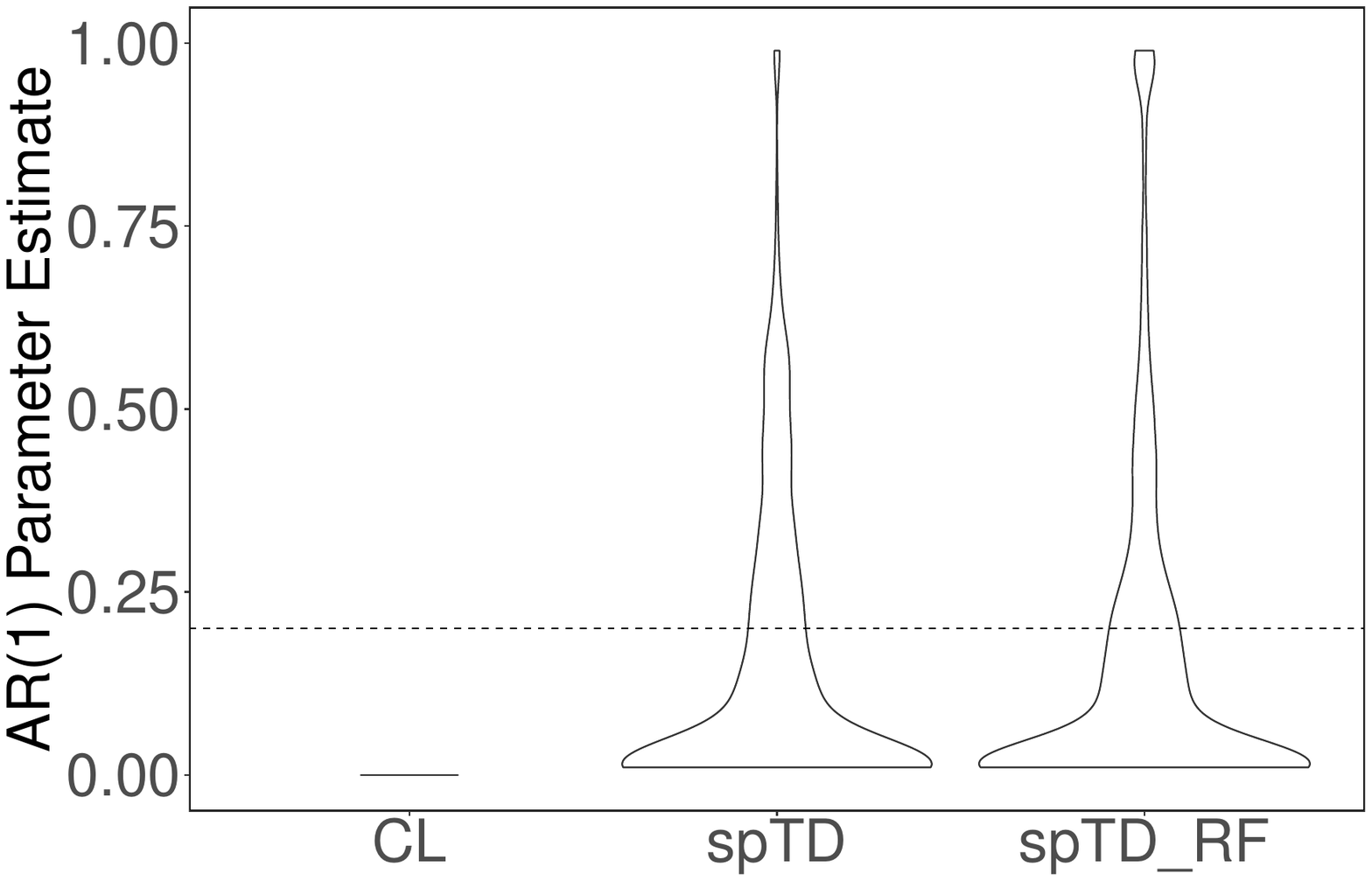}}}
    \end{minipage}
    \begin{minipage}{.22\linewidth}
    \centering
    \subfloat[]{\makebox{\includegraphics[width=\linewidth]{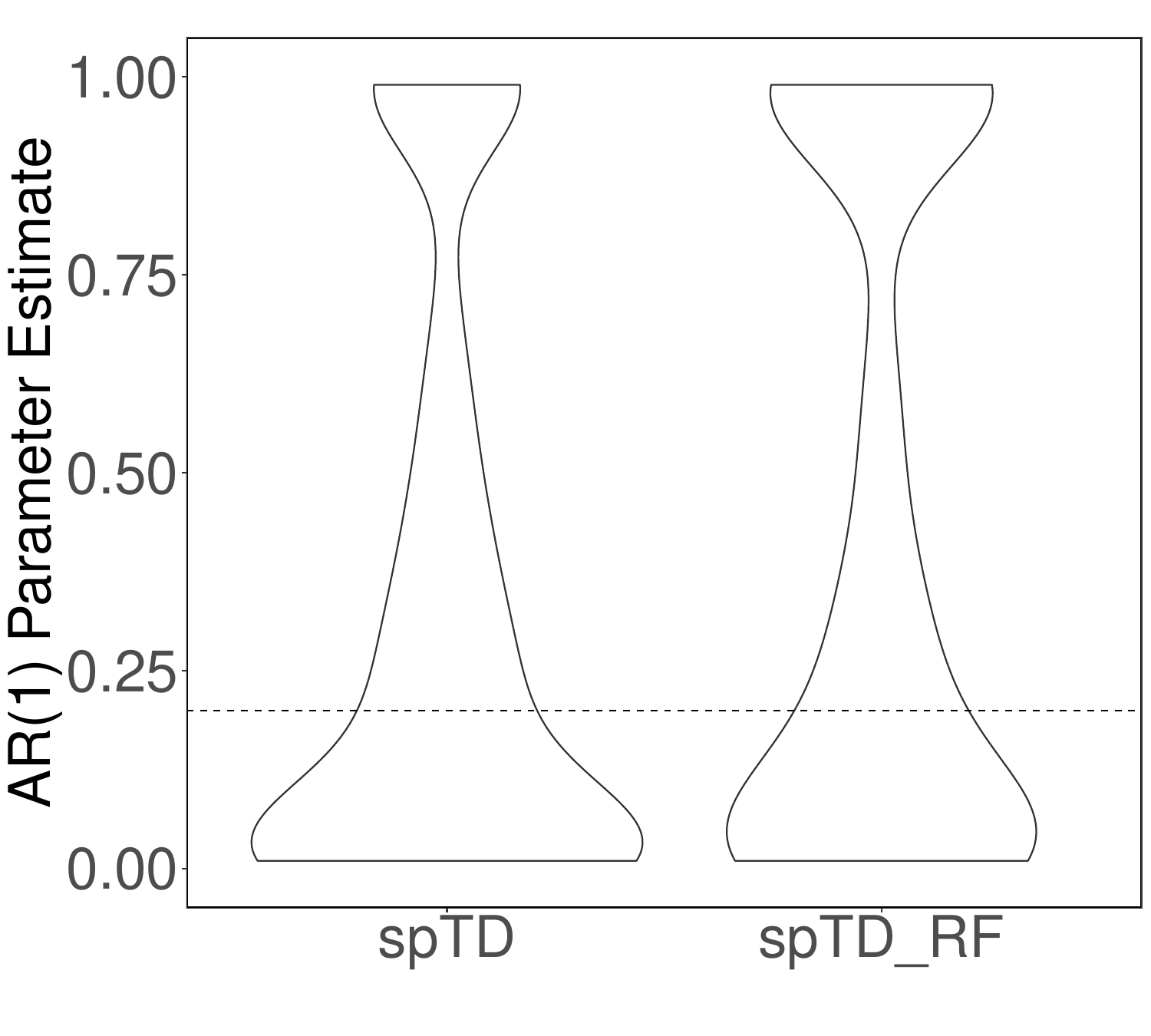}}}
    \end{minipage}\par\medskip
    \begin{minipage}{.3\linewidth}
    \centering
    \subfloat[]{\makebox{\includegraphics[width=\linewidth]{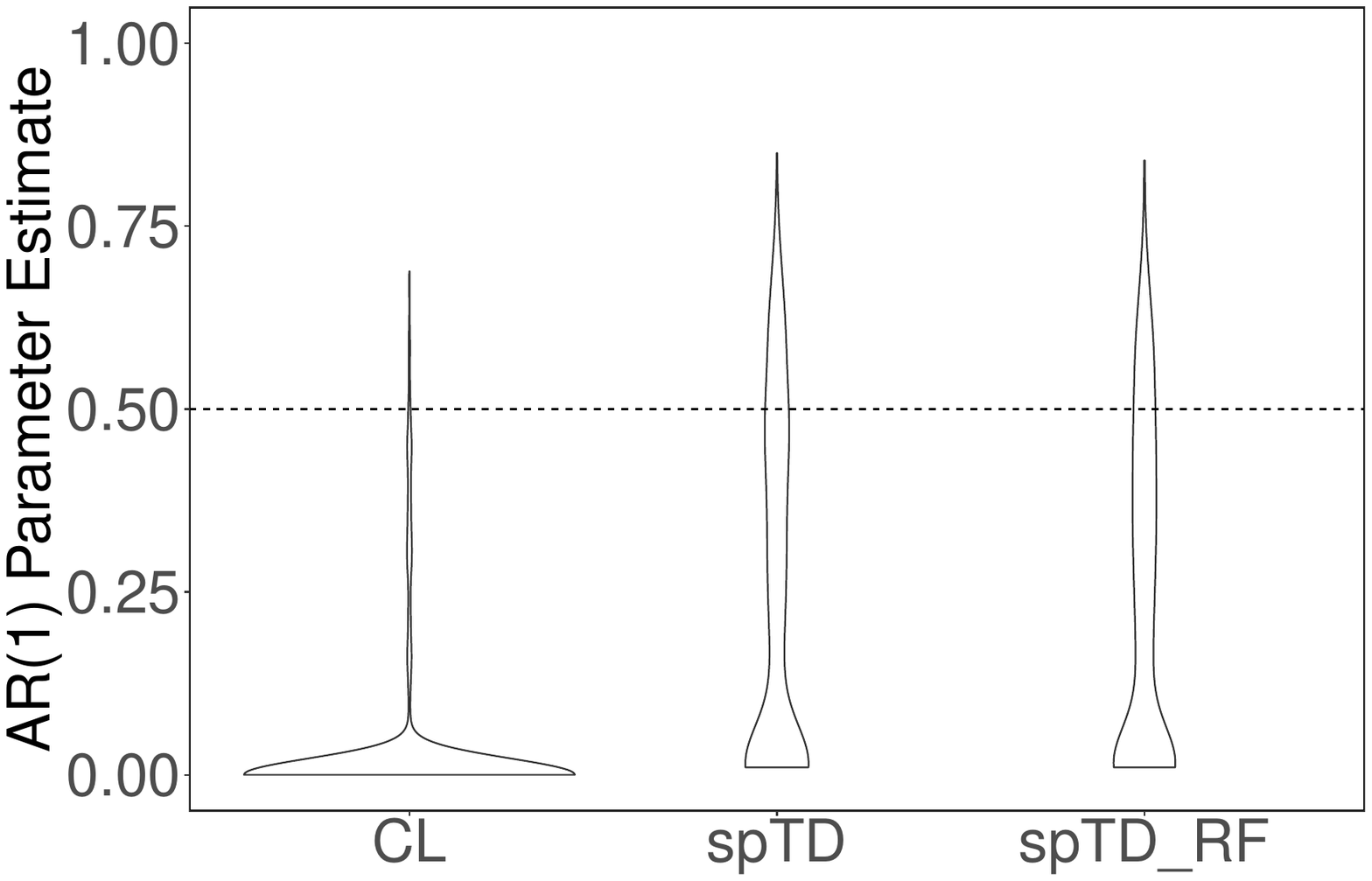}}}
    \end{minipage}
    \begin{minipage}{.3\linewidth}
    \centering
    \subfloat[]{\makebox{\includegraphics[width=\linewidth]{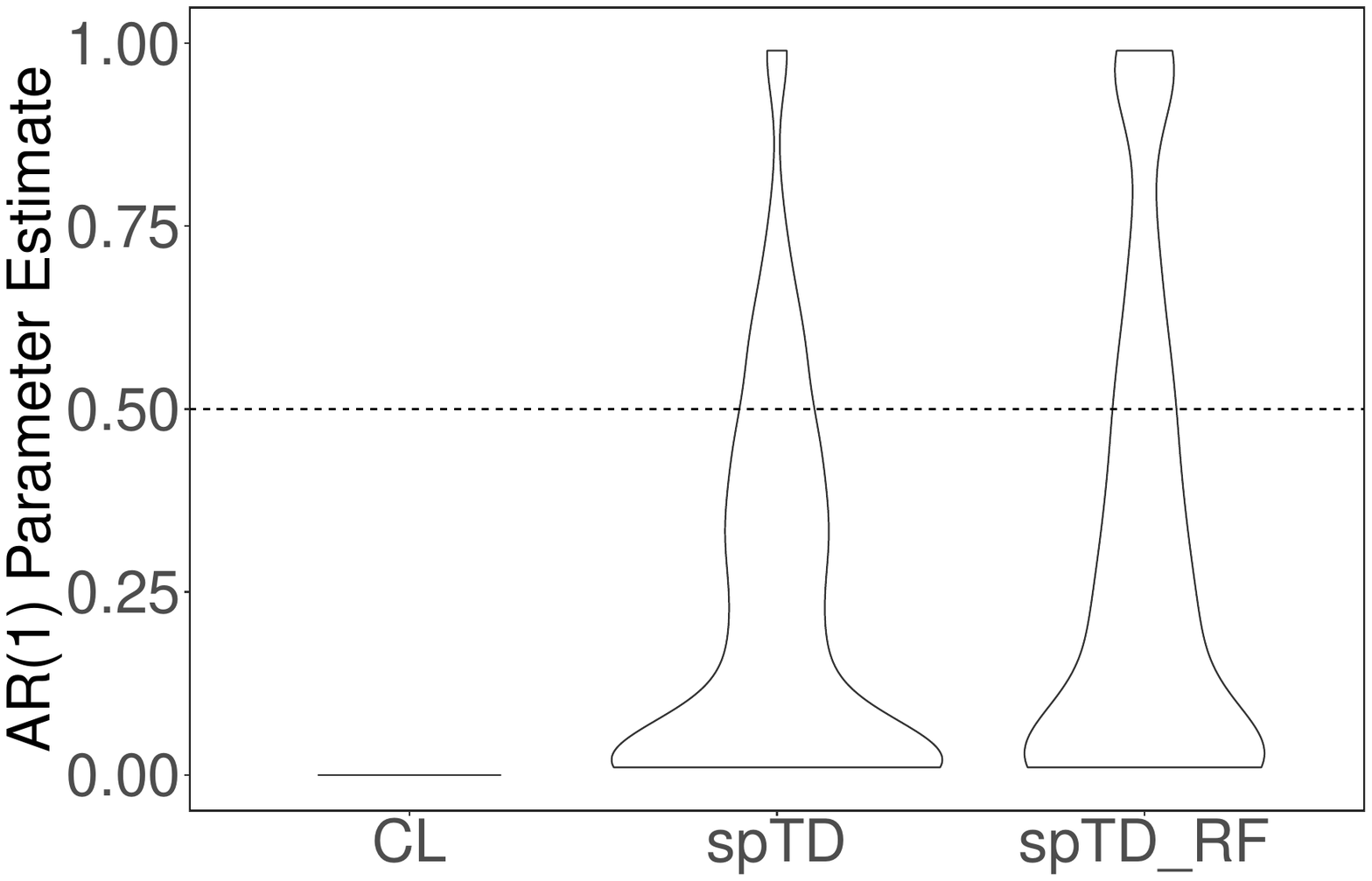}}}
    \end{minipage}
    \begin{minipage}{.22\linewidth}
    \centering
    \subfloat[]{\makebox{\includegraphics[width=\linewidth]{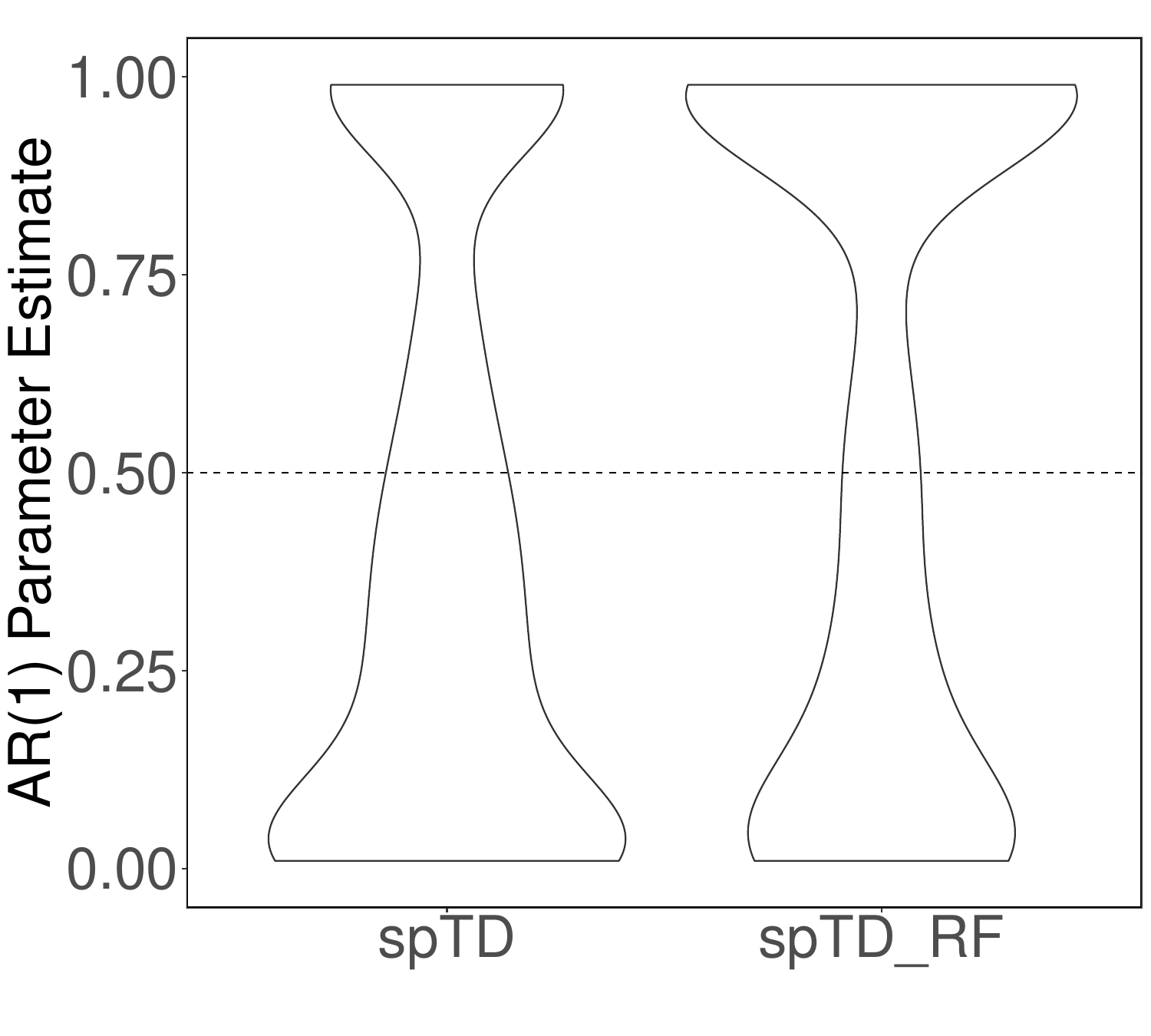}}}
    \end{minipage}\par\medskip
    \begin{minipage}{.3\linewidth}
    \centering
    \subfloat[]{\makebox{\includegraphics[width=\linewidth]{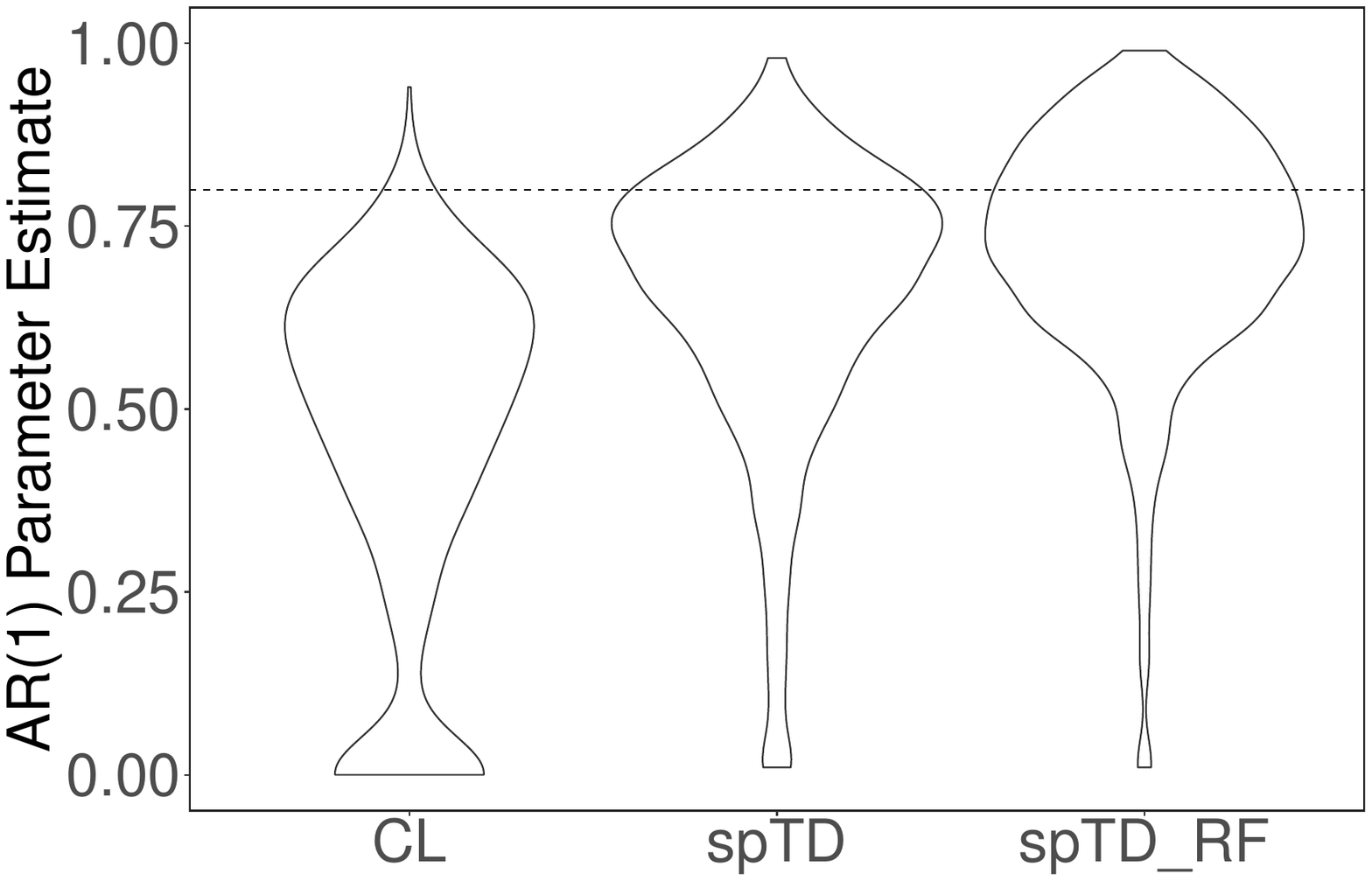}}}
    \end{minipage}
    \begin{minipage}{.3\linewidth}
    \centering
    \subfloat[]{\makebox{\includegraphics[width=\linewidth]{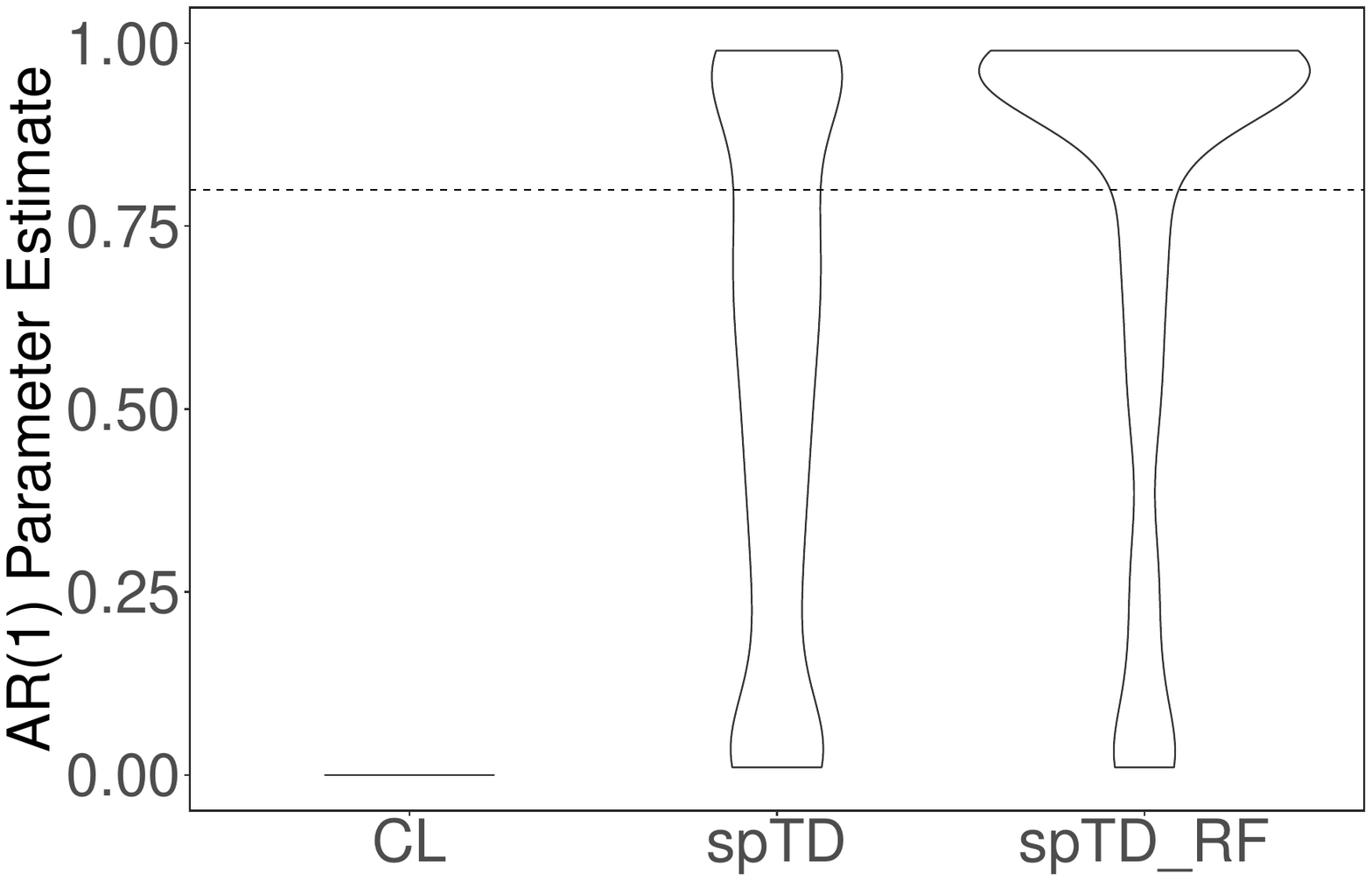}}}
    \end{minipage}
    \begin{minipage}{.22\linewidth}
    \centering
    \subfloat[]{\makebox{\includegraphics[width=\linewidth]{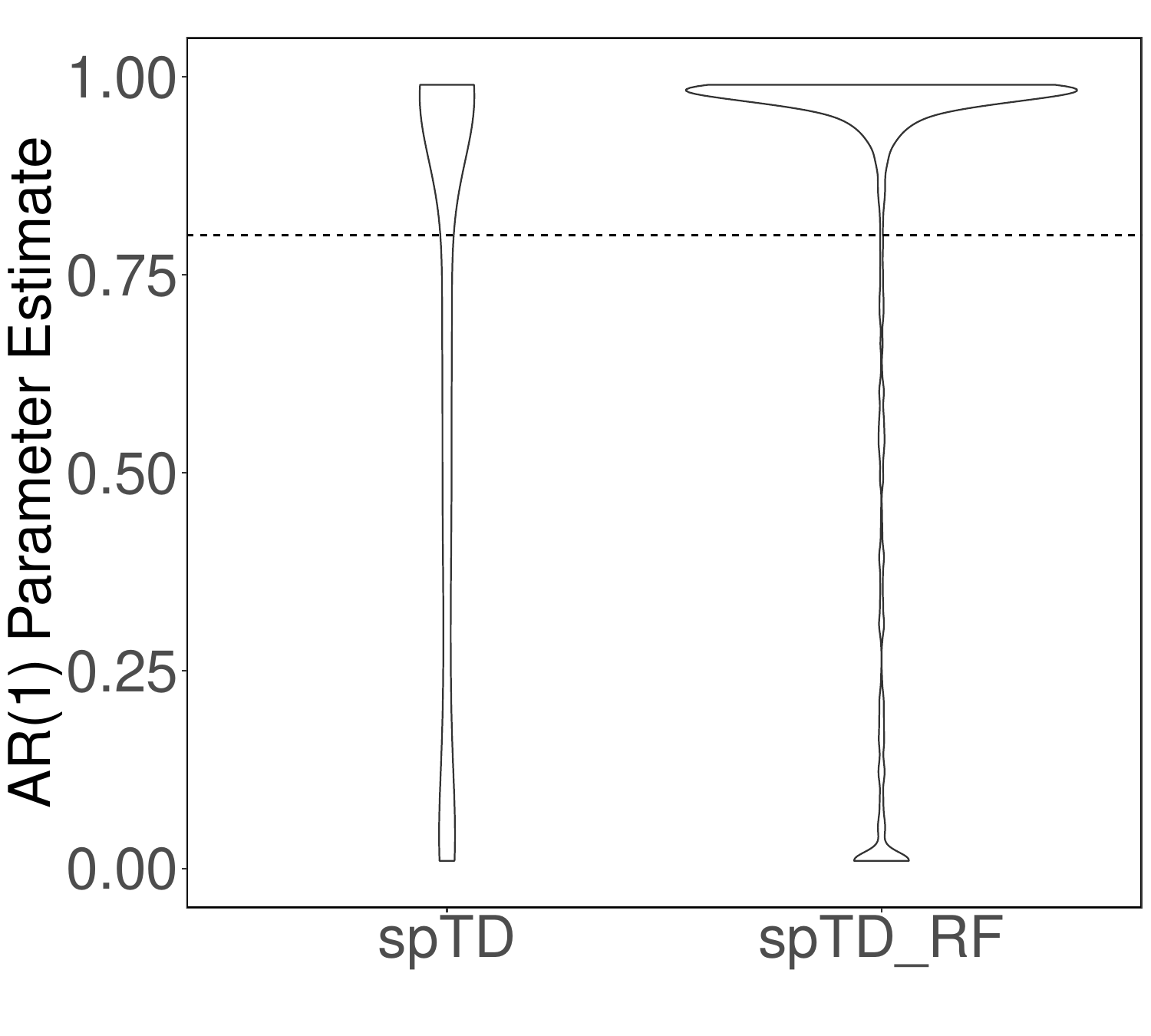}}}
    \end{minipage}\par\medskip
\caption{Violin plots showing the distribution of $\hat{\rho}$ for CL, spTD and spTD\_RF using non-stationary indicator series. Dimension $p=30$, $90$ and $150$ from left to right and true AR parameter $\rho$ indicated by horizontal dashed line. }
\label{fig:rwrho}
\end{figure}

Figure \ref{fig:rwrho} (Appendix) shows the distribution of $\hat{\rho}$ with non-stationary indicator series. It is evident from these plots that in this non-stationary setting, correctly identifying the true $\rho$ becomes a lot more challenging. In moderate dimensions ($p=90$) CL always selected 0 as the estimate, whereas, our methods spread the search space, tending to under-estimate the truth. This large variance in performance may be the result of the algorithm including a lot more false positives in the non-stationary setting. It is worth noting that even though estimating the true AR parameter is a lot less accurate in the non-stationary setting, the recovery of $\bbeta$ and $\by_q$ is more accurate in this setting. Thus, one may conclude that correctly identifying $\rho$ does not play that large of a factor in the quality of regression estimates. 

\subsection{Correlated Design Setting}

To investigate the effects of correlation present between indicator series, we extend the above i.i.d. setting for $\bX_q$ to now be generated from two correlated designs. The first has a block structure with 10 indicators in each block, where each block has a covariance matrix $\Sigma$ with $\Sigma_{ii}=1$ and $\Sigma_{ij}=\theta$ for $i \neq j$. We consider $\theta = \{0.2,0.6,0.9\}$ and let 
\[
\beta = (\underbrace{5,\dots,5}_{5},\underbrace{0,\dots,0}_{5})
\]
in the first 3 blocks and 0 everywhere after. This equicorrelated design does not violate the irrepresentability condition. Therefore, in order to assess an even more challenging situation we further consider a random covariance design setting for $\bX_q$ that does break this condition. 

\begin{figure}[htbp]
    \begin{minipage}{.32\linewidth}
    \centering
    \subfloat[]{\makebox{\includegraphics[width=\linewidth]{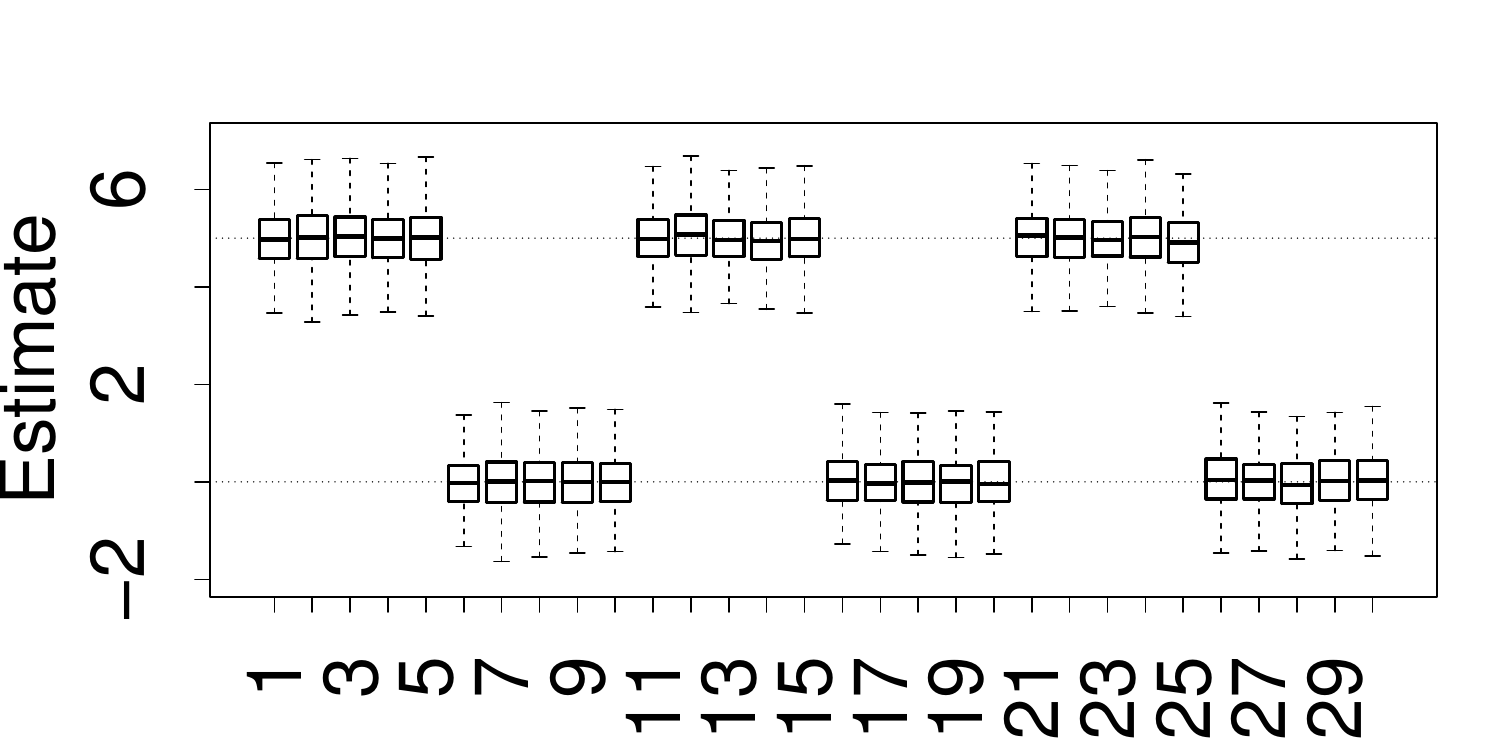}}}
    \end{minipage}
    \begin{minipage}{.32\linewidth}
    \centering
    \subfloat[]{\makebox{\includegraphics[width=\linewidth]{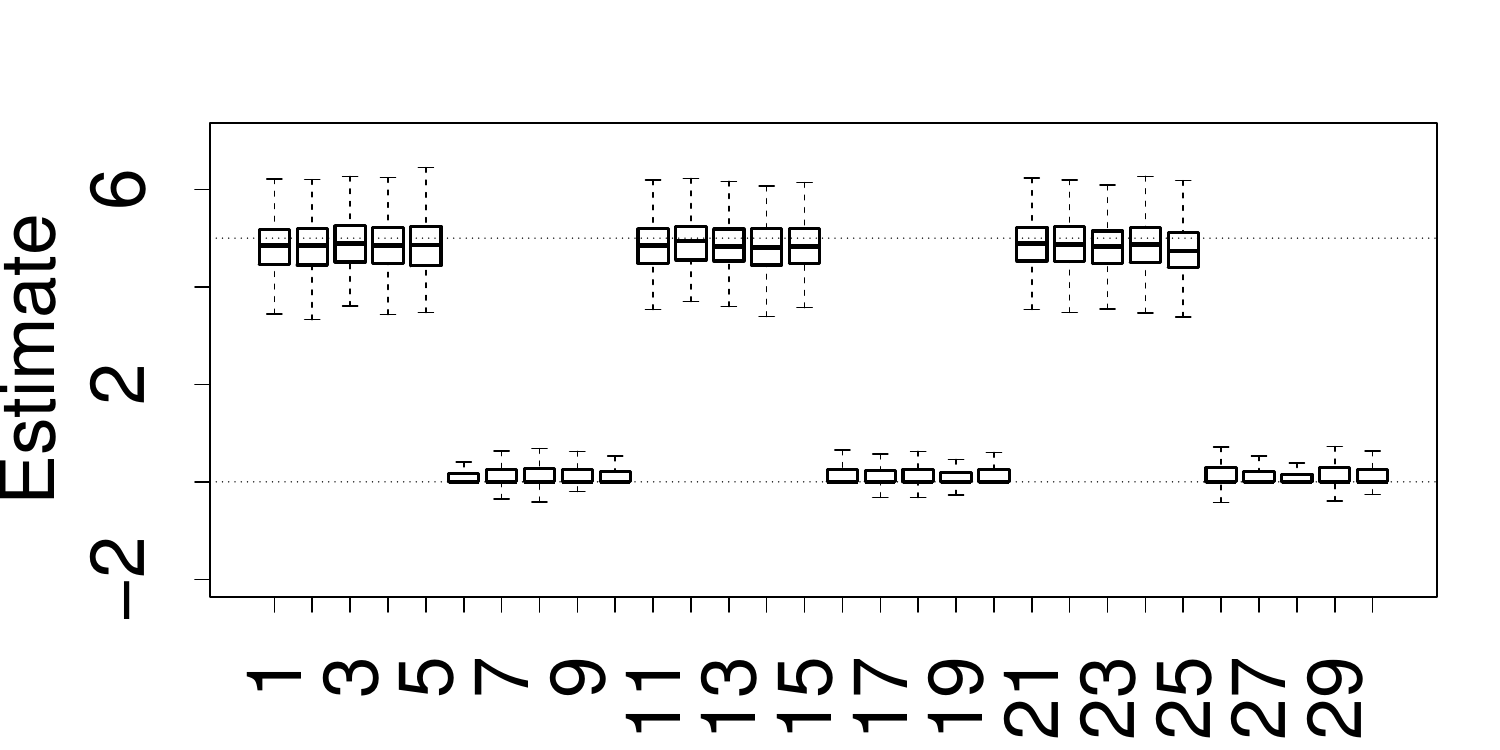}}}
    \end{minipage}
    \begin{minipage}{.32\linewidth}
    \centering
    \subfloat[]{\makebox{\includegraphics[width=\linewidth]{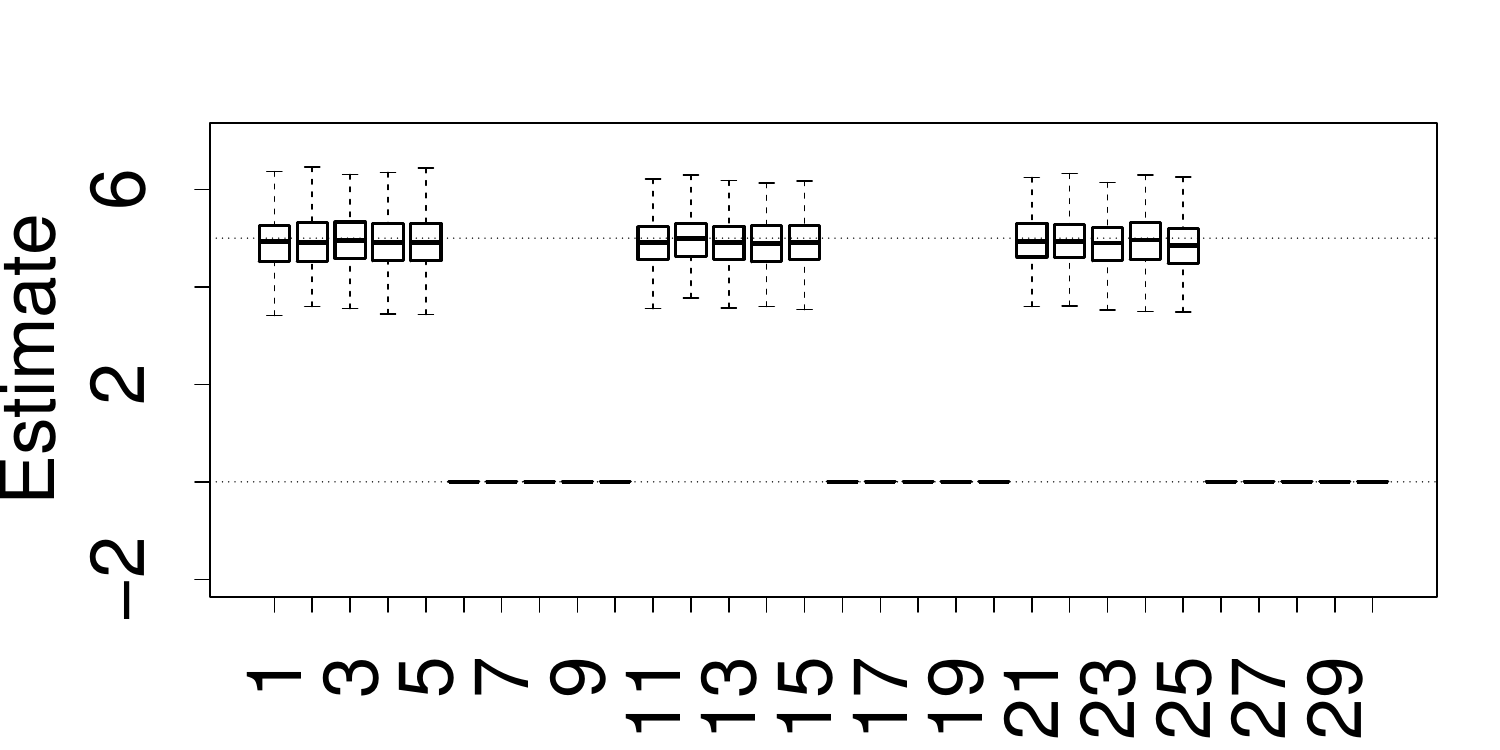}}}
    \end{minipage}\par\medskip
    \begin{minipage}{.32\linewidth}
    \centering
    \subfloat[]{\makebox{\includegraphics[width=\linewidth]{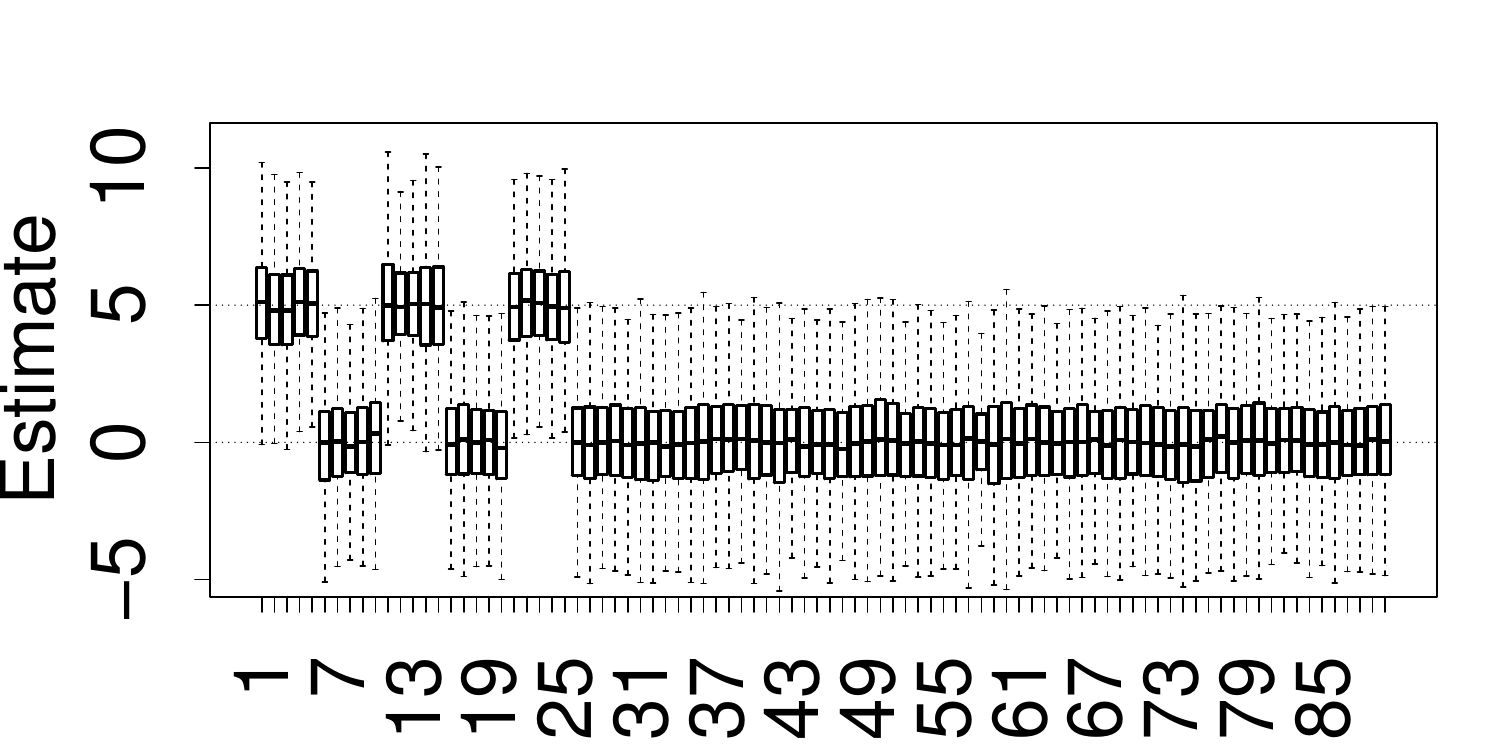}}}
    \end{minipage}
    \begin{minipage}{.32\linewidth}
    \centering
    \subfloat[]{\makebox{\includegraphics[width=\linewidth]{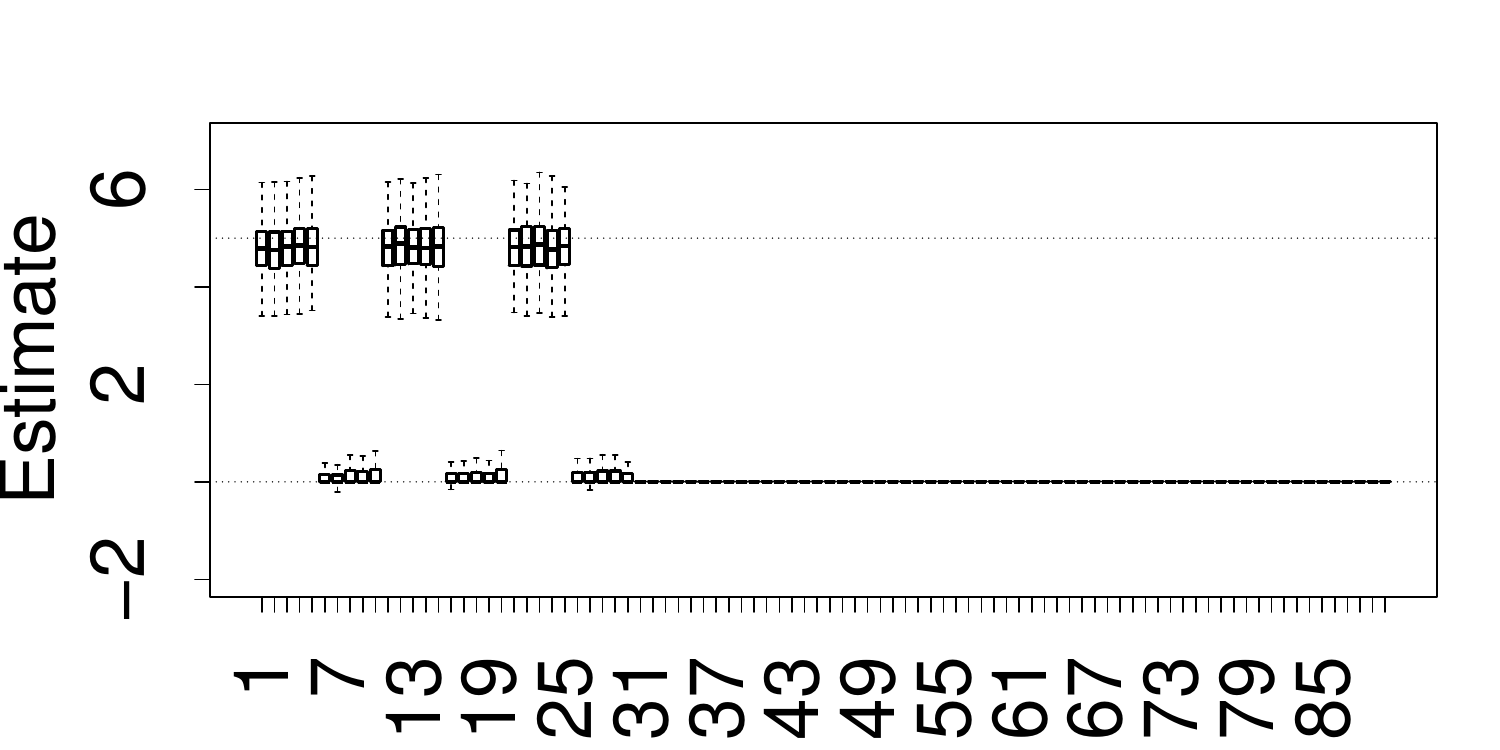}}}
    \end{minipage}
    \begin{minipage}{.32\linewidth}
    \centering
    \subfloat[]{\makebox{\includegraphics[width=\linewidth]{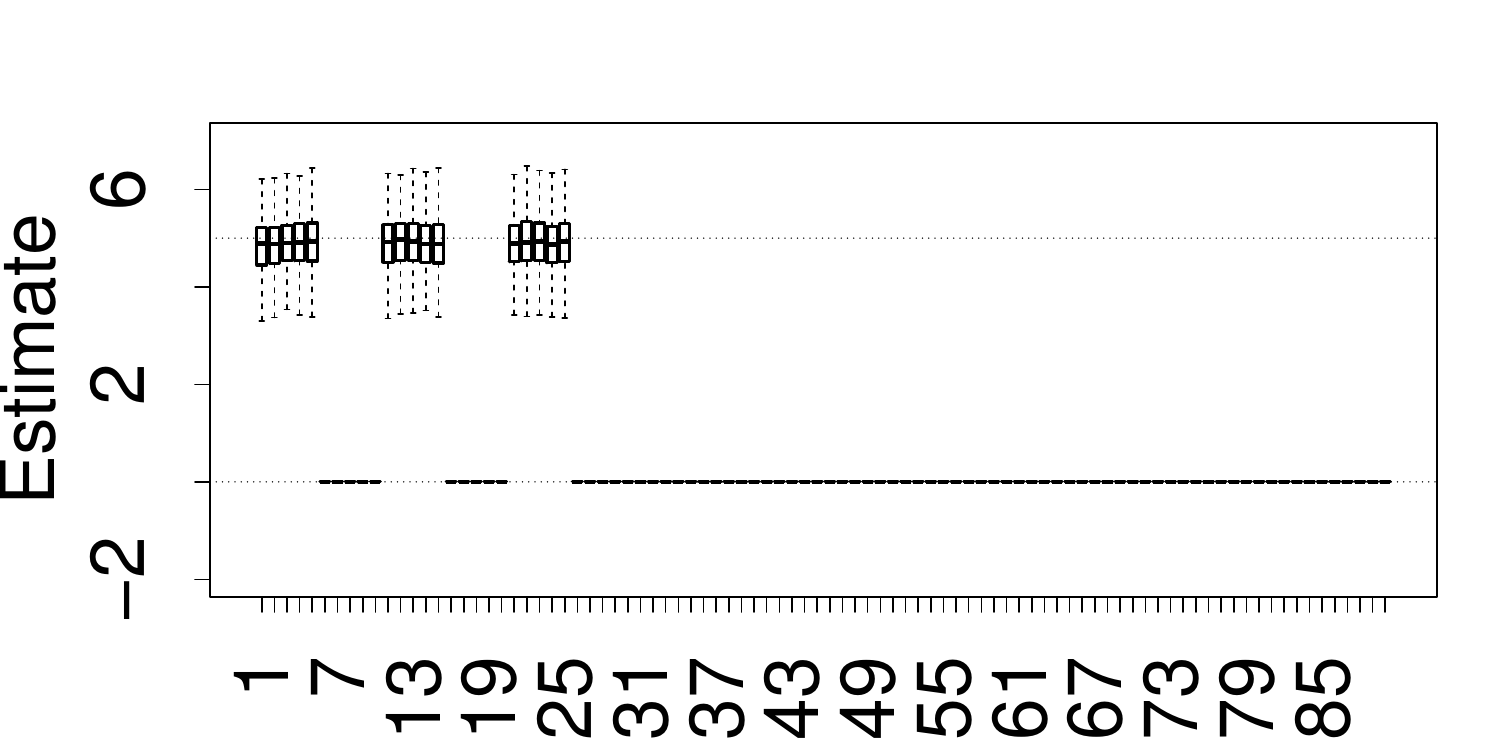}}}
    \end{minipage}\par\medskip
\caption{Boxplots for the estimates $\hat{\beta}_1,\dots,\hat{\beta}_p$ using CL, spTD and spTD\_RF (left to right respectively) with $\rho=0.5$ and $p=30$ (top row) and $p=90$ (bottom row) with correlated (0.9) indicator series. }
\label{fig:corbetas}
\end{figure}

We found the spTD\_RF method handles the block correlation design setting very well, presenting accurate predictions for $\by_q$ and recovery of the correct support for $\beta$ in all scenarios. Figure \ref{fig:corbetas} shows the distribution of each $\beta_j$ for the most difficult case of $\theta = 0.9$. The variance using Chow-Lin is very large in this setting as shown in \ref{fig:corbetas}(d). We notice some variance around 0 in spTD without re-fit, whereas spTD\_RF does an excellent job in recovering the true support. 

In the second experiment considering the design that breaks the irrepresentability conditions, we use the coefficients 
\[
\beta = (\underbrace{-2,2,-2\dots,2}_{10}, \underbrace{0,\dots,0}_{80})
\]
to allow a change of sign and a lower signal-to-noise ratio than previously. In this design setting we observe spTD\_RF is still able to provide accurate estimates for the high-frequency series $\hat{\by}_q$ but struggles to obtain the correct support for $\beta$. Out of the 1000 simulations, spTD\_RF produced on average $4.26$ false positives with a large standard deviation of $4.03$. By re-fitting the obtained estimates using the adaptive re-weighting scheme outlined in Section 3.3, we see significant improvements in the support recovery performance with $91\%$ of the 1000 iterations achieving exact support recovery and the remainder having at most $3$ false positives. 

\section{Temporal Disaggregation of Gross Domestic Product}

There are many measures we can use to assess the state of the economy. These are often disaggregated to different resolution levels, temporally, geographically, and across business sectors. To illustrate the methods developed, we consider the most popular measure of a nations production, its GDP as the output series of interest, and a range of high frequency indicator series as inputs. In our analysis here we deliberately include indicators which are direct inputs into the deterministic calculation of GDP, for instance, reports from the monthly business survey (MBS). However, to supplement this information, we also include a set of relatively novel \emph{fast indicators} produced by the ONS, based on VAT receipts, and traffic flows. A summary of the indicators we use in our analysis is given in Table \ref{tab:indicators_used}. The use of sensor data for traffic flows is an indicator of particular interest here as this aligns with the ONS' endeavour to incorporate and gain insight from alternative data sources into the production of official statistics. This data provides mean 15 minute counts, aggregated to the monthly level, of the number of vehicles on roads across England and within 10km of ports with details on the size of the vehicle observed. This has potential to be a key indicator of supply and demand across England and trade activity at ports. 

\begin{table}
\caption{Summary of indicators used in GDP analysis}
    \label{tab:indicators_used}
    \centering
    \fbox{%
    \begin{tabular}{p{0.4\textwidth}|p{0.55\textwidth}}
         \textbf{Indicator (No. of series)} & \textbf{Description} \\
         \hline
         MBS Turnover in Production (38) & Total turnover of production industries in £million \\ 
         MBS Turnover in Services (29) & Total turnover of service industries in £million \\
         Retail Sales Index (4) & Value of retail sales by commodity at current prices as an index \\
         VAT Diffusion Index (18) & Diffusion index tracking industry turnover from VAT returns \\
         Traffic Flow on Roads (4) & Mean 15 minute traffic count for vehicles on roads across England \\
         Traffic Flow at Ports (4) & Mean 15 minute traffic count for vehicles within 10km of a port in England
    \end{tabular}
    }
    
\end{table}

While it may seem counter-intuitive to try and recover the GDP series from quantities that directly drive this, there is still significant interest in how temporal disaggregated estimates align simply using monthly level data. In general, much more information goes into the estimation of the quarterly GDP results, and these are often considered the gold standard in terms of calibrated measurements of economic activity. Thus, disaggregating to a monthly level whilst maintaining consistency with the calibrated quarterly output is of significant interest. In this example, the output series we aim to disaggregate is the published quarterly (seasonally adjusted) GDP at chained volume measure from 2008 Q1 to 2020 Q2. To illustrate the behaviour of our monthly disaggregated estimates we make comparisons with the published monthly GDP index that represents a percentage growth-rate based on the average of 2016 being 100\%. To make these comparison meaningful we adjust the published monthly index to be a value in £million by benchmarking it onto  the quarterly output using Denton-Cholette benchmarking \citep{cholette1983adjusting}.     

As we are using a total count of 97 monthly indicator series and output data from 2008 Q1 to 2020 Q2 ($n = 50$ quarters) we are firmly in the high dimensional setting, and thus hope to capitalise on the benefits of sparse temporal disaggregation to produce a monthly estimate. Additionally, for comparison in the standard dimensional setting we produce an estimate using only 10 indicator series by aggregating our indicator set such that we have a single series each for MBS Turnover in Production , and MBS Turnover in Services, alongside four each for Retail Sales, and the VAT Diffusion Indices. This allows us to make a comparison with the traditional Chow and Lin temporal disaggregation procedure. 

Before discussing our results, it is worth considering our data-quality checks and pre-processing routines. Generally, releases from the ONS are of high-quality and based on robust production pipelines, however, in this case there are a couple of missing observations from January to March 2015 for the road traffic indicator data. This is due to a pause in the data being published throughout this period, as it is only 3 months in total we simply apply imputation based on linearly interpolation for these series. Beyond issues of data-quality, we need to ensure indicator series are of comparable form, for instance, if we want to benchmark with respect to a seasonally adjusted output, it would be prudent to use seasonally adjusted input series. To this end, and to enable better comparison we therefore seasonally adjust each of the indicator series before performing temporal disaggregation by making use of the \emph{seasonal} package in R that calls the automatic procedures of X-13ARIMA-SEATS to perform seasonal adjustment \citep{sax2018seasonal}. The final pre-processing step we perform is to re-scale our indicator series such that they are comparable in the eyes of the penalised regression problem. For example, the MBS results and road traffic counts are flow data, whereas VAT diffusion and retail sales are indexes. We therefore scale both the inputs and output to have zero mean and unit variance in order to obtain more interpretable results. To re-scale the monthly estimate back to the original £million scale we use the following:
$\text{Final Estimate} = \text{Model Estimate} \;\times\; \text{SD(Quarterly Output)} \;+\; \text{mean(Quarterly Output)}/3$.

\begin{figure}
        \centering
        \subfloat[]{\makebox{\includegraphics[width=0.8\linewidth]{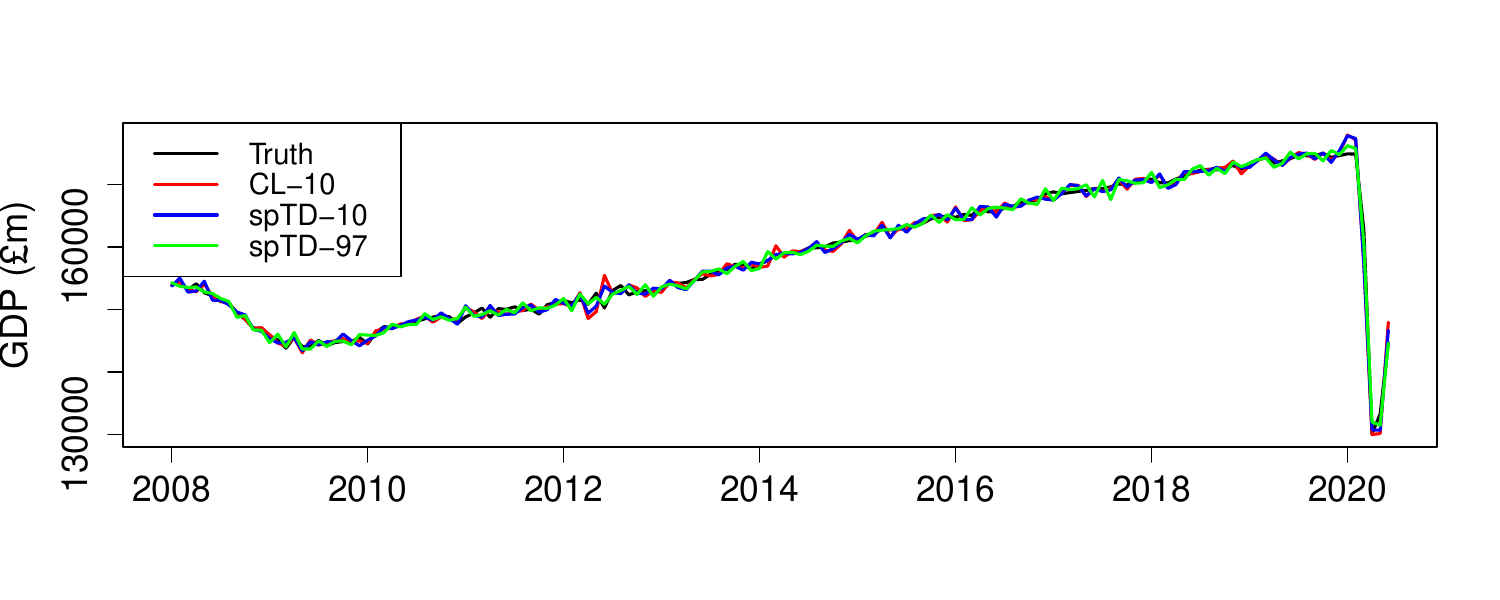}}}\\
        \vspace{-1em}
        \subfloat[]{\makebox{\includegraphics[width=0.8\linewidth]{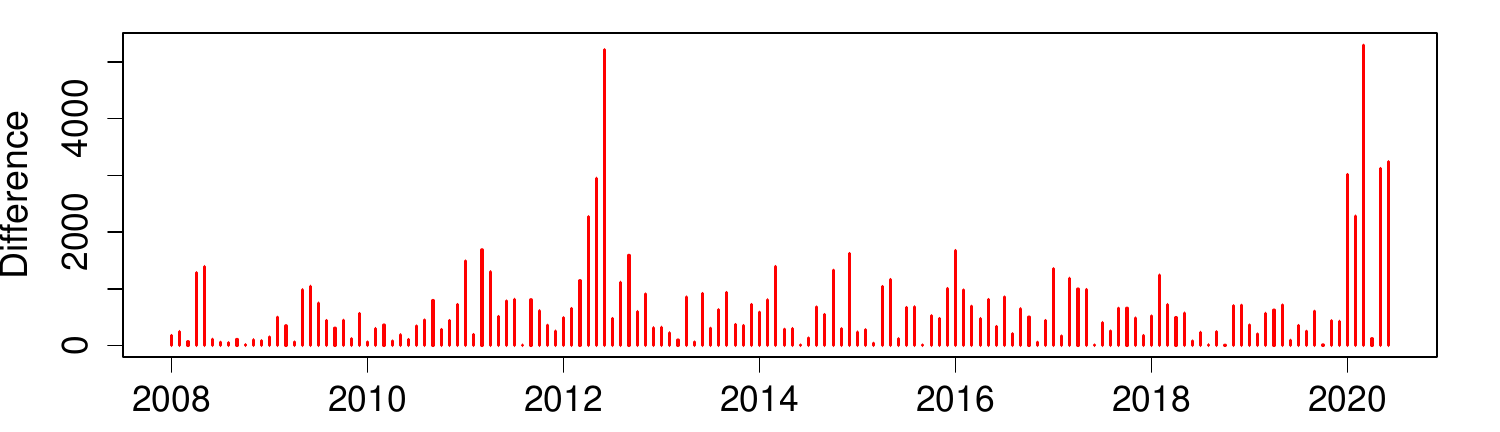}}}\\
        \centering
        \vspace{-1em}
        \subfloat[]{\makebox{\includegraphics[width=0.8\linewidth]{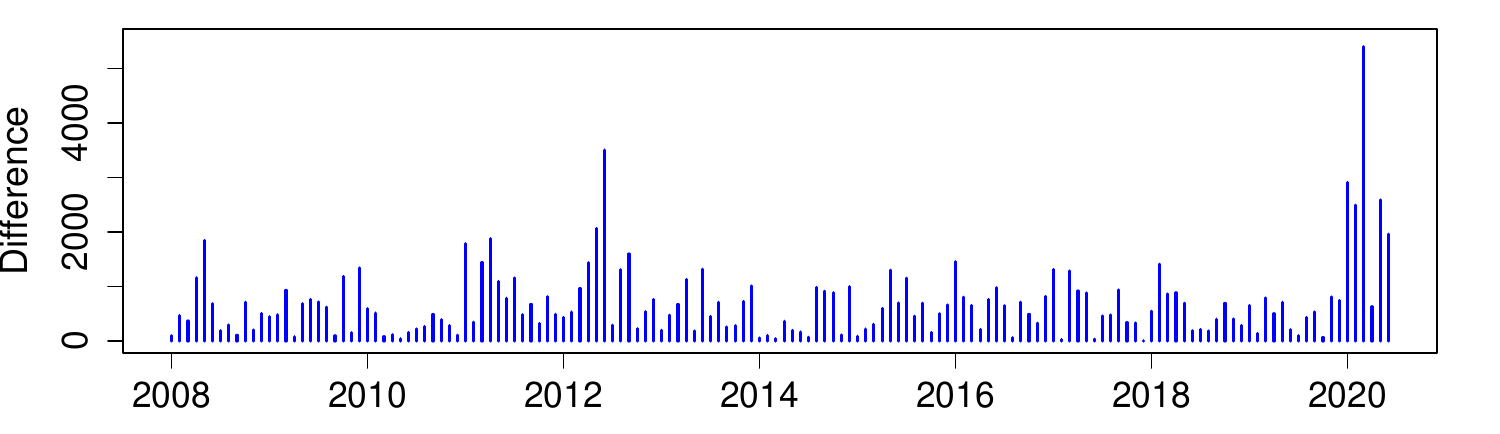}}}\\
        \vspace{-1em}
        \subfloat[]{\makebox{\includegraphics[width=0.8\linewidth]{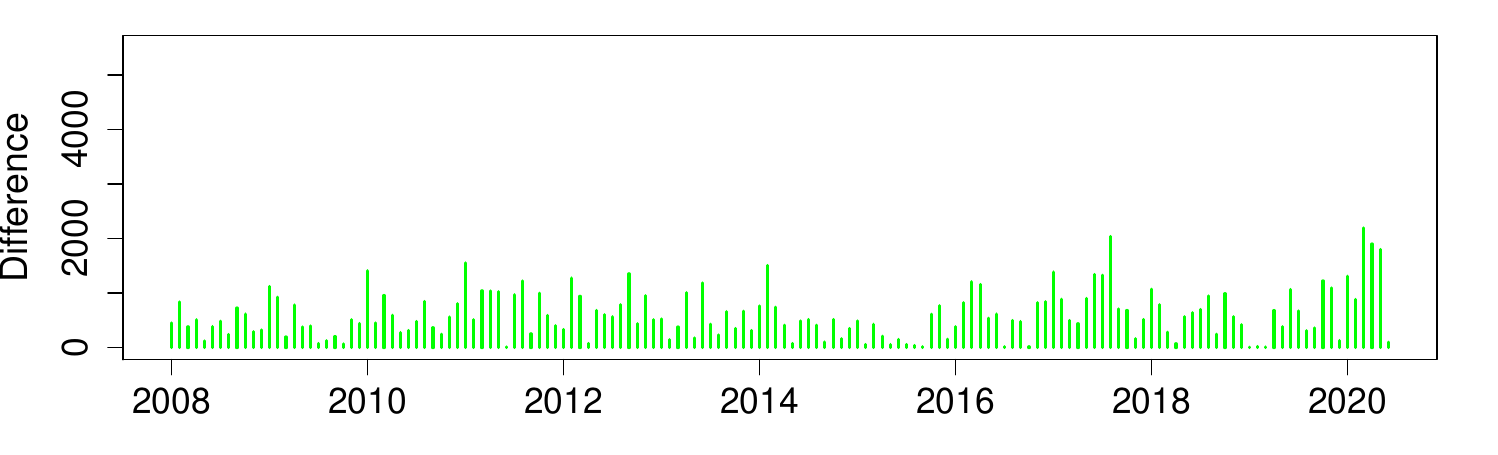}}}
    \caption{Comparison of estimated sequences and the ONS' monthly GDP index. a) Estimated high-frequency (monthly) series for 2008-2020. b-d) Absolute difference between the index and the estimate for each month (millions of pounds), b) CL with 10 indicators, c) $\ell_1$-spTD with 10 possible indicators, d) $\ell_1$-spTD with 97 possible indicators.}
\label{fig:GDPplots}
\end{figure}
\subsection{Results}

Given the automatic tuning procedure we have developed both $\ell_1$-spTD and the Chow-Lin method can be straightforwardly applied to the data. In the synthetic experiments, the refitting procedure improved performance over the pure lasso approach, as such we generally recommend the refitting strategy, and focus on this method in this application.
Figure \ref{fig:GDPplots} shows the monthly estimates of disaggregating quarterly seasonally adjusted GDP at chained volume measures using Chow-Lin and $\ell_1$-spTD with 10 indicators (red and blue respectively) and also using $\ell_1$-spTD with 97 indicators (green line). Further it shows the index based monthly GDP series described above to compare against (black). For the purposes of this discussion, we will refer to this index based series as the true series. 
In general, all methods seem to be doing a good job of estimating the true trend in Fig. \ref{fig:GDPplots}a) with $\ell_1$-spTD using 97 indicators being the smoothest and most accurate. This advantage is demonstrated in panels (Fig. \ref{fig:GDPplots}b-d) where the absolute error is shown to be both lower in magnitude, and more consistent over time. For instance, with the high-dimensional example we do not see large errors in the periods around early 2012, and 2020. This behaviour is further captured by the RMSE errors between the true monthly series and the estimates. For the Chow-Lin and $\ell_1$-spTD methods with 10 indicator series we obtain respectively RMSE scores of 1055.74 and 985.13. Given its access to a wider array of indicator series, the $\ell_1$-spTD method applied to 97 series gives significantly improved performance on average with an RMSE of 749.63. 

Even though it may seem the performance of the estimates between Chow-Lin and $\ell_1$-spTD with 10 indicators is close, the real advantage of $\ell_1$-spTD stems from the fact that our method selects a sparse subset of the most relevant indicators. This informs practitioners and users of this data about the main factors driving the economic variable of interest; assisting with understanding data revisions and analysis.
Table \ref{tab:GDP10} in the Appendix displays the coefficient estimates of the 10 indicator series used in the GDP analysis. $\ell_1$-spTD selects 3 out of the 10. These are: the aggregate series for MBS Turnover in Production, the VAT Diffusion Index for Production and the Retail Sales Index for Clothing and Footwear. Unsurprisingly, MBS for Production appears as one of the main components that goes into GDP. Perhaps somewhat more surprisingly is the large weight given to the clothing and footwear sales index. Indeed, if one plots the published monthly GDP index alongside this indicator it is clear they share very similar dynamics. Table \ref{tab:GDP97} displays the indicator selected out of 97 in the high-dimensional task and their corresponding coefficient estimate. They are, 4 MBS Production industries, 7 MBS Services industries, Vehicles over 11.66m on roads, VAT Diffusion Index for Services, and the Retail sales Index for Clothing and Footwear. The benefits of using this many indicators is that we get information on exactly which industries in production and services are most informative across MBS surveys, alongside the inclusion of faster, alternative indicators. Interestingly, the fact that vehicles over 11.66m on roads is selected gives evidence that these relatively novel traffic indicators can be of use for rapid economic measurement. More large vehicles/lorries on the motorways could be a sign of increased supply and demand of goods, positively impacting the economy. Cars and aircraft drive manufacturing movements so it is no surprise they are selected. Similarly, mining and quarrying are significant in wider production.

Since many of these indicators may be correlated, as a final check we also run the adaptive lasso extension outlined in Section 3.3 to assess the impact on the variables selected. Figure \ref{tab:GDP10} and \ref{tab:GDP97} shows the weightings from this adaptive extension. In the case of 10 indicators the method selects exactly the same variables, whereas in the 97 application, a subset of the indicators is returned as active. We can see in this case, the adaptive lasso acts to further prune out potential false positives, however, the RMSE score of 799.11 is very similar to that achieved without using adaptive weighting. Future work may look to extend our analysis to include more informative structured priors, for instance via the group lasso \citep{yuan2006model} based on pre-identified economic factors, or using knowledge of the aggregation hierarchy \citep{hecq2021hierarchical}.

\section{Conclusion}

The process of using high frequency indicators to construct high frequency renditions of low frequency information, known as temporal disaggregation, is widely used in official statistics. However, traditional methods \citep{chow1971best,dagum2006benchmarking} have become somewhat outdated and may not be compatible with the large-scale, high-frequency administrative and alternative data sources NSIs increasingly seek to utilise. 

In this article, we have introduced a novel Sparse Temporal Dissagregation (spTD) framework that can operate in settings where the number of indicators used can exceed the low frequency sample size. The method builds estimators that incorporate regulariser penalty functions into the well-established Chow and Lin cost function to promote more parsimonious modelling. By focusing on the $\ell_1$-norm (LASSO) penalty \citep{tibshirani1996regression}, we have produced methodology that is able to simultaneously select important indicators and estimate their regression coefficients. Through extensive simulation studies, we have shown $\ell_1$-spTD to outperform Chow-Lin temporal disaggregation in all moderate dimensional scenarios and provide accurate and interpretable estimates in high dimensions. While recovery of the true high-frequency series and regression parameters becomes slightly more challenging with a smaller sample size ($n<100$) and smaller regression coefficient value ($\beta_j < 5$) than ones considered, we do find our method still outperforms Chow-Lin in these scenarios. Extra simulation studies not presented in this paper confirm this. As might be expected with a LASSO related method, we found that when indicators are highly correlated and break the irrepresentability condition, recovery of the correct support of $\bbeta$ is difficult. A further limitation of our proposed method is in scenarios when the true support of the indicator set considered is actually dense. As the $\ell_1$ variant of our method relies on the assumption of sparsity, it may perform poorly in this setting. Factor based approaches may be a better option in this setting, or utilising alternative penalisation strategies, i.e. ridge regression.

Applying this method to the disaggregation of quarterly GDP data also highlighted $\ell_1$-spTD's ability to achieve an accurate monthly estimate whilst informing us of the most relevant indicators. Unlike previous approaches there are no restrictions here on the number of indicators we can input to the model. This application can be easily extended to produce monthly (or more frequent) estimates of numerous headline variables and accurate disaggregated estimates can act as cost-efficient alternatives to sending out expensive surveys frequently. 

Interestingly, our synthetic analysis also verified the difficulty of estimating $\rho$ in the temporal disaggregation setting. This was an area of particular weakness for the classical Chow-Lin method \citep{ciammola2005temporal}. A similar issue is encountered within \citet{proietti2006temporal} where the author suggests using a diffuse prior on the regression coefficients may help. Our proposed method appears to significantly outperform Chow-Lin in the estimation of $\rho$, however, can still struggle in moderate and high-dimensional settings. Interpreting our $\ell_1$ penalised solution as a maximum a-posteriori estimator with a Laplacian prior on the regression coefficients, we provide evidence that weakly informative priors may help in both the recovery of the regression and autocorrelation coefficients. Whilst in our experiments estimation of the autocorrelation structure (via $\rho$) does not have much impact on the high-frequency estimates themselves, it could significantly impact the confidence intervals associated with inference on the regression parameters and their coverage properties. Technical verification of this claim requires further work, and is complicated by the nature of the $\ell_1$ regularisation. There is however a considerable literature on post-selection inference for the lasso \citep[c.f.][]{vdgpostselection}, and it is feasible that this can also be adapted to the temporal disaggregation setting, an exciting direction for future work.

We would like to note, that whilst developed in the context of official-statistics, the methods proposed here have application far beyond this domain. For instance, they may be used to help companies produce more granular estimates of consumer behaviour, or to improve the resolution at which researchers can assess complex environmental processes. 

\section*{Acknowledgements}
The authors would like to acknowledge the advice and support of the Methodology Team within the Office for National Statistics for discussions relating to this work which have helped shape this manuscript. In particular we would like to thank Duncan Elliot and Hannah Finselbach for their advice and support with this project. Luke Mosley gratefully acknowledges the financial support of the UK Engineering and Physical Sciences Research Council (Grant EP/L015692/1) and the Office for National Statistics. Finally, we would like to acknowledge the editors and anonymous referees whose comments and suggestions have significantly improved the paper.

\newpage

\section*{Appendix}

\begin{landscape}
\begin{table}
\caption{ \label{tab:beta_metrics} Metrics for the performance of $\hat{\bbeta}$ using CL, spTD and spTD\_RF across $p=30, 90, 150$ and $\rho=0.2, 0.5, 0.8$ in both the stationary and non-stationary indicators setting. Metrics include the mean (standard deviation) across 1000 iterations of RMSE, $\ell_\infty$ and FP.    
} 
    \centering
    
    \resizebox{\linewidth}{!}{%
    \fbox{%
    \begin{tabular}{c|c|c|c|c|c|c|c|c|c}
        & \multicolumn{9}{c}{\textbf{Stationary Indicators}}  \\ \hline 
        & \multicolumn{3}{c}{$p=30$} & \multicolumn{3}{|c|}{$p=90$} & \multicolumn{3}{c}{$p=150$}  \\ 
        \hline 
         & $\rho=0.2$ & $\rho=0.5$ & $\rho=0.8$ & $\rho=0.2$ & $\rho=0.5$ & $\rho=0.8$ & $\rho=0.2$ & $\rho=0.5$ & $\rho=0.8$  \\ \hline 
         \textbf{RMSE} & & & & & & & & & \\ \hline
        CL & 0.143 (0.0237) & 0.195 (0.0316)  & 0.268 (0.0436) & 0.464 (0.151)& 0.657 (0.211) & 1.114 (0.332)& - & - & -  \\ \hline 
        spTD & 0.118 (0.0256)& 0.160 (0.0344)& 0.215 (0.0467) & 0.095 (0.0195)& 0.129 (0.0260)& 0.177 (0.0375)& 0.084 (0.0167)& 0.114 (0.0227) & 0.160 (0.0340) \\ \hline
        spTD\_RF & 0.084 (0.0246)& 0.115 (0.0335)& 0.152 (0.0471)& 0.066 (0.0249)& 0.089 (0.0347)& 0.110 (0.0441)& 0.065 (0.0234)& 0.086 (0.0367)& 0.099 (0.0482)\\ \hline
        $\bm{\ell_\infty}$ & & & & & & & & & \\ \hline
        CL & 0.334 (0.0728)& 0.461 (0.100)& 0.632 (0.140)& 1.264 (0.446)& 1.783 (0.629)& 3.032 (1.010)& - & - & - \\ \hline
        spTD & 0.343 (0.0875)& 0.465 (0.121)& 0.623 (0.164)& 0.445 (0.106)& 0.604 (0.142)& 0.825 (0.205)& 0.502 (0.114)& 0.681 (0.156)& 0.945 (0.236) \\ \hline
        spTD\_RF & 0.275 (0.0789)& 0.379 (0.107)& 0.493 (0.147)& 0.327 (0.0898)& 0.444 (0.125)& 0.560 (0.172)& 0.363 (0.106)& 0.488 (0.137)& 0.604 (0.210) \\ \hline
        \textbf{FP} & & & & & & & & & \\ \hline
        spTD & 4.613 (2.746)& 4.462(2.767)&4.462 (2.864)&8.775 (6.077)&8.971 (6.298)&9.964 (6.299)&11.653 (8.341)&11.921 (8.467)&13.619 (8.881)\\ \hline
        spTD\_RF & 0.776 (1.149)& 0.839 (1.167)& 0.928 (1.321)&3.406 (4.545)&3.688 (4.902)& 3.252 (4.248)& 9.034 (10.926)& 8.325 (10.548)& 5.813 (8.066)\\ 
        \hline 
        \\
        & \multicolumn{9}{c}{\textbf{Non-stationary Indicators}} \\ 
        \hline 
        & \multicolumn{3}{c}{$p=30$} & \multicolumn{3}{|c|}{$p=90$} & \multicolumn{3}{c}{$p=150$} \\ \hline 
        & $\rho=0.2$ & $\rho=0.5$ & $\rho=0.8$ & $\rho=0.2$ & $\rho=0.5$ & $\rho=0.8$ & $\rho=0.2$ & $\rho=0.5$ & $\rho=0.8$ \\ \hline 
        \textbf{RMSE} & & & & & & & & & \\ \hline
        CL & 0.040 (0.00804)& 0.059 (0.0114)& 0.098 (0.0181)& 0.182 (0.0486)& 0.231 (0.0605)& 0.298 (0.0775)& - & - & - \\ \hline 
        spTD & 0.343 (0.00835)& 0.465 (0.0117)& 0.623 (0.0190)& 0.445 (0.00990)& 0.604 (0.0126)& 0.825 (0.0167)& 0.502 (0.0104)& 0.681 (0.0126)& 0.945 (0.0141)\\ \hline 
        spTD\_RF & 0.021 (0.00776)& 0.033 (0.0117)& 0.055 (0.0194)& 0.025 (0.0110)& 0.037 (0.0154)& 0.048 (0.0242)& 0.025 (0.0114)& 0.034 (0.0154)& 0.042 (0.0223)\\ \hline 
        $\bm{\ell_\infty}$ & & & & & & & & & \\ \hline 
        CL & 0.095 (0.0235)& 0.138 (0.0342)& 0.230 (0.0552)& 0.493 (0.144)& 0.626 (0.187)& 0.808 (0.240)& - & - & - \\ \hline 
        spTD & 0.081 (0.0269)& 0.118 (0.0375)& 0.202 (0.0587)& 0.148 (0.0489)& 0.202 (0.0631)& 0.308 (0.0844)& 0.201 (0.0682)& 0.265 (0.0857) & 0.365 (0.0964)\\ \hline
        spTD\_RF & 0.062 (0.0219)& 0.096 (0.0330)& 0.164 (0.0549)& 0.105 (0.0390)& 0.152 (0.0543)& 0.222 (0.0795)& 0.135 (0.0484)& 0.182 (0.0634)& 0.240 (0.0890)\\ \hline 
        \textbf{FP} & & & & & & & & & \\ \hline 
        spTD & 8.205 (2.803)&8.366 (3.059)&7.863 (3.153)&21.687 (7.056)&21.926 (7.444)& 18.931 (9.566)& 25.335 (9.225)& 24.262 (10.065)& 19.026 (11.069)\\ \hline 
        spTD\_RF &5.530 (2.188)& 5.470 (2.293)& 3.903 (2.547)&16.127 (6.617)& 15.080 (7.756)&8.001 (9.513)& 17.275 (10.903)& 14.687 (11.982)&9.061 (11.631) \\
    \end{tabular}
    }
    }
    
\end{table}
\end{landscape}

\begin{table}
\caption{Coefficient estimates of GDP analysis with 10 indicator series using Chow-Lin (CL), $\ell_1$-spTD and adaptive extension }
    \label{tab:GDP10}
    \centering
    \fbox{%
    \begin{tabular}{c|c|c|c}
         \textbf{Indicator} & \textbf{CL } & \textbf{$\ell_1$-spTD} & \textbf{Adaptive} \\
         \hline
         MBS Turnover in Production Aggregate &  0.0728 & 0.121 & 0.121\\ 
         MBS Turnover in Services Aggregate &  0.0505 & 0 & 0\\
         VAT Diffusion Index - Agriculture & 0.00327 & 0 &0\\
        VAT Diffusion Index - Production & -0.395 & -0.0206 & -0.0206 \\
        VAT Diffusion Index - Services & 0.00235 & 0 & 0\\
        VAT Diffusion Index - Construction & 0.0138 & 0 & 0\\
        Retail Sales Index - Clothing & 0.201 & 0.239 & 0.239 \\
        Retail Sales Index - Food & -0.0772 & 0 & 0  \\
        Retail Sales Index - Household & -0.00325 & 0 & 0\\
        Retail Sales Index - Non-foods & 0.0587 & 0 & 0
    \end{tabular}
    }
    
\end{table}

\begin{table}[htbp]
\caption{Non-zero coefficient estimates of GDP analysis with 97 indicator series using $\ell_1$-spTD and adaptive lasso extension}
    \label{tab:GDP97}
    \centering
    \fbox{%
    \begin{tabular}{c|c|c}
         \textbf{Indicator} & \textbf{$\ell_1$-spTD} & \textbf{Adaptive} \\
         \hline
         MBS Production - Mining and Quarrying &  0.0232 & 0.0366\\ 
         MBS Production - Fabricated Metal Products (excluding weapons) &  0.00546 & 0\\ 
         MBS Production - Motor Vehicles &  0.0305 & 0.0234\\ 
         MBS Production - Air and Spacecraft &  0.00355 & 0\\ 
         MBS Services -Wholesale Retail Trade Repair of Motor Vehicles &  0.0599 & 0.0591\\ 
         MBS Services - Rail and Land Transport &  0.0140 & 0\\ 
         MBS Services - Food and Beverage Serving &  0.107 & 0.124\\ 
         MBS Services - Legal Activities &  0.0231 & 0.0281\\
         MBS Services - Rental and Leasing  &  0.0344 & 0.0426\\
         MBS Services - Employment &  0.0198 & 0\\
         MBS Services - Creative Arts and Entertainment &  0.0187 & 0.0168\\
         VAT Diffusion Index - `S' Other Service Activities & -0.0201 & -0.0216\\
        Retail Sales Index - Clothing & 0.00273 & 0\\
        Vehicles Over 11.66m on Roads Across England & 0.0263 & 0.0397   
       
    \end{tabular}
    }
    
\end{table}

\bibliographystyle{apalike}
\bibliography{Arxiv2}

\begin{thebibliography}{}

\bibitem[Angelini et~al., 2006]{angelini2006interpolation}
Angelini, E., Henry, J., and Marcellino, M. (2006).
\newblock Interpolation and backdating with a large information set.
\newblock {\em Journal of Economic Dynamics and Control}, 30(12):2693--2724.

\bibitem[Bach et~al., 2011]{bach2011convex}
Bach, F., Jenatton, R., Mairal, J., Obozinski, G., et~al. (2011).
\newblock Convex optimization with sparsity-inducing norms.
\newblock {\em Optimization for Machine Learning}, 5:19--53.

\bibitem[Ba{\'n}bura and R{\"u}nstler, 2011]{banbura2011look}
Ba{\'n}bura, M. and R{\"u}nstler, G. (2011).
\newblock A look into the factor model black box: publication lags and the role
  of hard and soft data in forecasting gdp.
\newblock {\em International Journal of Forecasting}, 27(2):333--346.

\bibitem[Bean, 2016]{bean2016independent}
Bean, C.~R. (2016).
\newblock {\em Independent review of UK economic statistics}.
\newblock HM Treasury.

\bibitem[Belloni et~al., 2013]{belloni2013least}
Belloni, A., Chernozhukov, V., et~al. (2013).
\newblock Least squares after model selection in high-dimensional sparse
  models.
\newblock {\em Bernoulli}, 19(2):521--547.

\bibitem[Belloni et~al., 2011]{belloni2011square}
Belloni, A., Chernozhukov, V., and Wang, L. (2011).
\newblock Square-root lasso: pivotal recovery of sparse signals via conic
  programming.
\newblock {\em Biometrika}, 98(4):791--806.

\bibitem[Bournay and Laroque, 1979]{bournay1979reflexions}
Bournay, J. and Laroque, G. (1979).
\newblock R{\'e}flexions sur la m{\'e}thode d'elaboration des comptes
  trimestriels.
\newblock In {\em Annales de l'INSEE}, pages 3--30. JSTOR.

\bibitem[B{\"u}hlmann and Mandozzi, 2014]{buhlmann2014high}
B{\"u}hlmann, P. and Mandozzi, J. (2014).
\newblock High-dimensional variable screening and bias in subsequent inference,
  with an empirical comparison.
\newblock {\em Computational Statistics}, 29(3-4):407--430.

\bibitem[B{\"u}hlmann and Van De~Geer, 2011]{buhlmann2011statistics}
B{\"u}hlmann, P. and Van De~Geer, S. (2011).
\newblock {\em Statistics for high-dimensional data: methods, theory and
  applications}.
\newblock Springer Science \& Business Media.

\bibitem[Chen, 2007]{chen2007empirical}
Chen, B. (2007).
\newblock An empirical comparison of methods for temporal disaggregation at the
  national accounts.
\newblock {\em Office of Directors Bureau of Economic Analysis, Washington,
  DC}.

\bibitem[Chen and Chen, 2008]{chen2008extended}
Chen, J. and Chen, Z. (2008).
\newblock Extended bayesian information criteria for model selection with large
  model spaces.
\newblock {\em Biometrika}, 95(3):759--771.

\bibitem[Cholette, 1983]{cholette1983adjusting}
Cholette, P.-A. (1983).
\newblock {\em Adjusting sub-annual series to yearly benchmarks}.
\newblock Statistics Canada, Methodology Branch, Time Series Research and
  Analysis.

\bibitem[Chow and Lin, 1971]{chow1971best}
Chow, G.~C. and Lin, A.-l. (1971).
\newblock Best linear unbiased interpolation, distribution, and extrapolation
  of time series by related series.
\newblock {\em The review of Economics and Statistics}, pages 372--375.

\bibitem[Ciammola et~al., 2005]{ciammola2005temporal}
Ciammola, A., Di~Palma, F., and Marini, M. (2005).
\newblock Temporal disaggregation techniques of time series by related series:
  A comparison by a monte carlo experiment.
\newblock Technical report, Eurostat, Working Paper.

\bibitem[Dagum and Cholette, 2006]{dagum2006benchmarking}
Dagum, E.~B. and Cholette, P.~A. (2006).
\newblock {\em Benchmarking, temporal distribution, and reconciliation methods
  for time series}, volume 186.
\newblock Springer Science \& Business Media.

\bibitem[Denton, 1971]{denton1971adjustment}
Denton, F.~T. (1971).
\newblock Adjustment of monthly or quarterly series to annual totals: an
  approach based on quadratic minimization.
\newblock {\em Journal of the American Statistical Association},
  66(333):99--102.

\bibitem[Di~Fonzo, 1990]{di1990estimation}
Di~Fonzo, T. (1990).
\newblock The estimation of m disaggregate time series when contemporaneous and
  temporal aggregates are known.
\newblock {\em The Review of Economics and Statistics}, pages 178--182.

\bibitem[Efron et~al., 2004]{efron2004least}
Efron, B., Hastie, T., Johnstone, I., Tibshirani, R., et~al. (2004).
\newblock Least angle regression.
\newblock {\em The Annals of statistics}, 32(2):407--499.

\bibitem[Eurostat, 2017]{euandun2017}
Eurostat (2017).
\newblock Handbook on cyclical composite indicators.
\newblock Technical report, Eurostat.

\bibitem[Eurostat, 2018]{ess2018}
Eurostat (2018).
\newblock {European Statistical System (ESS)} guidelines on temporal
  disaggregation, benchmarking and reconciliation.
\newblock Technical report, Eurostat.

\bibitem[Fan and Li, 2001]{fan2001variable}
Fan, J. and Li, R. (2001).
\newblock Variable selection via nonconcave penalized likelihood and its oracle
  properties.
\newblock {\em Journal of the American statistical Association},
  96(456):1348--1360.

\bibitem[Fernandez, 1981]{fernandez1981methodological}
Fernandez, R.~B. (1981).
\newblock A methodological note on the estimation of time series.
\newblock {\em The Review of Economics and Statistics}, 63(3):471--476.

\bibitem[Friel et~al., 2017]{friel2017investigation}
Friel, N., McKeone, J.~P., Oates, C.~J., and Pettitt, A.~N. (2017).
\newblock Investigation of the widely applicable bayesian information
  criterion.
\newblock {\em Statistics and Computing}, 27(3):833--844.

\bibitem[Ghysels et~al., 2004]{ghysels2004midas}
Ghysels, E., Santa-Clara, P., and Valkanov, R. (2004).
\newblock The midas touch: Mixed data sampling regression models.

\bibitem[Guerrero and Mart{\'\i}nez, 1995]{guerrero1995recursive}
Guerrero, V.~M. and Mart{\'\i}nez, J. (1995).
\newblock A recursive arima-based procedure for disaggregating a time series
  variable using concurrent data.
\newblock {\em Test}, 4(2):359--376.

\bibitem[Hastie and Efron, 2013]{hastie2013lars}
Hastie, T. and Efron, B. (2013).
\newblock lars: Least angle regression, lasso and forward stagewise.
\newblock {\em R package version}, 1(2).

\bibitem[Hastie et~al., 2007]{hastie2007forward}
Hastie, T., Taylor, J., Tibshirani, R., Walther, G., et~al. (2007).
\newblock Forward stagewise regression and the monotone lasso.
\newblock {\em Electronic Journal of Statistics}, 1:1--29.

\bibitem[Hastie et~al., 2015]{hastie2015statistical}
Hastie, T., Tibshirani, R., and Wainwright, M. (2015).
\newblock {\em Statistical learning with sparsity: the lasso and
  generalizations}.
\newblock CRC press.

\bibitem[Hecq et~al., 2021]{hecq2021hierarchical}
Hecq, A., Ternes, M., and Wilms, I. (2021).
\newblock Hierarchical regularizers for mixed-frequency vector autoregressions.

\bibitem[Hesterberg et~al., 2008]{hesterberg2008least}
Hesterberg, T., Choi, N.~H., Meier, L., Fraley, C., et~al. (2008).
\newblock Least angle and l1 penalized regression: A review.
\newblock {\em Statistics Surveys}, 2:61--93.

\bibitem[Jarmin, 2019]{jarmin2019evolving}
Jarmin, R.~S. (2019).
\newblock Evolving measurement for an evolving economy: Thoughts on 21st
  century us economic statistics.
\newblock {\em Journal of Economic Perspectives}, 33(1):165--84.

\bibitem[Johansen, 1988]{johansen1988statistical}
Johansen, S. (1988).
\newblock Statistical analysis of cointegration vectors.
\newblock {\em Journal of economic dynamics and control}, 12(2-3):231--254.

\bibitem[Koop et~al., 2019]{koop2018regional}
Koop, G., McIntyre, S., Mitchell, J., and Poon, A. (2019).
\newblock Regional output growth in the united kingdom: More timely and higher
  frequency estimates, 1970-2017.
\newblock {\em Journal of Applied Econometrics}.

\bibitem[Kuzin et~al., 2013]{kuzin2013pooling}
Kuzin, V., Marcellino, M., and Schumacher, C. (2013).
\newblock Pooling versus model selection for nowcasting gdp with many
  predictors: Empirical evidence for six industrialized countries.
\newblock {\em Journal of Applied Econometrics}, 28(3):392--411.

\bibitem[Labonne and Weale, 2020]{labonne2020temporal}
Labonne, P. and Weale, M. (2020).
\newblock Temporal disaggregation of overlapping noisy quarterly data:
  estimation of monthly output from uk value-added tax data.
\newblock {\em Journal of the Royal Statistical Society: Series A (Statistics
  in Society)}, 183(3):1211--1230.

\bibitem[Lisman and Sandee, 1964]{lisman1964derivation}
Lisman, J. H.~C. and Sandee, J. (1964).
\newblock Derivation of quarterly figures from annual data.
\newblock {\em Journal of the Royal Statistical Society: Series C (Applied
  Statistics)}, 13(2):87--90.

\bibitem[Litterman, 1983]{litterman1983random}
Litterman, R.~B. (1983).
\newblock A random walk, markov model for the distribution of time series.
\newblock {\em Journal of Business \& Economic Statistics}, 1(2):169--173.

\bibitem[Meinshausen and B{\"u}hlmann, 2010]{meinshausen2010stability}
Meinshausen, N. and B{\"u}hlmann, P. (2010).
\newblock Stability selection.
\newblock {\em Journal of the Royal Statistical Society: Series B (Statistical
  Methodology)}, 72(4):417--473.

\bibitem[Miralles et~al., 2003]{miralles2003performance}
Miralles, J. M.~P., Lladosa, L.-E.~V., and Vall{\'e}s, R.~E. (2003).
\newblock On the performance of the chow-lin procedure for quarterly
  interpolation of annual data: Some monte-carlo analysis.
\newblock {\em Spanish Economic Review}, 5(4):291--305.

\bibitem[Mitchell et~al., 2005]{mitchell2005indicator}
Mitchell, J., Smith, R.~J., Weale, M.~R., Wright, S., and Salazar, E.~L.
  (2005).
\newblock An indicator of monthly gdp and an early estimate of quarterly gdp
  growth.
\newblock {\em The Economic Journal}, 115(501):F108--F129.

\bibitem[ONS, 2021]{ONS2021}
ONS (2021).
\newblock Measuring monthly and quarterly uk gross domestic product during the
  coronavirus (covid-19) pandemic.
\newblock
  \url{https://www.ons.gov.uk/economy/grossdomesticproductgdp/articles/measuringmonthlyandquarterlyukgrossdomesticproductduringthecoronaviruscovid19pandemic/2021-11-11}.

\bibitem[Pav{\'\i}a-Miralles et~al., 2010]{pavia2010survey}
Pav{\'\i}a-Miralles, J.~M. et~al. (2010).
\newblock A survey of methods to interpolate, distribute and extra-polate time
  series.
\newblock {\em Journal of Service Science and Management}, 3(04):449.

\bibitem[Pfeffermann et~al., 2015]{pfeffermann2015methodological}
Pfeffermann, D., Eltinge, J.~L., Brown, L.~D., and Pfeffermann, D. (2015).
\newblock Methodological issues and challenges in the production of official
  statistics: 24th annual morris hansen lecture.
\newblock {\em Journal of Survey Statistics and Methodology}, 3(4):425--483.

\bibitem[Proietti, 2006]{proietti2006temporal}
Proietti, T. (2006).
\newblock Temporal disaggregation by state space methods: Dynamic regression
  methods revisited.
\newblock {\em The Econometrics Journal}, 9(3):357--372.

\bibitem[Proietti and Giovannelli, 2020]{proietti2020nowcasting}
Proietti, T. and Giovannelli, A. (2020).
\newblock Nowcasting monthly gdp with big data: A model averaging approach.
\newblock {\em Journal of the Royal Statistical Society: Series A (Statistics
  in Society)}.

\bibitem[Quilis, 2018]{quilis2018temporal}
Quilis, E.~M. (2018).
\newblock Temporal disaggregation of economic time series: The view from the
  trenches.
\newblock {\em Statistica Neerlandica}, 72(4):447--470.

\bibitem[Reid et~al., 2016]{reid2016study}
Reid, S., Tibshirani, R., and Friedman, J. (2016).
\newblock A study of error variance estimation in lasso regression.
\newblock {\em Statistica Sinica}, pages 35--67.

\bibitem[Sax, 2018]{sax2018seasonal}
Sax, C. (2018).
\newblock seasonal: R interface to x-13-arima-seats.
\newblock {\em R package version}, 1.

\bibitem[Sax and Steiner, 2013]{sax2013temporal}
Sax, C. and Steiner, P. (2013).
\newblock Temporal disaggregation of time series.
\newblock {\em The R Journal}, 5(2).

\bibitem[Schwarz, 1978]{schwarz1978estimating}
Schwarz, G. (1978).
\newblock Estimating the dimension of a model.
\newblock {\em The annals of statistics}, 6(2):461--464.

\bibitem[Tibshirani, 1996]{tibshirani1996regression}
Tibshirani, R. (1996).
\newblock Regression shrinkage and selection via the lasso.
\newblock {\em Journal of the Royal Statistical Society: Series B
  (Methodological)}, 58(1):267--288.

\bibitem[van~de Geer et~al., 2014]{vdgpostselection}
van~de Geer, S., Bühlmann, P., Ritov, Y., and Dezeure, R. (2014).
\newblock {On asymptotically optimal confidence regions and tests for
  high-dimensional models}.
\newblock {\em The Annals of Statistics}, 42(3):1166 -- 1202.

\bibitem[van~de Geer et~al., 2011]{vdg2011}
van~de Geer, S., Bühlmann, P., and Zhou, S. (2011).
\newblock {The adaptive and the thresholded Lasso for potentially misspecified
  models (and a lower bound for the Lasso)}.
\newblock {\em Electronic Journal of Statistics}, 5:688 -- 749.

\bibitem[Wei and Stram, 1990]{wei1990disaggregation}
Wei, W.~W. and Stram, D.~O. (1990).
\newblock Disaggregation of time series models.
\newblock {\em Journal of the Royal Statistical Society: Series B
  (Methodological)}, 52(3):453--467.

\bibitem[Yu and Bien, 2019]{yu2019estimating}
Yu, G. and Bien, J. (2019).
\newblock Estimating the error variance in a high-dimensional linear model.
\newblock {\em Biometrika}, 106(3):533--546.

\bibitem[Yuan and Lin, 2006]{yuan2006model}
Yuan, M. and Lin, Y. (2006).
\newblock Model selection and estimation in regression with grouped variables.
\newblock {\em Journal of the Royal Statistical Society: Series B (Statistical
  Methodology)}, 68(1):49--67.

\bibitem[Zheng et~al., 2014]{zheng2014high}
Zheng, Z., Fan, Y., and Lv, J. (2014).
\newblock High dimensional thresholded regression and shrinkage effect.
\newblock {\em Journal of the Royal Statistical Society: Series B: Statistical
  Methodology}, pages 627--649.

\bibitem[Zou, 2006]{zou2006adaptive}
Zou, H. (2006).
\newblock The adaptive lasso and its oracle properties.
\newblock {\em Journal of the American statistical association},
  101(476):1418--1429.

\end{thebibliography}
\end{document}